\documentclass[trackchanges, twocolumn]{aastex701} 

\usepackage{booktabs}
\usepackage{amsmath}

\usepackage{siunitx}
\DeclareSIUnit{\parsec}{pc}
\DeclareSIUnit{\jansky}{Jy}
\DeclareSIUnit{\steradian}{sr}

\usepackage{appendix}
\usepackage{placeins}
\usepackage{tabularx}   
\usepackage{makecell}   
\usepackage{array}      

\makeatletter
\def\NAT@open{\textup{(}\nobreak\hskip\z@skip}
\def\NAT@close{\unskip\textup{)}}
\makeatother

\defcitealias{HI4PI2016}{HI4PI Collaboration}

\newcommand{\citealtaliasy}[1]{\citetalias{#1}~\citeyear{#1}}

\begin{document}

\title{Constraining properties of dust formed in Wolf-Rayet binary WR~112 using mid-infrared and millimeter observations}

\correspondingauthor{Donglin Wu}
\email{donglin.wu@yale.edu}

\author[0009-0008-8224-3140]{Donglin Wu}
\affiliation{Department of Astronomy, Yale University, New Haven, CT 06511, USA}
\email{donglin.wu@yale.edu}

\author[0000-0002-2106-0403]{Yinuo Han}
\affiliation{Division of Geological and Planetary Sciences, California Institute of Technology, 1200 E. California Blvd., Pasadena, CA 91125, USA}
\email{yinuo@caltech.edu}

\author[0000-0002-8092-980X]{Peredur M. Williams}
\affiliation{Institute for Astronomy, University of Edinburgh, Royal Observatory Edinburgh, EH9 3HJ, UK}
\email{pmw@roe.ac.uk}

\author[0000-0002-8234-6747]{Takashi Onaka}
\affiliation{Department of Astronomy, Graduate School of Science, University of Tokyo, Tokyo 113-0033, Japan}
\email{onaka@astron.s.u-tokyo.ac.jp}

\author[0000-0002-7167-1819]{Joseph R. Callingham}
\affiliation{ASTRON, Netherlands Institute for Radio Astronomy, Oude Hoogeveensedijk 4, Dwingeloo, 7991\,PD, The Netherlands}
\affiliation{Anton Pannekoek Institute for Astronomy, University of Amsterdam, Science Park 904, 1098\,XH, Amsterdam, The Netherlands}
\email{callingham@astron.nl}

\author[0000-0001-9315-8437]{Matthew J. Hankins}
\affiliation{Arkansas Tech University, 215 West O Street, Russellville, Arkansas 72801, USA}
\email{mhankins1@atu.edu}

\author[0000-0001-7026-6291]{Peter Tuthill}
\affiliation{Sydney Institute for Astronomy, School of Physics, The University of Sydney, NSW 2006, Australia}
\email{peter.tuthill@sydney.edu.au}

\author[0000-0003-0778-0321]{Ryan M. Lau}
\affiliation{IPAC, Mail Code 100-22, Caltech, 1200 E. California Blvd., Pasadena, CA 91125, USA}
\email{ryanlau@ipac.caltech.edu}

\author[0000-0001-9754-2233]{Gerd Weigelt}
\affiliation{Max Planck Institute for Radio Astronomy, Auf dem H\"ugel 69, 53121 Bonn, Germany}
\email{weigelt@mpifr.de}

\author[orcid=0000-0003-2595-9114]{Benjamin~J.~S.~Pope}
\affiliation{School of Mathematical \& Physical Sciences, Macquarie University, 12 Wally's Walk, Macquarie Park, NSW 2113, Australia}
\email{benjamin.pope@mq.edu.au}

\author[0000-0002-2806-9339]{Noel D. Richardson}
\affiliation{Department of Physics and Astronomy, Embry-Riddle Aeronautical University, 
3700 Willow Creek Rd, 
Prescott, AZ 86301, USA}
\email{noel.richardson@erau.edu}

\author[0000-0002-4333-9755]{Anthony Moffat}
\affiliation{Département de physique, Université de Montréal, 1375 avenue Thérèse-Lavoie-Roux, Montréal, QC, H2V 0B3, Canada}
\email{anthony.f.j.moffat@umontreal.ca}

\begin{abstract}
    Binaries that host a carbon-rich Wolf-Rayet (WC) star and an OB-type companion can be copious dust producers. Yet the properties of dust, particularly the grain size distribution, in these systems remain uncertain. We present Band 6 observations of WR~112 by the Atacama Large Millimeter/submillimeter Array telescope (ALMA), which are the first millimeter observations of a WC binary system capable of resolving its dust emission. By combining ALMA observations with James Webb Space Telescope (JWST) images, we were able to analyze the spatially resolved spectral energy distribution (SED) of WR~112. We found that the SEDs are consistent with emissions from hydrogen-poor amorphous carbon grains. Notably, our results also suggest that the majority of grains in the system have radii below one micrometer, and the extended dust structures are dominated by nanometer-sized grains. Among four parameterizations of the grain radius distribution that we tested, a bimodal distribution, with abundant nanometer-sized grains and a secondary population of 0.1-micron grains, best reproduces the observed SED. This bimodal distribution helps to reconcile the previously conflicting grain size estimates reported for WR~112 and for other WC systems. We hypothesize that dust destruction mechanisms such as radiative torque disruption and radiative-driven sublimation are responsible for driving the system to the bimodal grain size distribution. 
\end{abstract}


\section{Introduction}
\label{sec:intro}
Carbon-rich Wolf-Rayet (WC) stars are a subclass of classical Wolf-Rayet (WR) stars. These massive stars have completely lost their hydrogen envelopes and drive powerful stellar winds with velocities of $\gtrsim$\SI{1000}{\kilo\meter\per\second} and mass loss rates of over $10^{-5}$ $M_{\odot}~\mathrm{yr^{-1}}$ \citep{Crowther2007, Sander2019}.
When WC stars are in binary orbits with another massive O- or B-type star, the collisions of stellar winds from the two stars can create a dense shock front. In systems with the appropriate orbital geometry (a counterexample being Gamma Velorum, see \citealt{Lau2020a}), carbonaceous dust can nucleate when the shock front cools, and is subsequently driven away by radiation pressure from the luminous stars \citep{Williams1990, Usov1991}. 

The emission from circumstellar dust in WC systems has been observed and modeled since the 1970s \citep{Gehrz1974, Williams1987}. 
\citet{Tuthill1999} showed with WR~104 that as dust forms while the WC star and its companion orbit each other, a spiral, or “pinwheel,” structure is naturally created.
More recently, many studies presented geometric models for the three-dimensional spatial morphology of dust structures around WC binaries that build upon this idea and demonstrate their consistency with various observations \citep{Callingham2019, Han2020, Lau2020b, Han2022, Lau2022, White2024}. Some of these models also account for the initial acceleration of dust due to the radiation pressure from the star before reaching the terminal speed \citep{Han2022, Monnier2025}. 

While the morphology of dust structures is well characterized, the properties of dust in these systems remain largely uncertain. In particular, there have been debates over the possible grain size distribution of dust formed in these systems, which nevertheless is important for understanding the contribution of these systems to the interstellar carbonaceous dust budget.
Using observations of 17 WR stars, \citet{Zubko1998} suggested the balance between grain destruction through kinetic sputtering and growth through collisions restricts the final grain radii of dust in WC systems to be around $10$--$20~\mathrm{nm}$. This small grain size is consistent with the radiative transfer modeling of dust in WR~104 by \citet{Harries2004}. \citet{Harries2000} studied the polarization of light from WR~137 and derived the maximum grain size in the system to be $\sim20~\mathrm{nm}$, supporting conclusions of \citet{Zubko1998}. On the other hand, \citet{Chiar2001} analyzed the \SI{6.2}{\micro\meter} absorption feature in the spectra of several WC stars and concluded that larger grains with radius $\sim$\SI{1}{\micro\meter} should be present. Infrared observations by \citet{Marchenko2002} suggest the characteristic grain size of dust in WR~112 is \SI{0.49}{\micro\meter}. 

Studies that consider more complex grain size distribution also yield mixed results. \citet{Yudin2001} presented \SI{2.18}{\micro\meter} speckle interferometric observations of WR~118 and found that two populations of grains with \SI{0.05}{\micro\meter} and \SI{0.38}{\micro\meter} respectively or a power law with index $-3$ ranging between radius \SI{5}{\nano\meter} and \SI{0.6}{\micro\meter} can both account for the data. 
Most recently, \citet{Lau2023} modeled the spectral energy distribution of WR~140 and their results suggest that \SI{1}{\nano\meter} grains and \SI{50}{\nano\meter} grains are both present in WR~140. \citet{Han2025} modeled the SED of three resolved dust features in Apep, a binary system that hosts two WR stars. They assumed a population of \SI{0.1}{\micro\meter} grains, which appears broadly consistent with the observations. Nevertheless, the lack of millimeter observations limited their ability to constrain the grain size distribution. 

The reason behind the disagreement among these studies is unknown. Some believe that it arises partly due to variability of dust properties across different WC systems, as different studies focus on different WC systems. The key to this debate might be systems like WR~112 where the presence of both small grains and large grains has been proposed \citep{Zubko1998, Chiar2001, Marchenko2002}. 

WR~112 is also an important system to understand the contribution of WC systems to interstellar dust, as it is one of the most significant WC dust producers. The estimated dust production rate of WR~112 is comparable to the dust production rate of all asymptotic giant branch (AGB) stars in the Small Magellanic Cloud \citep{Boyer2012, Lau2020b}. The grain size distribution in the system is crucial to further understanding this contribution. On one hand, the dust production rate is sensitive to the grain size distribution adopted, as we will discuss in Section \ref{sec:dust_rate}. On the other hand, knowing the grain size distribution is essential for assessing whether dust can survive and reach the interstellar medium, either by directly tracing its evolution as dust travels or by indirectly inferring it from radius-dependent destruction mechanisms (e.g., thermal sputtering \citealt{Tielens1994}). 

WR~112 is an ideal system for constructing models to constrain the dust properties: it is a continuous dust producer that hosts a relatively uniform dust structure and lacks excessive geometric complexity \citep{Lau2020b}. WR~112 has been observed by the James Webb Space Telescope (JWST) by \citet{Richardson2025}. This work presents Band 6 observations of WR~112 by the Atacama Large Millimeter/submillimeter Array telescope (ALMA), which are the first millimeter observations of a WC binary system capable of spatially resolving its dust emission. The infrared-to-millimeter observations combined are sensitive to the emissions of both small and large grains. This enables us to more accurately constrain the grain size distribution of dust in WR~112, and make an important step in resolving the conflicts in inferred grain sizes of the previous works.

This paper is organized as follows. In Section \ref{sec:obs}, we describe the ALMA observations we made and the archival observations to be used in this work. In Section \ref{sec:results}, we present the main results, including direct results from ALMA observations in Section \ref{sec:ALMA_results}, stellar properties of WR~112 in Section \ref{sec:stellarsed} and comprehensive dust emission models in Section \ref{sec:dust_sed}. Finally, in Section \ref{sec:disc}, we discuss the implications of our results, including possible mechanisms that can drive the system to the inferred grain size distribution.

\section{Observations}
\label{sec:obs}
\subsection{ALMA Observations and Data Reduction}
ALMA Band 6 ($\sim1.3~\mathrm{mm}$) observations of WR~112 were carried out using the 12-m array on three nights in September and October 2024, as Cycle 10 and 11 programs (2023.1.00999.S, 2024.1.00803.S, PI: Yinuo Han). Details of the observations, including number of antennas, on-source integration time, and minimum and maximum baselines are listed in Table~\ref{tbl:alma_obs}. We constructed a mosaic with four pointings around the star with a total field of view of over \SI{49}{\arcsecond} to capture any extended emission.

\begin{table}[h]
\centering
\caption{Summary of ALMA Band 6 ($\sim1.3~\mathrm{mm}$) observations of WR~112 in 2024.}
\label{tbl:alma_obs}
\begin{tabular}{lccc}
\toprule
Parameter &  Sep\ 26   & Sep\ 30   & Oct\ 27  \\
\midrule
No.\ antennas & 49 & 48 & 44  \\
On-source time (hrs) & 1.06 & 1.08 & 1.08  \\
Baseline (m) & 15.0--499.8 & 15.0--499.8 & 15.0--313.7  \\
Mean PWV (mm) & 1.4 & 0.8 & 1.2  \\
\bottomrule
\end{tabular}
\end{table}

We adopted a spectral setup with four 1.875-GHz wide spectral windows, centered at 217.899, 219.849, 230.899 and 232.899 GHz. The second spectral window contains the molecular transition $^{12}$CO(2-1) at 230.538 GHz. Two other transitions of CO isotopologues, $^{13}$CO(2-1) (220.399 GHz) and C$^{18}$O(2-1) (219.560 GHz), are in the third spectral window. 
Each spectral window is divided into 960 channels, with spectral resolution of 1.953 MHz, which corresponds to a velocity resolution of \SI{2.5}{}--\SI{2.7}{\kilo\metre\per\second}.

The raw data were processed using CASA version 6.6.1. We calibrated the data using the standard ALMA pipeline. The data from the three nights were then concatenated into a single dataset. The combined dataset was then imaged using the \texttt{tclean} task in CASA. Two continuum images, using channels without line emissions, were created with different weightings: one with uniform weighting, or Briggs weighting with a robustness of -2, to achieve the highest resolution possible, and the other with natural weighting, or Briggs weighting with a robustness of 2, and a $u-v$ taper of \SI{2}{\arcsecond} to increase sensitivity towards extended emissions. 
The synthesized beam of the former image has dimensions \SI{0.83}{\arcsecond}$\times$\SI{0.51}{\arcsecond} and a position angle of \SI{-89.8}{\degree}, while the latter has a \SI{2.54}{\arcsecond}$\times$\SI{2.21}{\arcsecond} beam with a position angle of \SI{-76.4}{\degree}. The RMS noise is roughly $0.06~\mathrm{mJy~beam}^{-1}$ and $0.03~\mathrm{mJy~beam}^{-1}$, respectively. 

The spectral cubes of the molecular transitions of the three CO isotopologues were imaged using natural weighting after continuum subtraction through \texttt{uvcontsub}. The synthesized beam on average has dimensions \SI{1.4}{\arcsecond}$\times$\SI{0.9}{\arcsecond} and a position angle of \SI{-77}{\degree}. The RMS noise is roughly $0.5$--$1.0~\mathrm{mJy~beam}^{-1}$ per channel. 
The deconvolution algorithm `multi-scale' was used for the images and cubes to reconstruct both point-like and extended emissions. Masks of the imaging were first roughly drawn by the \texttt{auto-multithresh} algorithm and then corrected to finer details by hand for each channel. All images and cubes presented and used in this study are primary beam corrected.

\subsection{Archival Observations}

\label{sec:archival_obs}
\begin{table*}
\centering
\caption{Summary of Archival Observations}
\label{tbl:other_obs}
\begin{tabular}{l c c}
\toprule
\textbf{Telescope} & \textbf{Wavelength} (\SI{}{\micro\meter}) & \textbf{Reference} \\
\midrule
JWST & 7.7, 15, 21 & \citet{Richardson2025} \\
Pan-STARRS & 0.481, 0.617, 0.752, 0.866, 0.962 & \citet{Chambers2016} \\
Skymapper & 0.502, 0.608, 0.773, 0.912 & \citet{Onken2024} \\
DENIS & 0.791, 1.228, 2.145  & \citet{Kimeswenger2004, Denis2005} \\
2Mass & 1.235, 1.662, 2.159 & \citealt{Cutri2003} \\
WISE & 22 & \citet{Wright2010} \\ 
MSX & 4.25, 4.29, 8.28, 12.13, 14.64, 21.34 & \citet{Egan2003} \\ 
IRAS & 12, 25, 60  & \citet{IRAS1988} \\ 
AKARI & 65, 90, 140 & \citet{Akari2020} \\ 
ESO/UKIRT & 1.228, 1.651, 2.216, 3.771, 4.772, 8.38, 9.69, 12.89, 19 & \citet{Williams1987} \\
\bottomrule
\end{tabular}
\end{table*}

JWST observations of WR~112 were able to resolve the extended spiral dust arcs surrounding the star \citep{Richardson2025}. The observations were done using the Mid-Infrared Instrument (MIRI) and filters F770W, F1500W, and F2100W, which correspond to wavelengths of \SI{7.7}{\micro\meter}, \SI{15}{\micro\meter} and \SI{21}{\micro\meter}. The JWST images have obvious diffraction spikes from the point spread function (PSF) of MIRI. To model the extended dust arcs, we obtained the background-subtracted and PSF-subtracted JWST images of WR~112 from \citet{Richardson2025} that use the PSF subtraction method described by \citet{Lieb2025}.

There have also been numerous spatially unresolved observations of WR~112. This includes the Panoramic Survey Telescope and Rapid Response System (Pan-STARRS), Skymapper Southern Survey, Deep Near-Infrared Survey of the Southern Sky (DENIS), Two Micron All Sky Survey (2Mass), Wide-field Infrared Survey Explorer (WISE), Midcourse Space Experiment (MSX), the Infrared Astronomical Satellite (IRAS), European Southern Observatory (ESO), and United Kingdom Infrared Telescope (UKIRT).
We excluded the following filters due to low quality and low signal-to-noise ratio: 3.4, 4.6, \SI{12}{\micro\meter} WISE observations, and \SI{100}{\micro\meter} IRAS observations. The filters that are included are summarized in Table~\ref{tbl:other_obs}. 

The flux calibration of DENIS and ESO/UKIRT is derived from \citet{Fouque2000} and \citet{vanderBliek1996}, respectively. The rest is derived from \citet{Rodrigo2012}. Uncertainties are not given for the following filters: WISE \SI{22}{\micro\meter}, Pan-STARRS \SI{0.866}{\micro\meter} and all ESO and UKIRT observations. We adopted an uncertainty of 5.7$\%$ recommended by the documentation for WISE. The standard deviation of pixels in the PSF for Pan-STARRS \SI{0.866}{\micro\meter} filter is over 1 magnitude, indicating potential issues with the measurement. We chose that to be the uncertainty in the observed magnitude. For ESO/UKIRT observations, we assumed an uncertainty of 0.03 magnitude for the 1.228-\SI{3.771}{\micro\meter} filters, 0.05 magnitude for the \SI{4.772}{\micro\meter}, 0.1 magnitude for the 8.38-\SI{12.89}{\micro\meter} filters and 0.2 magnitude for the \SI{19}{\micro\meter} filters, based on discussions of \citet{vanderBliek1996}. 
To account for additional uncertainties from background contamination by dust and stars, confusion with foreground stars and variability in the flux, we added $10\%$ uncertainties to all data points.

\section{Results}
\label{sec:results}

\subsection{Dust and gas structures in the millimeter observations}
\label{sec:ALMA_results}

\subsubsection{Continuum emission}
\begin{figure*}
    \centering
    \includegraphics[width=0.8\linewidth]{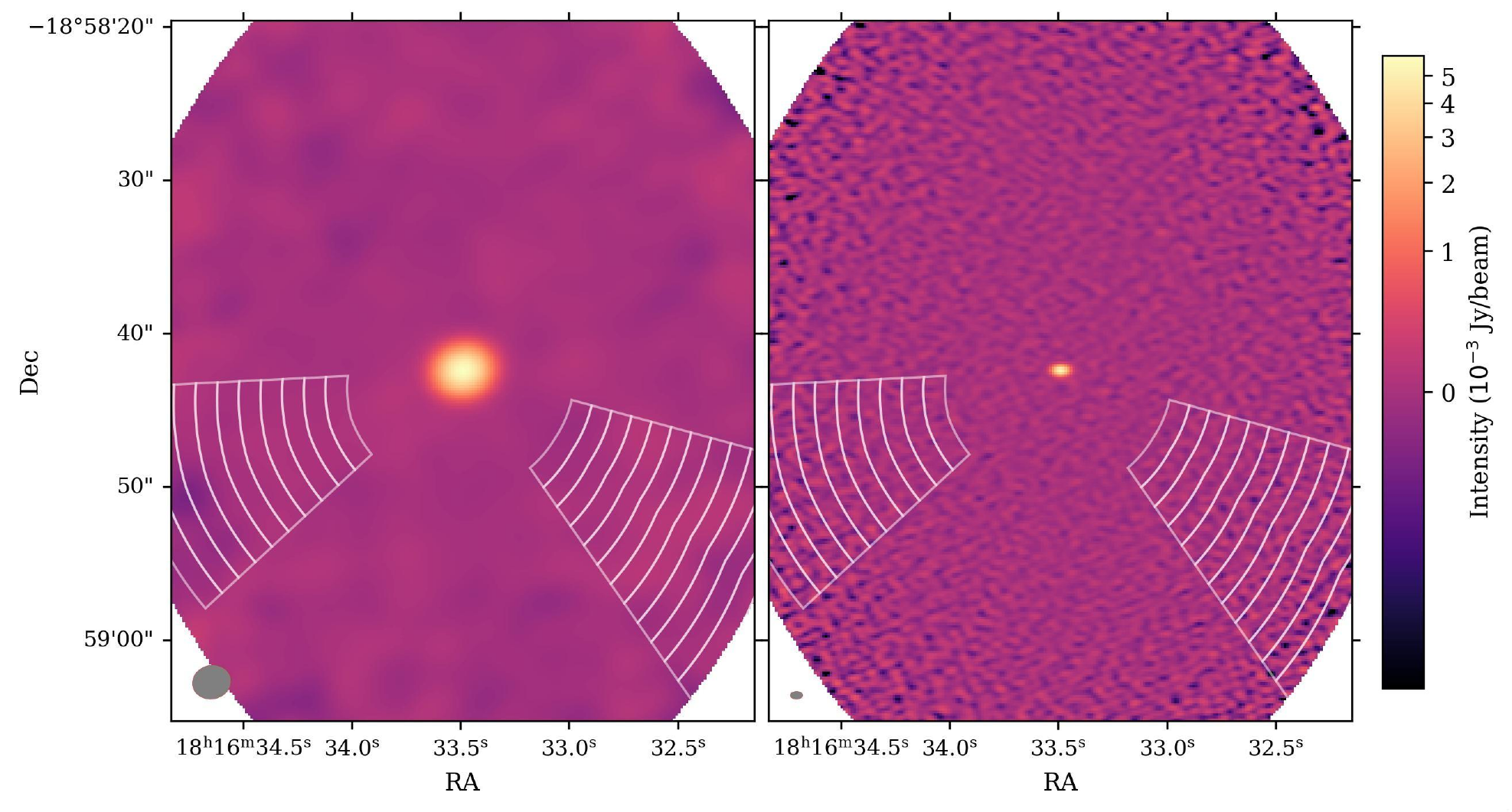}
    \caption{ALMA continuum images of WR~112 created using natural weighting and a $u-v$ taper of \SI{2}{\arcsecond} (left) and created using uniform weighting (right). Both panels share a common color scale. The white contours indicate the apertures used for the SED described in Section \ref{sec:dust_sed_aperture}, corresponding to the locations of the dust arcs seen in the JWST images. The light gray ellipse in the left bottom corner of each panel indicates the beam size.}
    \label{fig:ALMA_continuum}
\end{figure*}

The continuum image with the higher spatial resolution, created using uniform weighting, only shows one unresolved central emission at the location of WR~112, with a size and shape consistent with the synthesized beam, as shown in Figure~\ref{fig:ALMA_continuum}. No substructure is resolved with the \SI{0.83}{\arcsecond}$\times$\SI{0.51}{\arcsecond} beam. This suggests any substructure, if it exists, would have a size less than \SI{1700}{\astronomicalunit}, assuming a distance of \SI{3.39}{\kilo\parsec} to WR~112 \citep{Lau2020b}. In contrast, the outer dust arcs resolved by JWST images are each separated by $\sim5200$ au. 

Similarly, only a point-like emission at the location of the star is shown in the continuum image created using natural weighting and a $u-v$ taper of \SI{2}{\arcsecond}, as shown in Figure~\ref{fig:ALMA_continuum}. The total flux from the point-like emission is $F_\nu=8.5\pm$\SI{0.1}{\milli\jansky}. 
No extended structure is detected beyond the $3\sigma$ level. The flux from the dust structures near WR~112 at $\sim$\SI{1.3}{\milli\meter} must be below the level of \SI{1.53e6}{\jansky\per\steradian}. 

\subsubsection{CO emission}
The channel maps of emissions from CO isotopologues are shown in Appendix \ref{Appendix:channel_map}.
The $^{12}$CO(2-1) spectral cube shows emission at $v_{\text{LSR}} = 25-$\SI{100}{\kilo\meter\per\second}. We also detect $^{13}$CO(2-1) and C$^{18}$O(2-1) emission at $v_{\text{LSR}} = 25-$\SI{50}{\kilo\meter\per\second}. Outside these velocity channels, there is no significant emission within the field of view, even though stellar winds moving directly toward us would, in principle, be observable in the $^{12}$CO(2–1) line given the spectral window width and terminal wind speed of $1230\pm$\SI{260}{\kilo\meter\per\second} \citep{Lau2020b}. 

The emissions from the three CO isotopologues have similar morphologies, as shown in Figures \ref{fig:channel_3iso_set1} and \ref{fig:channel_3iso_set2} in Appendix \ref{Appendix:channel_map}. The spectra of the three isotpologues all peak at $\sim$\SI{35}{\kilo\meter\per\second}, as shown in Figure~\ref{fig:spectrum_CO}. The $^{12}$CO(2-1) and $^{13}$CO(2-1) emissions, which trace gas with low-to-medium density, are extended in several channels, spanning tens of arcseconds, and show filamentary structures. Strong and more localized $^{13}$CO(2-1) and C$^{18}$O(2-1) emissions are present directly north of WR~112.

Given the terminal wind speed of $1230\pm$\SI{260}{\kilo\meter\per\second}, if the CO emissions are tracing the stellar wind, they would be tracing winds traveling almost within the plane of the sky. However, the morphology of CO emissions observed does not exhibit any symmetry or ring-like structure, which would be expected if they are from spherically symmetric winds. A large-scale CO survey (\citealt{Dame2001}, \SI{8.5}{\arcminute} FWHM) and neutral hydrogen gas survey (\citealtaliasy{HI4PI2016}, \SI{8.1}{\arcminute} beam), both with a velocity resolution of \SI{1.3}{\kilo\meter\per\second}, show a peak at $v_{\text{LSR}} \approx$ \SI{37.5}{\kilo\meter\per\second} at the location of the star, as shown in Figure \ref{fig:spectrum_CO} in Appendix \ref{Appendix:COspectra}. Given the close match of the radial velocity between the CO detected by ALMA and that in the interstellar medium, and the lack of a correspondence between the CO morphology that we observe and the dust structures seen in the infrared and the structural symmetries expected, the CO emissions we observe are likely from interstellar gas rather than stellar winds of WR~112. However, whether the observed CO structures are in the environment of WR~112 or aligned by chance along the line of sight only in projection is unknown from current data.

\subsection{Stellar SED of the WR~112 system}
\label{sec:stellarsed}
Despite the lack of resolved dust structures in the continuum images, the ALMA observations provide upper limits on the dust surface brightness at $\sim1.3~\mathrm{mm}$. The ALMA upper limits can help constrain the grain properties of dust in WR~112, particularly the abundance of large millimeter-sized grains that emit efficiently at long wavelengths. When combined with JWST observations, the distribution of smaller grains can also be constrained. 
Nevertheless, models constraining the grain properties require knowledge of the properties of WR~112 and its companion which are not well characterized in previous works. In particular, the spectral luminosities are important, as photons from the stars are primary energy sources of dust and determine the temperatures of grains and consequently their emissions (see more in Section \ref{sec:tempgrid}).  
In this section, we attempt to constrain the stellar spectra of WR~112 and its companion by fitting stellar models to the stellar SED compiled from the archival observations described in Section \ref{sec:archival_obs}. 

Since the observations are spatially unresolved, emissions from both stars as well as circumstellar dust are captured by the observations. To constrain the stellar spectra of the stars, we only use observations with wavelength $\lambda <$ \SI{0.9}{\micro\meter}, where stellar light should dominate over dust thermal emissions. 

To compare observations with stellar models, the high extinction of WR~112 has to be corrected. We used the Milky Way extinction curve \texttt{G23} from \citet{Gordon2024}. 
\citet{Roche1984} examined extinction of six WC stars and found the extinction in the visible to be well correlated with the optical depth of WR~112 at \SI{9.7}{\micro\meter}, with $A_V/\tau_{9.7}= 18.5 \pm 1.5$. They found that the absorption is well represented by extinction curve derived from $\mu$ Cepheid, from which they derived the optical depth of WR~112 to be $\tau_{9.7} = 0.65$. A subsequent study by \citet{vanderHucht1996} found a similar optical depth ($\tau_{9.7} = 0.64$) for WR~112. Using optical depths and extinction of a few WC stars, \citet{vanderHucht2001} derived $A_V/\tau_{9.7}=17.2$, in agreement with \citet{Roche1984}. This allows extinction of WR~112 to be estimated, with $A_V \approx 11.0$. To account for potential uncertainties in extinction, we assume an uncertainty in $A_V$ of $0.3$. We added such uncertainties of extinction correction to the SED data points. 

\citet{Roche1984} deduced that the extinction of WR~112 and other WC stars in the sample is mostly interstellar. \citet{Chiar2001} also argued that the circumstellar extinction for WC systems is not significant ($A_V \sim 1$ mag), since the V-band variability of WR 104 is roughly 1 mag \citep{Crowther1997}. The presence of CO$_2$ ice in the ISO spectrum of WR~112 suggests that a dense molecular cloud is in the line of sight \citep{Schutte1998}. If the dense molecular cloud is primarily responsible for the extinction, an extinction law with larger $R_V$ than that of the interstellar medium ($R_V = 3.1$) should be adopted. Therefore, $R_V$ is set to be a free parameter when we fit the stellar models to the observed stellar SED, with possible values between $3.1$ and $4.1$, in increments of $0.1$. 
A different value of $R_V$ would not affect the measured $A_V$ value, since $A_V$ is obtained from the optical depth at \SI{9.7}{\micro\meter}, independently from color excess or $R_V$. 

The Potsdam Wolf-Rayet Models (PoWR, \citealt{Grafener2002}) computed and published stellar spectra for WC \citep{Sander2012, Sander2019} and OB stars \citep{HainichR2019}. We fit these model spectra to the observed SED. The stellar spectra are given in model grids labeled ``MW WC'' and ``GAL-OB-Vd3'', respectively, both assuming Galactic metallicity. 
The OB grids are two-dimensional, parameterized by the surface gravity $g_{\text{OB}}$ and temperature $T_{\text{OB}}$. Both the luminosity and the mass of the OB star are determined by these two parameters. 
The WC model grids are given as two-dimensional grids, parameterized by the temperature $T_{\text{WC}}$ and ``transformed" radius $R_t$. Unlike the OB stars, the luminosity of the WC star, $L_{\text{WC}}$, is independent of these two parameters and fixed at $2\times10^5$ $L_{\odot}$ for the model spectra. According to the documentation of PoWR\footnote{https://www.astro.physik.uni-potsdam.de/~wrh/\\PoWR/powrgrid1.php} and \citealt{Grafener2002}, the stellar spectra of WC stars with different luminosities can be obtained by scaling the spectra proportional to $L_{\text{WC}}$. Properties of the stars have to be scaled correspondingly, with the radius being proportional to $L_{\text{WC}}^{1/2}$ and the mass loss rate being proportional to $L_{\text{WC}}^{3/4}$. In our fits, we limit the range of $L_{\text{WC}}$ to be between $1\times 10^5$ and $1\times 10^6$ $L_{\odot}$, which is the observed range of luminosities for WC stars \citep{Sander2019}.

The stellar spectra observed for WR~112 and its companion are thus specified by six free parameters: a five-dimensional parameter grid, parameterized by $R_V$, $g_{\text{OB}}$, $T_{\text{OB}}$, $T_{\text{WC}}$, and a continuous free parameter $L_{\text{WC}}$ at every point in the grid. Ideally, we can derive the stellar spectra of WR~112 and its companion by fitting the sum of the spectra of the WC star and OB star to the observed stellar SED and finding the best point in the parameter grid and the corresponding best-fit $L_{\text{WC}}$. However, we are not able to do so, since we only have 8 observations with $\lambda <$\SI{0.9}{\micro\meter}, and there are degeneracies between the six free parameters (e.g. temperature or luminosity of the OB and luminosity of the WC star). 

Consequently, we placed additional physical constraints on the properties of the stars. WR~112 has been identified as a WC8 or WC9 stars \citep{vanderHucht2001, Williams2015}. Based on the observed range of mass loss rates of WC8 and WC9 stars \citep{Sander2019}, we only included WC stars whose mass loss rate is greater than $1.0 \times 10^{-5}$ $M_{\odot}~\mathrm{yr^{-1}}$. 
In addition, \citet{Lau2020b} estimated, from the geometry of the dust structures, that the ratio of the momentum of stellar wind from the OB star to that from WR~112 is $\eta \sim 0.13$. Since this estimate is dependent on the model of \citet{Canto1996}, we assume a conservative uncertainty in $\eta$ of one order of magnitude. We then included $\log \eta$ as an additional data point to be fitted. 

\begin{figure}[!ht]
    \centering
    \includegraphics[width=0.99\linewidth]{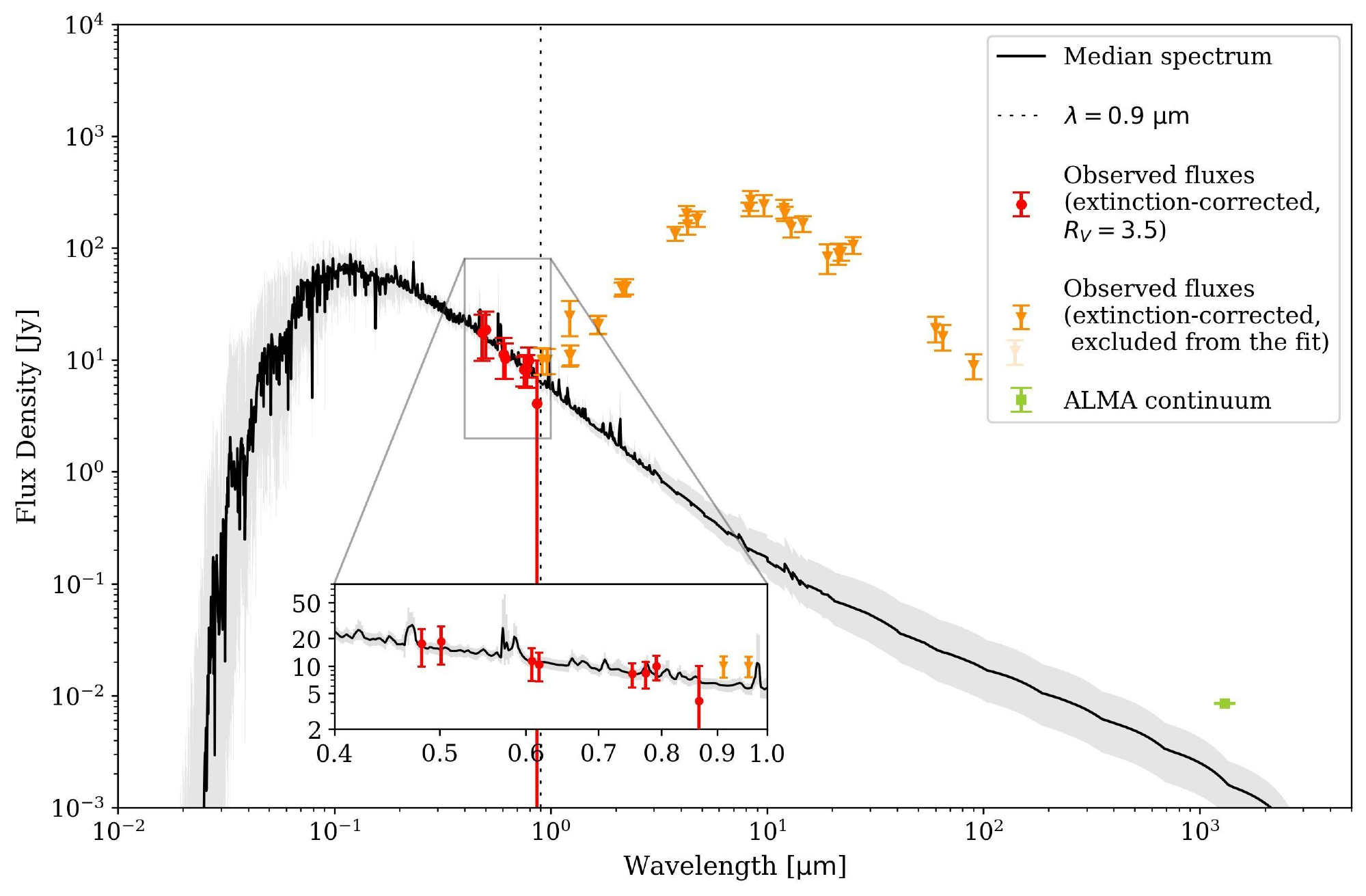}
    \caption{The stellar SED of WR~112 and its companion. The red and orange data points are from the spatially unresolved observations described in Section \ref{sec:archival_obs} and Table \ref{tbl:other_obs}. They are corrected for extinction with best-fit $R_V = 3.5$. While the red points are included in the fit, the orange points are excluded as they predominantly reflect dust rather than stellar emission. The green point at \SI{1.3}{\milli\meter} is the flux from the point source in ALMA continuum image. The black line indicates the median total stellar spectrum of WR~112 and its companion among the acceptable models. The gray region indicates the 16th and 84th percentile spectra. The black dotted line indicates $\lambda = $ \SI{0.9}{\micro\meter}. }
    \label{fig:stellar_spec}
\end{figure}

We then optimized $L_{\text{WC}}$ for every point in the five-dimensional parameter grid that satisfies the physical constraint, and computed the $\chi^2$. With a degree of freedom of 3 (8 photometric data points plus one constraint on $\log \eta$ and 6 free parameters), models with $\chi^2 < 6.25$ fall within the 90$\%$ confidence interval, and are considered acceptable. We computed the median total stellar spectrum of the two stars across all acceptable models, along with the corresponding 16th and 84th percentile spectra. These spectra are shown in in Figure~\ref{fig:stellar_spec}.

Despite the limited data points, we are able to constrain the stellar spectra of the two stars in the optical relatively well. The properties of the stars, given by the PoWR models, can also be computed among the acceptable models. The mass loss rate of WR~112 is $2.69^{+2.40}_{-1.33} \times 10^{-5}$ $M_{\odot}~\mathrm{yr^{-1}}$, while that of its companion is $7.94^{+17.18}_{-5.95} \times 10^{-7}$ $M_{\odot}~\mathrm{yr^{-1}}$. 
The luminosity of WR~112, corrected for terminal wind speed \citep{Sander2019}, is $5.04^{+4.13}_{-3.13} \times 10^5 ~L_{\odot}$. The luminosity of its OB companion is $4.57^{+7.73}_{-2.91} \times 10^5 ~L_{\odot}$. The mass of the companion is $42.4^{+43.9}_{-18.2} ~M_{\odot}$. According to the mass-luminosity relation of WR stars \citep{Langer1989}, the mass of WR~112 should be $19.8^{+9.5}_{-8.4} ~M_{\odot}$, if we assume a helium mass fraction of $55\%$ \citep{Sander2012}. This is consistent with observations, as several WC8 and WC9 stars have estimated masses of around $20~M_{\odot}$ \citep{Sander2019}.

Our estimate for the wind momentum ratio ($\eta=0.049^{+0.182}_{-0.041}$) is consistent with the estimate of \citet{Lau2020b} at $1\sigma$ level. On the other hand, our estimate for the mass loss rate of WR~112 is a factor of 4 lower than the estimates from radio emission ($1.1 \times 10^{-4}~M_{\odot}~\mathrm{yr^{-1}}$, \citealt{Monnier2002, Lau2020b}). Their estimate is subject to large uncertainties due to possible contamination from non-thermal synchrotron radiation. The disagreement is thus inconclusive.


\subsection{Dust SED Modeling Approaches}
\label{sec:dust_sed}

In this section, we constrain the grain size distribution of dust in WR~112 through modeling of the dust SED. Section~\ref{sec:dust_sed_aperture} describes the construction of a spatially resolved dust SED using JWST and ALMA observations. Sections~\ref{sec:Tfree}–\ref{sec:stochastic} present dust emission models of increasing complexity and the corresponding grain size distributions inferred by fitting these models to the spatially resolved SED. Finally, in Section~\ref{sec:dust_stellar_SED}, we model the integrated emission from dust structures in an attempt to infer the grain size distribution of unresolved inner dust from the spatially unresolved data introduced in Section~\ref{sec:archival_obs}.

\subsubsection{Dust SED for the extended dust arcs}
\label{sec:dust_sed_aperture}
According to the geometric model (e.g. \citealt{Han2022}), the dust around WR stars is in three-dimensional structures that spiral outward. The dust arcs observed in the JWST images are projections of these structures onto the plane of the sky. The bright dust arcs in the images are regions where the dust column density is high. The regions between these bright edges are regions with low dust column density, which we refer to as cavities in this paper.

To model the properties of these extended dust arcs, we derived and analyzed the SED for each arc. Each SED includes three photometric data points, which are derived from the \SI{7.7}{\micro\meter}, \SI{15}{\micro\meter} and \SI{21}{\micro\meter} JWST images. Since the dust arcs are not detected by ALMA, each SED also includes an upper limit for the flux from the dust structures at \SI{1.3}{\milli\meter}.

\begin{figure*}[!ht]
    \centering
    \includegraphics[width=0.8\linewidth]{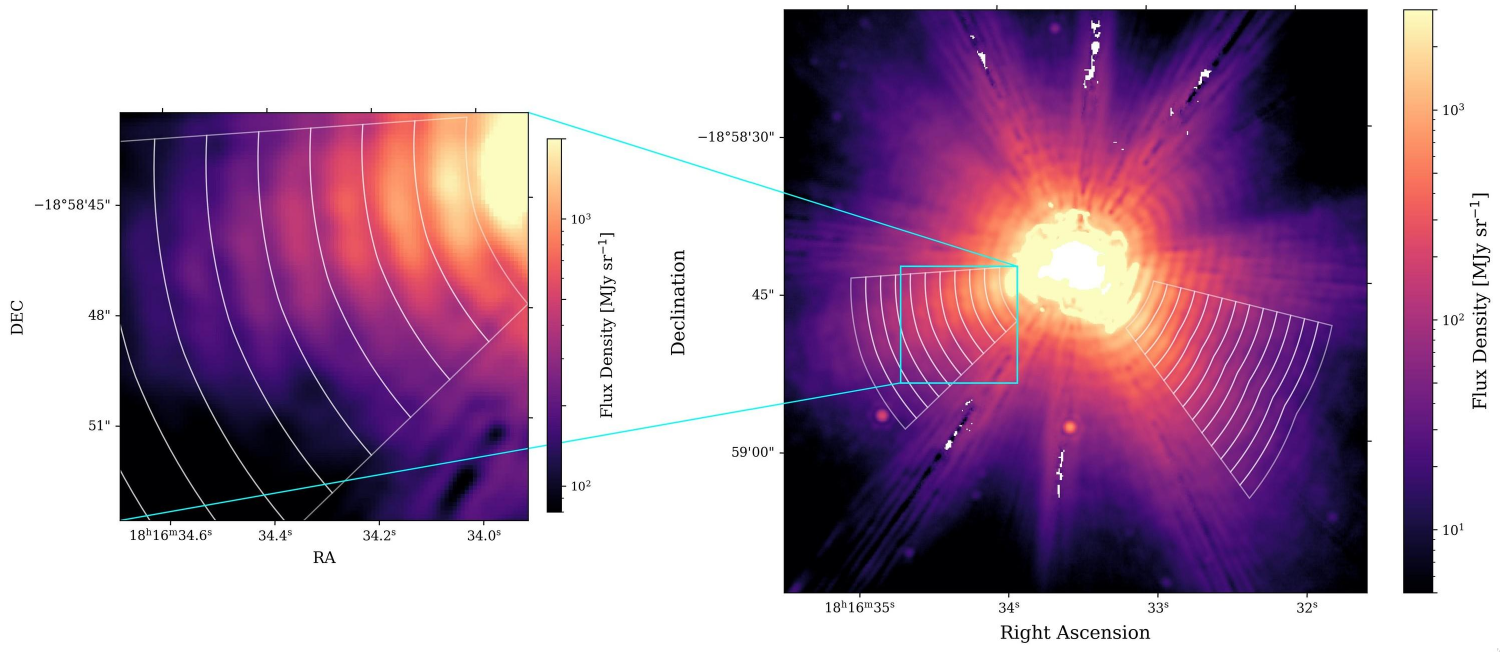}
    \caption{JWST \SI{21}{\micro\meter} image (right) with a zoomed-in view of the region outlined by the cyan box (left). White contours mark the apertures used for the SED described in Section \ref{sec:dust_sed_aperture}, 13 of which are in sector 1 and 10 are in sector 2.}
    \label{fig:JWST_apertures}
\end{figure*}

While PSF subtraction was performed on the JWST images, artifacts due to the PSF and the diffraction spikes remain prominent. To avoid contamination, we chose two sector regions on the image where there is no diffraction spike or obvious PSF artifact. The sectors are defined by position angles: $216^{\circ} \leq \mathrm{PA} \leq 256^{\circ}$ for the first and $94^{\circ} \leq \mathrm{PA} \leq 134^{\circ}$ for the second.
We then created apertures with shapes similar to annulus sectors, as shown in Figure~\ref{fig:JWST_apertures}, in each of these two sectors. However, since the projected dust arcs are not perfect arcs or spirals, the apertures have to be computed from the \SI{21}{\micro\meter} image in the following way. 

For each sector, the radial brightness profile is computed for three positions angles ($\mathrm{PA} = 216^{\circ}, 235^{\circ}, 256^{\circ}$ for the first, $\mathrm{PA} = 94^{\circ}, 114^{\circ}, 134^{\circ}$ for the second). The troughs in the brightness profile correspond to the cavities between dust arcs. The distance of the cavities from the star is found to be well-described by linear relations for the position angles. This suggests that the dust structures are expanding almost uniformly. The locations of the cavities at other position angles are computed by interpolating the predicted locations of the cavities at the three selected position angles. The aperture for each dust arc is then given by the region enclosed by the locations of the neighboring cavities and the radial boundaries of the sector. 
There are 13 apertures in the first sector and 10 apertures in the second. The further inner regions are avoided because JWST becomes saturated, while regions further away from the star are avoided when the distinction between cavities and dust arcs is not obvious. 

To avoid contamination from bright background sources, we masked the background sources for the PSF-subtracted and background-subtracted images. This is particularly important for \SI{7.7}{\micro\meter} image in which there are many visible background stars. The masks were computed by repeatedly applying the $3\sigma$ level mask estimated from \texttt{photutils.background}. The masked pixels in each aperture are then filled with the average of the rest of the pixels in the aperture. Then, the JWST photometric data points for each dust arc are obtained by integrating the total flux in each aperture for each image. The upper limit at \SI{1.3}{\milli\meter} for each aperture is given by the $3\sigma$ level of emission of the ALMA continuum image multiplied by the square root of the number of pixels in the aperture.

The geometric model of WR~112 suggests the deprojected distance from the star to each aperture can be estimated as the distance (to the midpoint of the aperture along the radial direction) in the plane of sky at $\mathrm{PA} = 216^{\circ}$ for the first sector and $\mathrm{PA} = 134^{\circ}$ for the second \citep{Lau2020b}.

The uncertainties of the JWST photometric points due to read and Poisson noise of the instrument are estimated to be much less than $1\%$. The uncertainties caused by background subtraction and masking of background stars is also found to be less than $1\%$.
According to the documentation, the absolute flux calibration of MIRI imaging introduces a few percent ($\sim 5 \%$) of errors. Although PSF subtraction is carried out and diffraction spikes are avoided, there could still be significant uncertainties due to the PSF, given the brightness and spatial extent of the circumstellar dust. There are perhaps also other less well-characterized sources of uncertainties. To account for these uncertainties conservatively, we assume a $20 \%$ uncertainty for every JWST photometric data point. 

\subsubsection{Temperatures as Free Parameters}
\label{sec:Tfree}
As a first step toward constraining the dust grain properties, we model the dust SED using the simplest approach, in which the temperature of each dust shell is treated as an independent free parameter. 

The emission model considers only the thermal emission from dust, which is expected to dominate over scattered light at mid-infrared wavelengths in WR~112, given the limited amount of mid-infrared photons from the hot WR and OB stars. This thermal component of the emission can be described using a modified blackbody spectrum. If we assume that all the dust grains around WR~112 have the same grain radius $a$ and are at radiative equilibrium (RE) with the stellar radiation field, the flux density from the thermal emission of dust at wavelength $\lambda$ in the $i^{\text{th}}$ aperture is given by:
\begin{equation}
\label{eq:Fnu1}
    F_\nu^{(i)}(\lambda, a) = \Omega_{\text{eff}}^{(i)} B_\nu(T_d^{(i)}, \lambda) Q_{\text{abs}}(a, \lambda),
\end{equation}
where $\Omega_{\text{eff}}^{(i)}$ is the total effective solid angle subtended by the dust in the $i^{\text{th}}$ aperture, $B_\nu$ is the Planck function, $T_d^{(i)}$ is the temperature of the dust in the $i^{\text{th}}$ aperture, and $Q_{\text{abs}}(a, \lambda)$ is the absorption efficiency factor, or the ratio of absorption cross section to actual cross section, for dust with radius $a$ at wavelength $\lambda$. Technically, the emission efficiency factor $Q_{\text{em}}(a, \lambda)$ should be used here instead of $Q_{\text{abs}}(a, \lambda)$. However, according to Kirchhoff's Law, $Q_{\text{abs}}(a, \lambda) = Q_{\text{em}}(a, \lambda)$ for a body in equilibrium. 

Since the stellar winds of WC stars are carbon-rich but hydrogen-poor \citep{Sander2012}, it is commonly believed that the dust around WC stars consists of amorphous carbon (\citealt{Cherchneff2000, Lau2020b, Lau2023}, discussed in detail in Section \ref{sec:disc_chem}). We adopted optical constants of ACAR-type amorphous carbon prepared by \citet{Colangeli1995} in laboratory and measured by \citet{Zubko1996}. The ACAR-type grains are adopted over other laboratory analogues of amorphous carbon grain because they were created in hydrogen-poor (argon) environment, which is what we expected the stellar winds of WR stars to be. A more comprehensive discussion of the grain chemistry is presented in Section \ref{sec:disc_chem}. 
The absorption efficiency factor $Q_{\text{abs}}$ is then calculated using Mie scattering code from \citet{Dullemond2012}. 

Since the optically thin dust structure is expanding uniformly and the apertures are drawn between fixed position angles, the dust in one aperture will expand into the dust in the next aperture. Therefore, the effective solid angle of dust should be the same for all the apertures in each sector. We denote the effective solid angle of dust in each aperture in the first sector ($216^{\circ} \leq \mathrm{PA} \leq 256^{\circ}$) by $\Omega_{\text{eff}}^{\text{I}}$ and that in the second sector ($94^{\circ} \leq \mathrm{PA} \leq 134^{\circ}$) by $\Omega_{\text{eff}}^{\text{II}}$. 

To constrain the parameters that are shared across apertures in a robust way, we performed a joint Markov chain Monte Carlo (MCMC) fit to the SEDs from the apertures using the \texttt{emcee} package \citep{ForemanMackey2013}. There are 26 free parameters in total, including three parameters shared over apertures $\Omega_{\text{eff}}^{\text{I}}, \Omega_{\text{eff}}^{\text{II}}, a$ as well as temperatures that are specific to each aperture, $T_d^{(1)}, \cdots ,T_d^{(23)}$. For the grain radius $a$, the bounds are chosen to be $-9.0 \leq \log [a/\textrm{meter}] \leq -4.0$, since it is unrealistic for grains to be smaller than 1 nanometer. 

\begin{figure}[ht]
    \centering
    \includegraphics[width=0.9\linewidth]{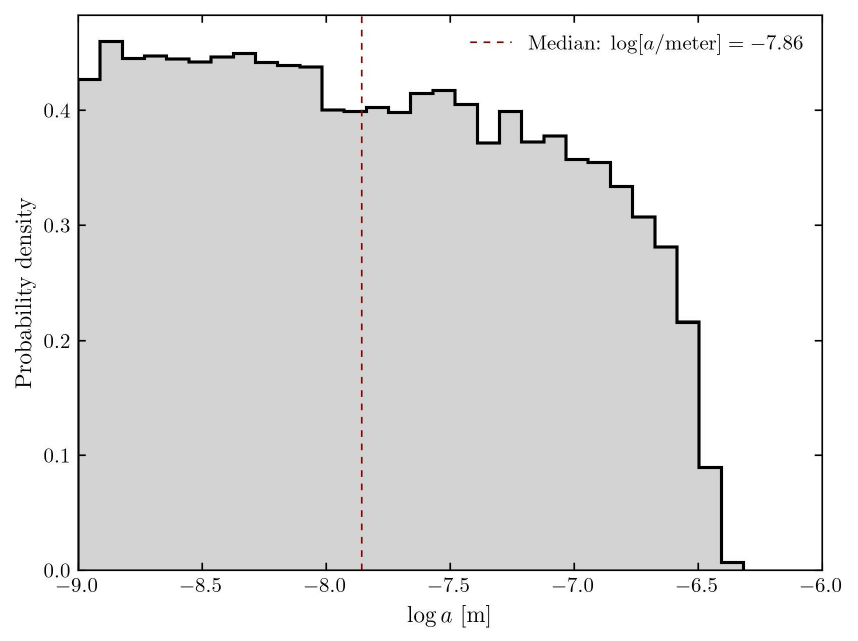}
    \caption{Probability density of grain radius $a$ from the MCMC joint fit of SEDs from the 23 apertures. The dark red dashed line indicates the median of the distribution, $\log [a/\textrm{meter}] = -7.86$.}
    \label{fig:grain_Tfree}
\end{figure}

The model SEDs predicted by the model using medians of the posterior distribution of the MCMC is shown in Figure~\ref{fig:SED_Tfree}.
Figure~\ref{fig:grain_Tfree} shows the probability density of grain radius $a$ from the MCMC joint fit. We see that the ALMA upper limits and the JWST data points are able to place an upper limit on the grain size. Most of the grains have to be smaller than $a \approx $ \SI{0.5}{\micro\meter}, as the probability density of grain radius drops to zero for $\log [a/\textrm{meter}] \geq -6.3$. The median is $\log [a/\textrm{meter}] = -7.86$ or $a = $ \SI{13.8}{\nano\meter}. However, the probability density remains roughly constant for $\log [a/\textrm{meter}] \leq -7.0$. This suggests the lower limit of grain radius is unconstrained. Because of that, the effective solid angles for the two sectors are not strongly constrained, ranging over two orders of magnitudes. This contributes to the large credible interval for the model SEDs.

\begin{figure}[ht]
    \centering
    \includegraphics[width=0.9\linewidth]{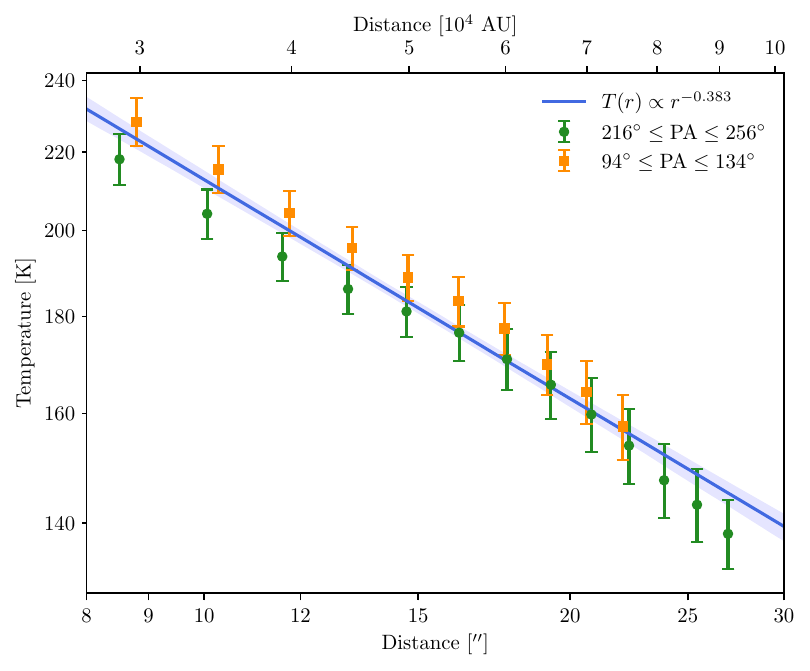}
    \caption{The temperature profile of dust in the two sectors, $216^{\circ} \leq \mathrm{PA} \leq 256^{\circ}$ (green) and $94^{\circ} \leq \mathrm{PA} \leq 134^{\circ}$ (orange), obtained from the MCMC joint fit. The bottom axis shows the deprojected distance from the star in arcsecond (see Section \ref{sec:dust_sed_aperture}), while the top axis shows the deprojected distance in au, assuming a distance of \SI{3.39}{\kilo\parsec} to WR~112 \citep{Lau2020b}. The blue line indicates the best-fit power law to the temperature profile, with power law index $\gamma=-0.384$. The narrow light blue region shows the credible interval of the fit.}
    \label{fig:Tprofile_Tfree}
\end{figure}

Figure~\ref{fig:Tprofile_Tfree} shows the temperature profile of dust obtained from the MCMC joint fit. The temperature profiles from both sectors are similar and are relatively well constrained. They can be well described by a power law,
$ T_d(r) = T_0 ~ (r/r_0)^{\gamma} $, 
where $T_0$ is the normalization and $\gamma$ is the power law index. The best-fit power law for the temperature profile from all the apertures has power law index $\gamma = -0.384 \pm 0.021$. This is consistent with $\gamma = -0.4$ inferred from observations of WR~112 by \citet{Marchenko2002}. For Apep, a WR+WR binary system, \citet{Han2025} derived a shallower temperature profile ($\gamma = -0.31 \pm 0.02$) compared to WR~112, when the grains are assumed to be \SI{0.1}{\micro\meter} in size. It remains uncertain whether WC systems share a common temperature profile. 
The dust in the first sector appears cooler than the dust in the second sector at similar distance away from the star, but such a difference is smaller than the estimated uncertainties.

The $\chi^2$ for the median posterior model is 227.2. With 26 free parameters and 69 data points from 23 apertures, the reduced $\chi^2$ is $\chi_\nu^2 = 5.28$. As shown in Figure~\ref{fig:SED_Tfree} in Appendix \ref{sec:figures_SED_Tfree}, the predicted flux is lower than the observed flux at \SI{21}{\micro\meter} for apertures closest to the star and at \SI{7.7}{\micro\meter} for apertures furthest away from the star. This suggests that the assumptions for this model might not hold. For example, the assumption that there is a single population of dust grains with the same grain radius is highly unrealistic and could not be true. To better constrain the properties of dust, some assumptions have to be relaxed.

\subsubsection{Equilibrium Temperature Grid}
\label{sec:tempgrid}
Due to different optical properties or $Q_{\text{abs}}(a, \lambda)$, grains with different sizes have different temperatures at the same distance away from the star. Therefore, to model the SED from grains with more than one different radii, it is better to predict the temperatures of grains from models that account for grain sizes rather than having tens of free parameters for temperatures. 

If we assume that the dust grains are at RE, the temperature of a dust grain with radius $a$ can be derived from the following equation:
\begin{equation}
\label{eq:Tgrid}
\begin{aligned}
    \pi a^2 & \int_{0}^{\infty}  S_{\ast}(\lambda) \frac{1}{4\pi r^2} Q_{\text{abs}}(a, \lambda) \, d\lambda \\ & = 4\pi^2 a^2 \int_{0}^{\infty} B_\lambda(T_d(a,r), \lambda) Q_{\text{abs}}(a, \lambda) \, d\lambda,
\end{aligned}
\end{equation}
which equates the power the grain receives from the stellar photon flux (left side) to the power the grain emits as thermal emissions (right side). $S_{\ast}(\lambda)$ is the spectral luminosity of the stellar photon flux, for which we adopt the median stellar spectrum of WR~112 and its companion derived in Section \ref{sec:stellarsed}, and $r$ is the distance of the dust grain from the star. $T_d(a,r)$ is the equilibrium temperature of dust grains derived numerically from Equation \ref{eq:Tgrid} for every grain radius $a$ at every distance $r$ away from the stars. 

If we ignore scattering, the spectral luminosity of the photon flux equals the sum of stellar spectra of WR~112 and its companion. Using the stellar spectra obtained in Section \ref{sec:stellarsed} using archival observations of WR~112, we computed the 2-dimensional grid of dust temperature $T_d(a,r)$ around WR~112. 

\begin{figure*}[ht]
    \centering
    \includegraphics[width=0.78\linewidth]{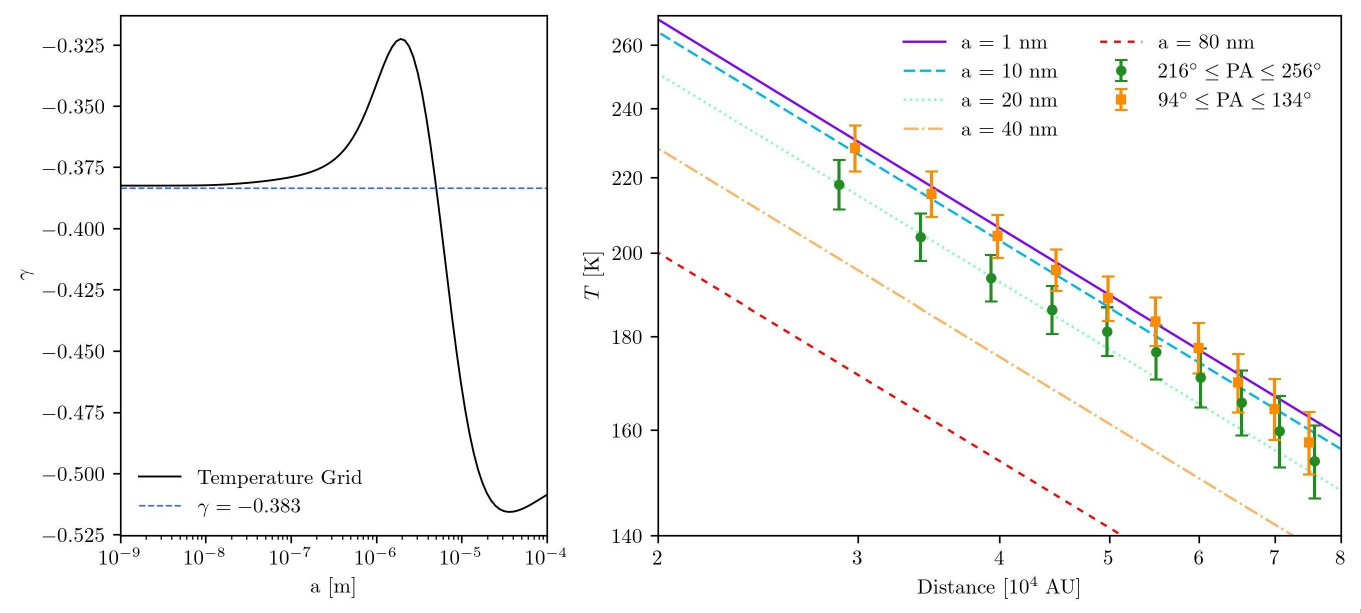}
    \caption{Left panel: the power law index $\gamma$ for the temperature profile of dust as a function of grain radius $a$ derived from Equation \ref{eq:Tgrid}. The dashed blue line shows $\gamma = -0.384$, the best-fit power law index from Section \ref{sec:Tfree} when temperatures of dust are free parameters. Right: the temperature profile of dust. The data points show the temperature profile of dust in the two sectors obtained from the MCMC joint fit in Section \ref{sec:Tfree}. The lines show the model-predicted temperature profile derived from Equation \ref{eq:Tgrid} for dust with different grain radii, from \SI{1}{\nano\meter} to \SI{80}{\nano\meter}.}
    \label{fig:Tprofile_Tgrid}
\end{figure*}

The temperature profile of dust with a specific radius $a$ derived from Equation \ref{eq:Tgrid} is also well described by a power law, $T_d(a,r) = T_0(a) r^{\gamma(a)}$, where $T_0(a)$ is the normalization and $\gamma(a)$ is the power law index. The left panel of Figure~\ref{fig:Tprofile_Tgrid} shows the power law index $\gamma(a)$ for different grain radii $a$. Grains that are large enough, with $a \gtrsim 1 \times 10^{-5}~\mathrm{m}$, behave almost like a blackbody and hence have a power index $\sim 0.5$. The index increases as the grains become smaller. This is because the grains become less effective at emission at longer wavelength with $Q_{\text{abs}}(a, \lambda) \sim \lambda^{-1}$. A small decrease in temperature causes a large reduction in the amount of power it emits, leading to a more shallow temperature profile. 

For $a \lesssim 3 \times 10^{-8}~\mathrm{m}$, the power law index is approximately $-0.38$, which agrees with similar computation using the optical constants by \citet{Williams2009}. This value is also consistent with the $\gamma = -0.384$ derived in Section \ref{sec:Tfree} when temperatures of dust are free parameters. Since the temperature for each aperture is a free parameter in Section \ref{sec:Tfree}, the temperature profile from that Section primarily depends on the observed SED instead of the optical properties of the grains. The agreement between the results from the two Sections suggest that the observed SED is at least consistent with the emission from amorphous carbon dust, and the assumption that the dust is in RE with stellar radiation is justified to first order. It also narrows down the constraint on the radius of most grains to $a \lesssim 3 \times 10^{-8}~\mathrm{m}$. 

The right panel of Figure~\ref{fig:Tprofile_Tgrid} shows the temperature profile derived from Equation \ref{eq:Tgrid} for grains with a few different radii compared to the temperature profile from Section \ref{sec:Tfree}. Larger grains which emit more effectively are at lower temperatures. The temperatures of smaller grains are generally hotter but gradually converge as $a$ approaches \SI{1}{\nano\meter}. The temperature profile for grains with radius $a \approx 10$--$20~\mathrm{nm}$ is able to roughly reproduce the temperature profile from Section \ref{sec:Tfree}. This agreement again lends confidence to the approach of computing dust temperatures using Equation \ref{eq:Tgrid}. 


Using the model-predicted dust temperature $T_d(a,r)$, we are able to consider emission from populations of grains with different sizes. The flux density emitted by dust with a continuous grain size distribution in aperture $i$ is given by:
\begin{equation}
\label{eq:Fnu2}
    F^{(i)}_\nu(\lambda) = \int B_\nu(T_d(a, r^{(i)}), \lambda) Q_{\text{abs}}(a, \lambda) \frac{\pi a^2}{d^2} N^{(i)}(a) \,da,
\end{equation}
where $r^{(i)}$ is the deprojected distance to the $i^{\text{th}}$ aperture, $d$ is the distance to WR~112 and $N^{(i)}(a)$ is the grain size distribution, or the total number of grains in the aperture per unit grain radius. The total solid angle subtended by the dust grains in the $i^{\text{th}}$ aperture is given by:
\begin{equation}
\label{eq:Omegaeff}
    \Omega^{(i)}_{\text{eff}} = \int \frac{\pi a^2}{d^2} N^{(i)}(a) \,da.
\end{equation}
Similar to the case in Section \ref{sec:Tfree}, since the dust shells expand uniformly and we assume that the grain size distribution remains constant over time, the grain size distribution $N^{(i)}(a)$ is identical for apertures in the first sector ($1 \leq i \leq 13$) and identical for apertures in the second sector ($14 \leq i \leq 23$). The two distributions only differ by a normalization, $N^{(1)}(a)/N^{(14)}(a) = \Omega^{(1)}_{\text{eff}} / \Omega^{(14)}_{\text{eff}}$. 
We keep our notation $\Omega_{\text{eff}}^{\text{I}}$ and $\Omega_{\text{eff}}^{\text{II}}$ for the total solid angle of dust in each aperture in the first sector and the second sector. 
If the grain size distribution of dust is represented by a discrete set of grain radii, the integral in equations \ref{eq:Fnu2} and \ref{eq:Omegaeff} becomes summations.

We conducted MCMC fitting to the observed dust SEDs using the model-predicted temperatures with four parameterizations of the grain size distributions: a single grain radius, a (base-10) log-normal distribution, two grain radii (or two delta functions), and a continuous power law. The parameterizations and reasons behind our choices are as the following:
\begin{enumerate}
    \item A single grain radius (or a delta function): the possible grain radius is specified as $a_e$. It is the most simple grain size distribution.
    \item A (base-10) log-normal distribution: $\log a_c$ specifies the mean of $\log a$ and $\sigma_{\log a}$ is the standard deviation of $\log a$. It is chosen to take into account for possible dispersions in grain sizes around a single radius.
    \item Two grain radii: the two possible grain radii are $a_1$ and $a_2$, and the relative abundance of grains with radius $a_2$ compared to grains with radius $a_1$ is $f_{2}$. It is chosen because previous studies on the spectra of circumstellar dust near WC stars suggest that the emission can be modeled by a two-component grain distribution \citep{Lau2020b, Lau2023}. 
    \item Power law distribution: the power law index is $\beta$, and the lower and upper grain radius cutoffs are $a_l$ and $a_u$, respectively. It is chosen because the grain distribution for interstellar dust is well described by a power law \citep{Mathis1977}
\end{enumerate}
The form of $N^{(i)}(a)$ for each grain size distribution is summarized in Table~\ref{tbl:grain_Tgrid}. 

We use the total solid angle of dust in the two sectors, $\Omega_{\text{eff}}^{\text{I}}$ and $\Omega_{\text{eff}} ^{\text{II}}$, to specify the normalization of the grain size distribution $N^{(i)}(a)$ using Equation \ref{eq:Omegaeff}. 
All free parameters except $\sigma_{\log a}$ and $\beta$ are sampled from the $\log$ space during the MCMC fit. The sampling of all parameters related to the grain radius is restricted to the range \SI{1}{\nano\meter} to \SI{100}{\micro\meter}. 

\renewcommand{\arraystretch}{1} 
\setcellgapes{4pt}  
\makegapedcells      

\begin{table*}[ht]
\caption{Summary of MCMC fit results for models described in Equation \ref{eq:Fnu2} with four different grain size distributions.}
\label{tbl:grain_Tgrid}
\centering
\begin{tabular}{@{}>{\centering\arraybackslash}p{4cm} 
                >{\centering\arraybackslash}p{6.5cm} 
                >{\centering\arraybackslash}p{6.5cm}@{}}
\toprule
\makecell{} 
    & \makecell{One radius} 
    & \makecell{Base-10 log-normal} \\
\midrule
\makecell{Size distribution} 
    & \makecell{$N^{(i)}(a) \propto \delta(a-a_e)$} 
    & \makecell{$N^{(i)}(a) \propto a^{-1}\exp[-\frac{(\log a - \log a_c)^2}{2 \sigma_{\log a}^2} ]$} \\
\makecell{Parameters} 
    & \makecell{$\Omega_{\text{eff}}^{\text{I}} = 0.014^{+0.021}_{-0.002}~\mathrm{arcsec}^2$ \\[3pt]
    $\Omega_{\text{eff}}^{\text{II}} = 0.017^{+0.025}_{-0.003}~\mathrm{arcsec}^2$ \\[3pt]
    $a_e = 10.2^{+3.8}_{-6.6}~\mathrm{nm}$}
    & \makecell{$\Omega_{\text{eff}}^{\text{I}} = 0.016^{+0.003}_{-0.002}~\mathrm{arcsec}^2$ \\[3pt]
    $\Omega_{\text{eff}}^{\text{II}} = 0.018^{+0.004}_{-0.002}~\mathrm{arcsec}^2$ \\[3pt]
    $a_c = 1.79^{+2.43}_{-0.63}~\mathrm{nm}$ \\[3pt]
    $\sigma_{\log a} = 0.381^{+0.055}_{-0.112}$} \\
\makecell{Chi-squared} 
    & \makecell{$\chi^2 = 243.4$ \\ $\chi^2_\nu = 3.69$} 
    & \makecell{$\chi^2 = 240.5$ \\ $\chi^2_\nu = 3.70$} \\
\makecell{Note} 
    & \makecell{} 
    & \makecell[l]{PD of $a_c$ peaks at $a = 1 ~\mathrm{nm}$.} \\
\midrule
\makecell{} 
    & \makecell{Two radii} 
    & \makecell{Power law} \\
\midrule
\makecell{Size distribution}
    & \makecell[l]{
    $\begin{aligned}
    N^{(i)}(a) & \propto  \delta(a-a_1) + f_{2} \delta(a-a_2)
    \end{aligned}$}
    & \makecell{$N^{(i)}(a) \propto
    \begin{cases}
    a^\beta & \text{if } a_l \leq a \leq a_u \\
    0 & \text{if otherwise}
    \end{cases}$}  \\
\makecell{Parameter medians} 
    & \makecell{$\Omega_{\text{eff}}^{\text{I}} = 0.058^{+0.225}_{-0.049}~\mathrm{arcsec}^2$ \\[3pt]
    $\Omega_{\text{eff}}^{\text{II}} = 0.064^{+0.250}_{-0.054}~\mathrm{arcsec}^2$ \\[3pt]
    $a_1 = 2.08^{+1.74}_{-0.84}~\mathrm{nm}$ \\[3pt]
    $a_2 = 1.40^{+0.30}_{-0.36}$ \SI{}{\micro\meter} \\[3pt]
    $f_2 = 5.7^{+20.7}_{-4.5} \times 10^{-7}$}
    & \makecell{$\Omega_{\text{eff}}^{\text{I}} = 0.023^{+0.004}_{-0.004}~\mathrm{arcsec}^2$ \\[3pt]
    $\Omega_{\text{eff}}^{\text{II}} = 0.026^{+0.005}_{-0.004}~\mathrm{arcsec}^2$ \\[3pt]
    $a_l = 1.40^{+0.85}_{-0.34}~\mathrm{nm}$ \\[3pt]
    $a_u = 8.40^{+37.6}_{-7.20}$ \SI{}{\micro\meter} \\[3pt]
    $\beta = -3.57^{+0.12}_{-0.17}$} \\
\makecell{Chi-squared} 
    & \makecell{$\chi^2 = 179.2$ \\ $\chi^2_\nu = 2.80$} 
    & \makecell{$\chi^2 = 219.9$ \\ $\chi^2_\nu = 3.44$} \\
\makecell{Note} 
    & \makecell[l]{PD of $a_1$ peaks at $a = 1 ~\mathrm{nm}$.} 
    & \makecell[l]{PD of $a_l$ peaks at $a = 1 ~\mathrm{nm}$. 
    \\ PD of $a_u$ flattens out for $a \geq~$\SI{3}{\micro\meter}. } \\
\bottomrule
\end{tabular}
\end{table*}

The main results of the MCMC fits are summarized in Table~\ref{tbl:grain_Tgrid}, including the median parameters and errors, the $\chi^2$ and reduced $\chi^2$ values, and notes on the posterior distribution (PD) of the parameters. The $\chi^2$ when one single grain radius is assumed (one-radius case) is similar to the $\chi^2$ in Section \ref{sec:Tfree} when temperatures are free parameters, although the reduced $\chi^2_\nu$ improves due to great reduction in number of free parameters. Compared to the one-radius case, the $\chi^2$ is similar for the power law distribution and the log-normal distribution. 
The $\chi^2$ value is lowest for the grain size distribution with two possible grain radii, which improves by roughly $25\%$ compared to the one-radius case. 

Appendix \ref{sec:figures_SED_Tgrid} presents the median posterior models for the four grain distributions along with the observed SEDs. The median posterior models for the one-radius, log-normal and power law distribution do not reproduce the emission at \SI{21}{\micro\meter} for inner apertures and at \SI{7.7}{\micro\meter} for outer apertures. In comparison, by including a grain component that is larger in size and thus cooler in temperature, the grain distribution with two grain radii is able to reproduce the emission at \SI{21}{\micro\meter} for inner apertures. 
On the other hand, while the model attempts to reproduce the excess emission at \SI{7.7}{\micro\meter} in outer apertures by including smaller grains at high temperatures, which leads to the PD of $a_1$ peaking at $a=~$\SI{1}{\nano\meter}, it is not able to do so. 
Even grains as small as \SI{1}{\nano\meter} are not warm enough under RE conditions in outer apertures to produce enough emission at \SI{7.7}{\micro\meter} that is consistent with the observations. However, it is unreasonable for dust grains to be smaller than \SI{1}{\nano\meter}. This suggests that simply modifying the grain size distribution cannot account for the observed SEDs under the RE assumption.

\subsubsection{Stochastic heating}
\label{sec:stochastic}

Section \ref{sec:tempgrid} assumes the dust grains are at RE with the stellar radiation. The MCMC fit to the observed SEDs under that assumption suggests that there is a significant amount of small grains with radius $\sim 1-20~$ \SI{}{\nano\meter}. However, the assumption that such small grains are at RE conditions is likely not valid. Since the internal energies of such small grains is smaller than or comparable to the energies of photons from starlight, the temperature (or internal energy) of the grain spikes every time it absorbs a photon \citep{Draine2001, Camps2015}. This process is referred to as stochastic heating or transient heating.

Without detailed information of energy levels of individual grains, a canonical way to model emissions from stochastically heated small grains is to consider a probability distribution of temperature. The flux density emitted by a single dust grain with radius $a$ is given by
\begin{equation}
\label{eq:stoch_heating1}
\begin{aligned}
    F^{\text{grain}}_\nu&(a, \lambda;~S_{\ast}, r, \cdots)  = \frac{\pi a^2}{d^2} Q_{\text{abs}}(a, \lambda) \\ &\times  \int  B_\nu(T, \lambda) P(T;~a, S_{\ast}, r, \cdots)  \,dT,    
\end{aligned}
\end{equation}
where $T$ is the temperature the grain can be at, and $P(T;~a, S_{\ast}, r, \cdots)$ is the probability density of the temperature distribution. The variables after the semicolon represent conditions the dust is subject to. These conditions include the stellar luminosity $S_{\ast}$ and distance $r$ from the star, which determines the photon flux the dust receives, the optical properties of the grain and information about internal energies of the grains (e.g. specific heat capacities). The calculation of the probability density from these conditions using temperature grids and transition rates is summarized in \citet{Camps2015} (see more details in \citealt{Desert1986, Draine2001}). 

For a continuous distribution of grains with different sizes, the total flux density emitted by dust in aperture $i$ is given by:
\begin{equation}
\begin{aligned}
\label{eq:stoch_heating2}
    & F_\nu^{(i)}(\lambda;\cdots) \\ &  = \int N^{(i)}(a) F^{\text{grain}}_\nu(a, \lambda;~S_{\ast}, r^{(i)}, \cdots) \, da \\
     & = \int N^{(i)}(a) \frac{\pi a^2}{d^2} Q_{\text{abs}}(a, \lambda)\\ & ~~~~~~~~~~ \times \int  B_\nu(T, \lambda) P(T;~a, S_{\ast}, r^{(i)}, \cdots)  \,dT \, da.
\end{aligned}
\end{equation}
In this work, we used \texttt{DustEM} \citep{Compigne2011} to compute the probability density $P(T;~a, S_{\ast}, r^{(i)}, \cdots)$, which is based on the formalism of \citet{Desert1986}. Like before, we computed the optical properties of dust from the optical constants of amorphous carbon dust measured in laboratory by \citet{Zubko1996}. \citet{Michelsen2008} argued that the heat capacity of graphite is a good estimate for the heat capacity of amorphous carbon. Thus, we adopted the heat capacity of graphite from \citet{Draine2001}. We also set the density of dust to be $\rho = 1.81~\mathrm{g~cm^{-3}}$, which is in agreement with \citet{Compigne2011}. 

\begin{figure}[h]
    \centering
    \includegraphics[width=0.9\linewidth]{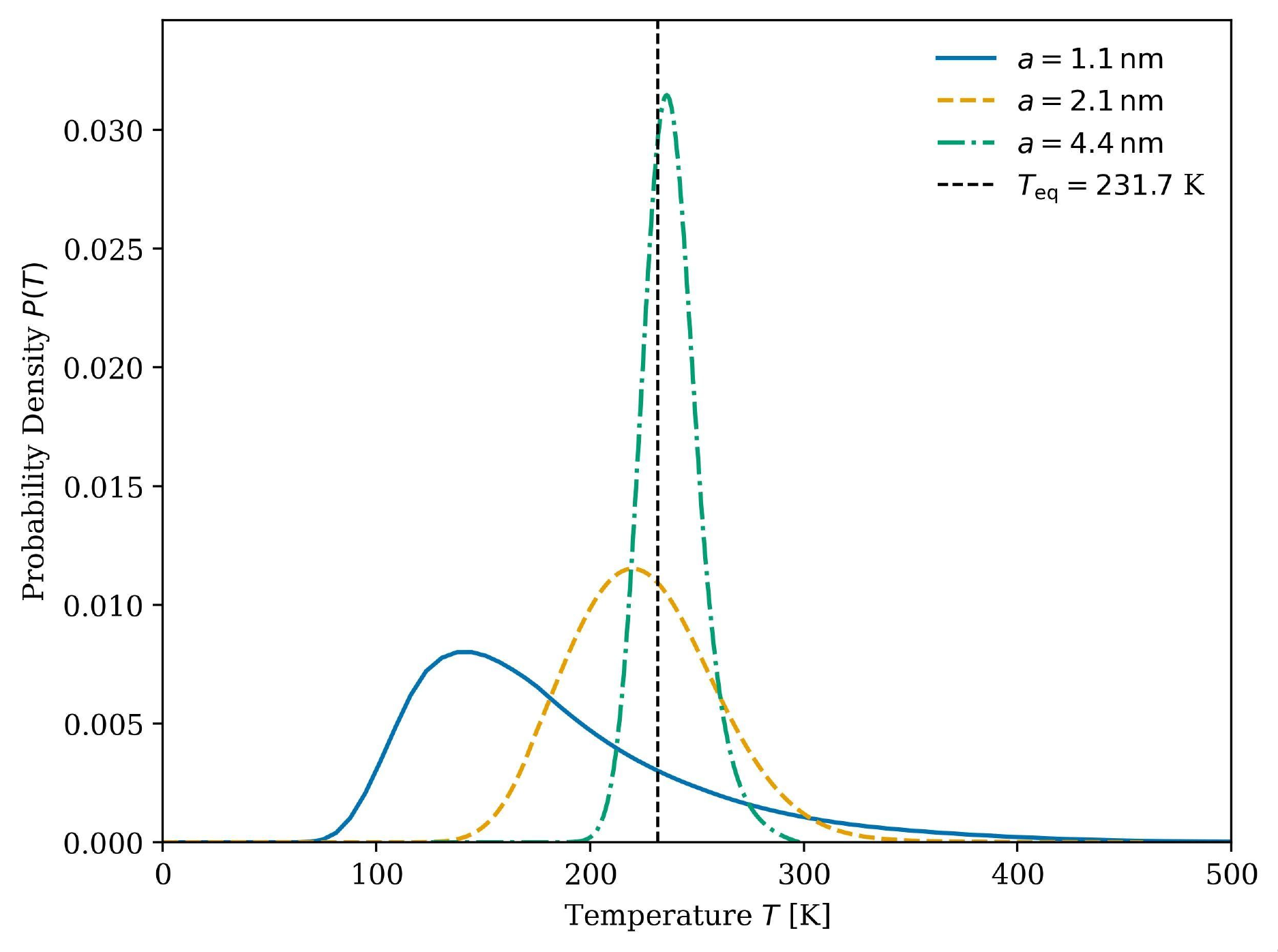}
    \caption{The probability density function of temperatures for grains with radii 1.1, 2.1 and \SI{4.1}{\nano\meter} in the first aperture ($r^{(1)} = 28870$ au). The equilibrium temperature of dust with these three radii computed from Equation \ref{eq:Tgrid} is almost the same at approximately \SI{232}{\kelvin} and shown as the black vertical line.}
    \label{fig:PDF_Temp}
\end{figure}

To speed up computation, we computed the probability density of temperature $P(T;~a, S_{\ast}, r^{(i)}, \cdots)$ for each aperture and a grid of 200 grain radii $a$ between \SI{1}{\nano\meter} and \SI{10}{\micro\meter}. Grains larger than \SI{10}{\micro\meter} are not included because we found that they are not necessary for reproducing the observed SED given the ALMA upper limits and shape of the SEDs. The probability density functions for grain radii not in the grid are then obtained by transportation of probability density, as described in Appendix \ref{sec:optimal_transport}. 
Figure~\ref{fig:PDF_Temp} shows the probability density of temperatures and the equilibrium temperature for grains with radii a few \SI{}{\nano\meter}. While the distribution peaks at temperatures lower than the equilibrium temperature for smaller grains, there is also a significant high-temperature tail that allows photons of shorter wavelengths to be emitted. The distribution becomes narrower for larger grains. For grains that are larger than \SI{40}{\nano\meter}, the probability distribution is narrower than \SI{1}{\kelvin} and the assumption that the grain is at RE becomes a good approximation again.

\begin{table*}
\caption{Summary of MCMC fit results for models considering stochastic heating of small grains and the ACAR-type amorphous carbon dust adopted.}
\label{tbl:grain_stoch_ACAR}
\centering
\begin{tabular}{@{}>{\centering\arraybackslash}p{4cm} 
                >{\centering\arraybackslash}p{6.5cm} 
                >{\centering\arraybackslash}p{6.5cm}@{}}
\toprule
\makecell{} 
    & \makecell{One radius} 
    & \makecell{Base-10 log-normal} \\
\midrule
\makecell{Parameter medians} 
    & \makecell{$\Omega_{\text{eff}}^{\text{I}} = 0.011^{+0.031}_{-0.001}~\mathrm{arcsec}^2$ \\[3pt]
    $\Omega_{\text{eff}}^{\text{II}} = 0.013^{+0.038}_{-0.001}~\mathrm{arcsec}^2$ \\[3pt]
    $a_e = 12.6^{+2.5}_{-9.9}~\mathrm{nm}$}
    & \makecell{$\Omega_{\text{eff}}^{\text{I}} = 0.014^{+0.001}_{-0.001}~\mathrm{arcsec}^2$ \\[3pt]
    $\Omega_{\text{eff}}^{\text{II}} = 0.017^{+0.002}_{-0.002}~\mathrm{arcsec}^2$ \\[3pt]
    $a_c = 1.160^{+0.320}_{-0.122}~\mathrm{nm}$ \\[3pt]
    $\sigma_{\log a} = 0.440^{+0.027}_{-0.033}$} \\
\makecell{Chi-squared} 
    & \makecell{$\chi^2 = 242.9$ \\ $\chi^2_\nu = 3.68$} 
    & \makecell{$\chi^2 = 218.5$ \\ $\chi^2_\nu = 3.36$} \\
\midrule
\makecell{} 
    & \makecell{Two radii} 
    & \makecell{Power law} \\
\midrule
\makecell{Parameter medians} 
    & \makecell{$\Omega_{\text{eff}}^{\text{I}} = 0.039^{+0.023}_{-0.013}~\mathrm{arcsec}^2$ \\[3pt]
    $\Omega_{\text{eff}}^{\text{II}} = 0.042^{+0.022}_{-0.013}~\mathrm{arcsec}^2$ \\[3pt]
    $a_1 = 1.247^{+0.247}_{-0.159}~\mathrm{nm}$ \\[3pt]
    $a_2 = 0.134^{+0.084}_{-0.026}$ \SI{}{\micro\meter} \\[3pt]
    $f_2 = 2.85^{+1.27}_{-0.82} \times 10^{-5}$}
    & \makecell{$\Omega_{\text{eff}}^{\text{I}} = 0.023^{+0.002}_{-0.001}~\mathrm{arcsec}^2$ \\[3pt]
    $\Omega_{\text{eff}}^{\text{II}} = 0.027^{+0.002}_{-0.002}~\mathrm{arcsec}^2$ \\[3pt]
    $a_l = 1.031^{+0.043}_{-0.021}~\mathrm{nm}$ \\[3pt]
    $a_u = 3.219^{+3.829}_{-1.964}$ \SI{}{\micro\meter} \\[3pt]
    $\beta = -3.52^{+0.08}_{-0.08}$} \\
\makecell{Chi-squared} 
    & \makecell{$\chi^2 = 30.1$ \\ $\chi^2_\nu = 0.470$} 
    & \makecell{$\chi^2 = 147.4$ \\ $\chi^2_\nu = 2.30$} \\
\bottomrule
\end{tabular}
\end{table*}

Using the probability density of temperatures and Equation \ref{eq:stoch_heating2}, we conducted MCMC fits to the observed SEDs. 
We adopted the four parameterizations of grain distributions used in Section \ref{sec:tempgrid}: one radius, the base-10 log-normal, two radii and a power law. Like in previous sections, we adopted the ACAR-type grains given the compositions of stellar winds of WR~112. The results of the MCMC fits are summarized in Table~\ref{tbl:grain_stoch_ACAR}. The comparison between the median posterior models for the four grain distributions and the observed SEDs is shown in Section \ref{sec:figures_SED_stoch}. 
In addition to the ACAR-type grains adopted, we also conducted the fits for the two other laboratory analogues of amorphous carbon grain prepared by \citet{Colangeli1995} and measured by \citet{Zubko1996}, the ACH2-type and the BE-type, to gain insights over the effect of grain chemistry. The results for these two types of grains are shown and discussed in Section \ref{Appendix:results_types}.

Comparison between the results for ACAR-type amorphous carbon dust in this and the previous Section (Table~\ref{tbl:grain_Tgrid} and \ref{tbl:grain_stoch_ACAR}) highlights the importance of stochastic heating to modeling the dust SED. For smaller grains, the wide temperature distribution makes the shape of the SED wider. In particular, the long tail at high temperature allows more photons with shorter wavelength to be emitted. As a result, the \SI{7.7}{\micro\meter} emission can be well accounted for by the presence of nanometer-size grains. This improves the fit for the model with two grain radii and a power law as the grain distribution. Particularly, the two grain radii model almost perfectly reproduces the observed SED, with a reduced chi-squared $\chi^2_\nu = 0.470$. We believe that $\chi^2_\nu < 1$ is not a sign of overfitting but instead suggests we overestimated the uncertainties in photometry when we assumed them to be $20\%$.

Except for the one-radius size distribution, the MCMC fits suggest that the dust in WR~112 mostly consists of small, nanometer-size grains. 
In particular, the two grain radii model predicts the presence of dust with radius $a \sim 1.2-1.3$ \SI{}{\nano\meter}. The posterior distribution for the radius of these small grains peaks at the lower bound $a\sim$ \SI{1}{\nano\meter}. We are, however, uncertain if there can be grains below \SI{1}{\nano\meter}. 

Notably, the distribution with two distinct possible grain radii best reproduces the observed SEDs. It is able to reproduce both the flux at  \SI{7.7}{\micro\meter} in the outer dust arcs and the slope of the SED from \SI{15}{\micro\meter} to \SI{21}{\micro\meter}, despite the conservative uncertainties adopted.
The power law distribution achieves an acceptable level of agreement with the data. The single grain radius distribution and the log-normal distribution fit the observed SEDs the worst, and should be unlikely. The implications of our results on the grain size distribution is discussed in details in Sections \ref{sec:disc_size_power} and \ref{sec:disc_size_two} . The robustness of the inferred grain size distribution is discussed in Section \ref{sec:robust}.

\subsubsection{Dust SED at the location of the star}
\label{sec:dust_stellar_SED}
Emissions from the circumstellar dust are also captured in the spatially unresolved observations of WR~112. This is obvious from figure~\ref{fig:stellar_spec}, where a second peak in flux density is observed around a few microns. 
Unlike the resolved ones, the unresolved observations have large apertures (e.g., $> 15^{\prime\prime}$ for MSX) and should capture emissions from most of the circumstellar dust around WR~112. 
In this section, we consider models of the combined emissions from all circumstellar dust and place additional constraints on the dust properties by fitting to the unresolved observations. 

When dealing with resolved observations, we modeled the emissions from dust in apertures around WR~112, each of which covers a $40^{\circ}$ azimuthal segment of a dust arc. The small difference in total solid angle between apertures at different position angles ($\sim 10-20\%$, in Table~\ref{tbl:grain_stoch_ACAR}) suggests the dust structure is largely uniform. The contrasting example would be WR~140, where the prominent ``bars'' with high dust density appear in the images, which are results of episodic dust production due to the high eccentricity in the binary orbit \citep{Lau2022}. The relative uniformity of dust around WR~112 indicates continuous dust production, suggesting that the separation between WR~112 and its companion can only vary moderately. This is consistent with the low eccentricity suggested by the low K-band variability of WR~112 \citep{Williams2015}. 

Given the overall spatial uniformity of the circumstellar dust around WR~112, we can approximate the dust structure to be continuous concentric spherical shells.
The emission from dust in all the shells is given by:
\begin{equation}
\begin{aligned}
\label{eq:Fnu_sphere1}
    & F^{\text{shell}}_\nu\big(\lambda; ~r_1, r_2, n(a) \big) \\ & = \int_{r_1}^{r_2} \int n(a)  F^{\text{grain}}_\nu(a, \lambda;~S_{\ast}, r, \cdots) \,da \, dr,
\end{aligned}
\end{equation}
where $r_1$ and $r_2$ are the inner and outer radius of the spherical shell. $F^{\text{grain}}_\nu(a, \lambda;~S_{\ast}, r, \cdots)$ is the emission from a single grain with radius $a$ from Equation \ref{eq:stoch_heating1}, where the temperature distribution of dust is likewise computed by DustEM. $n(a)=\frac{dN(a)}{dr}$ is the grain size distribution per unit radial distance, or the total number of grains with radius $a$ per unit grain radius per unit radial distance. This equation assumes that the emissions of all the dust shells are captured by the apertures of the instruments used.

Constraining the grain size distribution using only the unresolved SED is impractical due to the degeneracy between grain size and distance from the star. Larger grains closer to WR~112 produce similar emissions as smaller grains further away from the stars. 
Fortunately, the grain size distribution of circumstellar dust in the outer dust structures near WR~112 is already constrained by the MCMC fits to the spatially resolved SEDs in the previous section: the outer dust structures most likely have two populations of grains, one with radius $a_1 = $ \SI{1.247}{\nano\meter}, and another with radius $a_2 = $ \SI{0.134}{\micro\meter} and is $f_2 = 2.85 \times 10^{-5}$ less abundant than the smaller ones. The spatially resolved JWST observations also allow us to conclude that $n(a_1)$ should be constant for $r>25{,}000$ au, since the observed SEDs suggests that dust expands at constant speed. We can estimate $n(a_1)$ using the effective solid angles, by ignoring projection effects and equating the total number of dust in a spherical shell to the number of dust in a projected ring, which is 9 times the number of dust in a $40^{\circ}$ aperture with the same width. 

However, the grain size distribution is only known for dust between $r=25{,}000$ and $r=90{,}000$ au. The dust at $r>90{,}000$ au, if present, is too dim to be detected by JWST. Nevertheless, even the $1~\mathrm{nm}$ dust grains at those distances have relatively low temperature ($T \lesssim 150$ K). They should not contribute to the near- and mid-infrared emissions that the unresolved observations are most concerned with, and can be ignored. We hence set outer radius cutoff for shells to be $R=90{,}000$ au. 
The JWST images become saturated for $r<25{,}000$ au, where properties of the dust cannot be constrained from the resolved SEDs. This is a problem because the conditions of dust most close to the star are unknown. The grain size distribution can evolve shortly after dust formation, and dust is still undergoing acceleration due to radiation pressure \citep{Han2022}. 

We consider three different models to account for the unknown conditions of dust most close to the star. Details of the three models (A, B, and C) are summarized in Table~\ref{tbl:grain_stellar_sed}. Model A assumes that the properties of dust, including the grain size distribution per radial distance, remain the same even in the innermost regions.  
Model B assumes that the grain size distribution remains the same in the innermost regions; but it allows the number density of dust grains to be different in the innermost region. Model C assumes that there is a sudden transition in the properties of dust at a certain radius $r_s$: the grain size distribution for $r<r_s$ is completely different from the distribution for $r>r_s$. We parameterize the grain size distribution for $r<r_s$ as a delta function with one single grain radius $a_3$, because the unresolved data is unable to constrain more complicated grain size distributions. 
For all three models, there is a inner radius cutoff $r_{\rm in}$, which is where the dust is supposed to form. Since dust is driven outward by the radiation pressure, there should not be dust for $r<r_{\rm in}$. 

\begin{table*}
\caption{Summary of MCMC fit results for models considering stochastic heating of small grains and the ACAR-type amorphous carbon dust adopted.}
\label{tbl:grain_stellar_sed}
\centering
\begin{tabular}{@{}>{\centering\arraybackslash}p{3cm} 
                >{\centering\arraybackslash}p{4.5cm} 
                >{\centering\arraybackslash}p{4.5cm}
                >{\centering\arraybackslash}p{4.7cm}@{}}
\toprule
\makecell{} 
    & \makecell{Model A} 
    & \makecell{Model B}
    & \makecell{Model C} \\
\midrule
\makecell{$n(a)$ \\ $r_s\leq r<R$}
    & \multicolumn{3}{c}{\makecell{$n_{\rm out}(a) =  n_1 \big[\delta(a-a_1) +f_2\delta(a-a_2) \big]$}} \\
\makecell{$n(a)$ \\ $r_{\rm in}\leq r< r_s$} 
    & \makecell{$n_{\rm out}(a)$} 
    & \makecell{$f_{\rm in} n_{\rm out}(a)$}
    & \makecell{$n_3\delta(a-a_3)$} \\
\makecell{Parameters} 
    & \makecell{$r_{\rm in} = 281.11^{+13.17}_{-13.41}$ au} 
    & \makecell{$r_{\rm in} = 380.25^{+28.38}_{-30.18}$ au \\ [2.5pt]
    $r_s = 10255^{+1708}_{-3341}$ au \\ [2.5pt]
    $f_{\rm in}=1.63^{+0.14}_{-0.16}$}
    & \makecell{$r_{\rm in} =239.98^{+31.68}_{-25.09}$ au \\ [2.5pt]
    $r_s = 14846^{+3666}_{-7272}$ au \\ [2.5pt]
    $a_3=0.192^{+0.038}_{-0.044}$ \SI{}{\micro\meter} \\ [2.5pt]
    $\frac{n_3}{n_1} = 1.86^{+1.52}_{-0.65} \times 10^{-5}$}  \\
\makecell{Chi-squared} 
    & \makecell{$\chi^2 = 106$} 
    & \makecell{$\chi^2 = 76$}
    & \makecell{$\chi^2 = 88$}\\
\bottomrule
\end{tabular}
\end{table*}

\begin{figure*}
    \centering
    \includegraphics[width=0.58\linewidth]{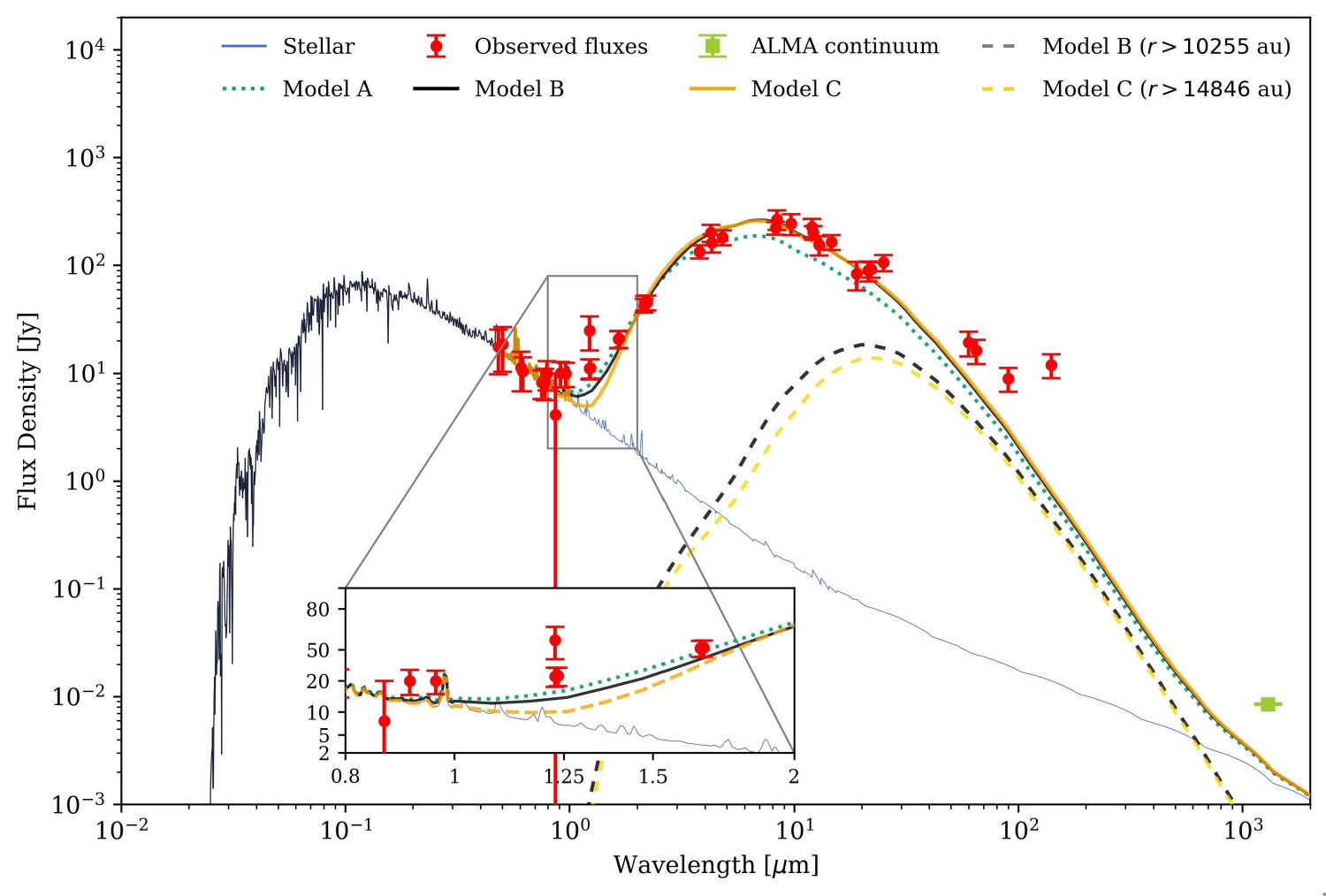}
    \caption{The SED of WR~112, with both the stellar components and the dust component. The red data points are for the spatially unresolved observations described in Section \ref{sec:archival_obs} and corrected for extinction with $R_V=3.5$ and the light green data point corresponds to the total flux from the point source in ALMA continuum image. The emissions from the model A (green, dotted), model B (black, solid) and model C (orange, solid) are also shown. To understand the contribution from the outer dust shell, the emission from the outer dust shell for model B and C are plotted as gray and yellow dashed lines. The blue curve shows the stellar emission from WR~112 and its companion.}
    \label{fig:stellar_sed_dust}
\end{figure*}

We conducted MCMC fits of the three models to the unresolved SED, and the results are summarized in table \ref{tbl:grain_stellar_sed}. The total emissions described by each model is shown in Figure \ref{fig:stellar_sed_dust}. 

The figure shows that all three models are able to reproduce the near-infrared and mid-infrared data relatively well, and they all underestimate the emissions in the far-infrared, which leads to the relatively large $\chi^2$. This underestimation could be due to contamination from dust emission along the line of sight. In particular, there could be a dense cloud along the line of sight (discussed in section \ref{sec:stellarsed}) which hosts cold dust that emits at $\lambda \gtrsim 100$ \SI{}{\micro\meter}. This could explain the unexpectedly high flux density observed at \SI{140}{\micro\meter}. It is also possible that we have underestimated the dust emission from the outer shell. The dust structures around WR~112 might be more extended than what is observed in JWST. The cold dust further away from the star does not emit strongly in mid-infrared but can emit at far-infrared. We could also be underestimating $n_1$ by ignoring projection effects. If $n_1$ is $2-5$ times larger than our current estimates, the emission from dust between $r=25{,}000$ and $r=90{,}000$ au is able to reproduce the far-infrared emissions at $\lambda \lesssim 100$ \SI{}{\micro\meter}. It is uncertain if projection effect can lead to such a large difference. 

In addition, underestimation of $n_1$ has negligible effect on the reproduction of the near- and mid-infrared observations. As shown in Figure \ref{fig:stellar_sed_dust}, the emission in those wavelengths is dominated by dust close to the star at $r<15{,}000$ au. Model B and C have already considered a different number of dust in those regions compared to the outer dust structures. The emission from dust close to the star being dominant also suggests that the exact apertures used for the unresolved observations are not important, as dust in the outskirts makes insignificant contribution.

It is interesting that the same bimodal grain size distribution, with nanometer-sized and \SI{0.1}{\micro\meter} grains, across all radii is able to reproduce the unresolved SED relatively well, and even better when the inner region is slightly denser. When the grain size distribution of dust in the inner region is not constrained, we infer that the emission is most consistent with emissions from grains with radius of $0.1-0.2$ \SI{}{\micro\meter}, as indicated by model C. Interestingly, the confidence interval of the inferred grain radius for model C overlaps with the confidence interval of the grain radius of larger grains in the bimodal grain size distribution. However, the exact grain radius of dust cannot be determined accurately with the spatially unresolved observations considering the degeneracy in distance from the star and grain radius. 

Although the inferred radius is smaller than the \SI{0.5}{\micro\meter} deduced by \citet{Marchenko2002} for the central $3^{\prime\prime}$ region ($r < 10{,}000$ au), this serves as another piece of evidence for the existence of the population of sub-micron grains. The dominance of this population of larger grains over smaller grains in emission also explains why the nanometer-sized grains are not detected in WR~112 in previous works. 
The apparent consistency between the grain size distribution of the inner unresolved dust and dust in the outer structures seems to suggest that there is little evolution in the grain size distribution from $r\sim 500$ au to $r\sim 25{,}000$ au. 
The system has to be driven into the bimodal grain size distribution shortly after formation (at $r\lesssim 500$ au). Mechanisms that could allow such rapid destruction are discussed in Section \ref{sec:disc_size_two}. 

We believe that uncertainties in the extinction or the stellar spectra have minimal impact on the infrared observations. While WR~112 exhibits variability in K band magnitude \citep{Williams2015}, the variability is relatively small ($\Delta M_K \lesssim 0.1$). The main source of uncertainties would be whether the excess far-infrared observations are due to cold dust along the line of sight or underestimation of our model. Another possible source of uncertainties is the contribution from light scattered by the dust grains in the near-infrared. 
These uncertainties, however, do not affect the main conclusions of this section that the unresolved SED is consistent with emission of $0.1-0.2$ \SI{}{\micro\meter} grains close to the star, whose contribution is primarily constrained by the mid-infrared observations.

\section{Discussion}
\label{sec:disc}

\subsection{Continuum and CO emissions from ALMA}
\subsubsection{Sources of continuum emission observed by ALMA}
The total flux from the unresolved central emission observed by ALMA is \SI{8.5}{}$\pm$\SI{0.1}{\milli\jansky}, as shown in Figure~\ref{fig:stellar_sed_dust} along with the derived spectra of stars and dust. 

\citet{Monnier2025} measured the flux of WR~112 at 8 GHz (3.6 cm) from 1995 to 2022. The flux attributed to free-free emissions at 8 GHz is $\sim$\SI{0.8}{\milli\jansky}. The spectral index of free–free emission at millimeter wavelengths is typically modeled to be about $2/3$ \citep{Wright1975}, but has been observed to be slightly higher ($0.69 \pm 0.02$) in WR stars \citep{Williams1990b}. This leads to an expected flux of $\sim$\SI{8}{\milli\jansky} at \SI{1.3}{\milli\meter}. This implies that the observed ALMA flux should be almost entirely from free-free emissions. 

Such free-free emission should come from a region with relatively small size \citep{Wright1975}, consistent with the fact that the emission is unresolved. This is also consistent with our models of dust emission in Section \ref{sec:dust_stellar_SED}. The results suggest that the thermal emission of circumstellar dust is responsible for at most $\sim 6\%$ of the observed emission. The actual contribution is likely lower, since the dust contributing to the unresolved SEDs at \SI{1.3}{\milli\meter} should be at $r>5000$ au, outside the region encompassed by the ALMA beam. 

The stellar spectra of WR stars from the PoWR models include the free-free emission from stellar winds. The spectrum of WR~112 constrained in Section \ref{sec:stellarsed} suggests that the free-free emission from the stellar winds of WR~112 is $\sim 20\pm10 \%$ of the emission at \SI{1.3}{\milli\meter}. The discrepancy between the constrained stellar spectrum and observed flux could arise from uncertainties in the stellar spectrum, which is relatively weakly constrained by only optical data points. It can also be explained by additional free-free emissions from the collision of stellar winds of WR~112 and its companion. The shock front formed by the collisions of the winds has considerably high electron density, which could lead to additional thermal free-free emissions.

\subsubsection{Lack of CO gas from stellar winds}
WR~112 is a WC star, so its stellar wind should be rich in carbon. Based on observations of oxygen lines from stellar winds (e.g. O IV at \SI{3411}{\angstrom}), \citet{Sander2012} suggested that the stellar wind of WC stars should have $\sim 5\%$ of oxygen. Thus, there should be enough material from which CO could form. UV photons from the hot WC star, however, would lead to rapid photodissociation of CO. If there is not enough dust to shield UV radiation for CO where it forms, photodissociation would be significant. This could explain the absence of detectable CO directly from the stellar winds. This suggests that a substantial amount of carbon and oxygen could remain in the atomic phase, which could alter the chemical pathways available for dust formation. Dust formation models should take this constraint into account. 

\subsection{Dust production rate}
\label{sec:dust_rate}
Using the grain size distribution of dust, we are able to estimate the dust production rate of WR~112. The total mass of dust in an aperture should be given by:
\begin{equation}
\label{eq:mass_d}
    M^{(i)}_d = \int \frac{4\pi}{3} \rho_d a^3 N^{(i)}(a) \, da,
\end{equation}
where $\rho_d \approx$ \SI{1.81}{\gram\per\centi\meter\cubed} is the mass density of the dust we assumed, and $N^{(i)}(a)$ is the abundance of dust. The form of $N^{(i)}(a)$ depends on the grain size distribution adopted, and the normalization of $N^{(i)}(a)$ can be derived from the total solid angle of dust with Equation \ref{eq:Omegaeff}. 

To estimate the dust production rate from the mass of dust in apertures, we can approximate the dust structure formed over in one orbital period to be projected into a dust spiral that spans $360^{\circ}$, as mentioned in Section \ref{sec:stellarsed}. Since each aperture spans $40^{\circ}$ in the azimuthal direction, the mass of dust formed in an orbital period of WR~112 is approximately $9 M^{\text{ap}}_d$, where $M^{\text{ap}}_d$ is the average of the mass of dust in an aperture in the first and the second sector. The dust production rate is then approximately $9 M^{\text{ap}}_d / P$, where P is the orbital period of WR~112.

Since the dust expands at a uniform speed at large radii from the star ($\sim$ an orbital period after production), the orbital period of WR~112 can be approximated as the radial separation of each dust arc divided by the terminal speed of the dust. The radial width of each dust arc, assuming a distance of $3.39$ kpc to WR~112, is \SI{5210}{au}. If we assume that the dust is coupled to the gas, the terminal speed of the dust equals the terminal wind velocity, which is estimated to be $v_{\infty} =$ \SI{1230}{\kilo\meter\per\second} \citep{Lau2020b}. The resulting orbital period is approximately $20.1$ yr. This is consistent with the $21.18^{+3.1}_{-2.4}$ yr derived by \citet{Lau2020b}. 

\begin{figure}
    \centering
    \includegraphics[width=0.95\linewidth]{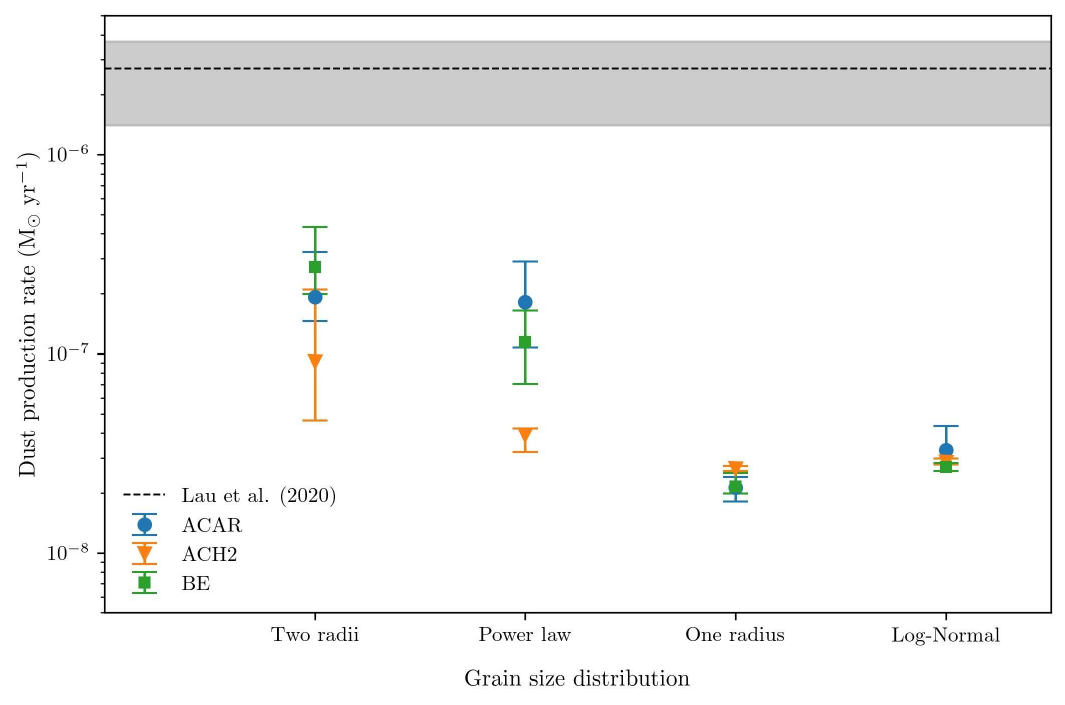}
    \caption{The dust production rate of WR~112 estimated for different assumed grain size distribution and types of amorphous carbon grain. The black dashed line and the grey region indicate the estimates and uncertainties from \citet{Lau2020b}.}
    \label{fig:dust_prod_rate}
\end{figure}

Our results suggest that the circumstellar dust around WR~112 should consist of ACAR-type amorphous carbon grains with two different possible grain radii. Under this condition, the dust production rate is $1.93^{+1.31}_{-0.41} \times 10^{-7}~\mathrm{M}_{\odot}~\mathrm{yr}^{-1}$. This corresponds to $\sim 0.7\%$ of the total mass loss rate from WR~112. However, the dust production rate is highly dependent on the grain size distribution adopted, as shown in Figure~\ref{fig:dust_prod_rate}. For the ACAR-type amorphous carbon grain in particular, the bimodal grain size distribution predicts the highest dust production rate. The power law  grain size distribution predicts a similar dust production rate, while the dust production rates for the other two grain size distributions are roughly one order of magnitude lower. 

The difference in dust production rate is mainly caused by the presence or absence of larger grains. For grains with a given radius $a$, the total solid angle subtended by the grains is proportional to $a^2$, while the mass of the grains is proportional to $a^3$. The population of larger grains, despite having smaller or comparable total emitting area as the smaller grains, can contain most of the mass. The number of larger grains is close to zero for the one-radius or log-normal grain radius distribution, leading to a lower dust production rate. 

\citet{Lau2020b} estimated the dust production rate to be $\dot{M}_d = 2.7^{+1.0}_{-1.3} \times 10^{-6}~\mathrm{M}_{\odot}~\mathrm{yr}^{-1}$, which is roughly an order of magnitude larger than our results. The difference between our results and theirs is partially caused by the difference in grain size distribution. They adopted a power law with $N(a) \propto a^{-3}$ for radii between $0.1-1.0$ \SI{}{\micro\meter}. The presence of much larger grains in this distribution leads to a higher $\dot{M}_d$. 

The sensitivity of dust production rates to the assumed grain size distribution highlights the importance of constraining the grain size distribution. If the adopted distribution in previous works is not representative, their estimates of the dust production rate may be systematically offset from the true values by $\gtrsim 1$ order of magnitude.

\subsection{Power law grain size distribution}
\label{sec:disc_size_power}
In Section \ref{sec:stochastic}, we conducted MCMC fits of the dust SED for four different parameterizations of grain size distributions, under the assumptions that the grain size distribution is relatively constant at $r> 25{,}000$ au and the grains are hydrogen-poor. The grain size distribution with two distinct grain radii best reproduces the observed SED, while the power law grain size distribution provides an acceptable fit. 
In this subsection, we explore whether the power law grain size distribution can be explained by collisions and whether permitting the distribution to vary for $r>25{,}000$ au results in a better reproduction of the observed SEDs.   

\subsubsection{Collisions as a potential origin}
Interestingly, we find that the power law index of the power law distribution is $\beta = -3.52^{+0.08}_{-0.08}$. This is consistent with the grain size distribution of the interstellar dust \citep{Mathis1977}, where $\beta \sim -3.5$. It could be the steady-state distribution after a series of collisional cascades between the large dust grains \citep{Dohnanyi1969}. If the grain size distribution is indeed a power law, it is possibly the result of large numbers of collisions of large dust grains shortly after their formation, since collisions become less frequent as dust travels away from the star and the number density rapidly decreases .

We found in Section \ref{sec:dust_stellar_SED} that the unresolved SED is consistent with emissions from $0.1-$\SI{0.2}{\micro\meter} grains close to star ($r<15{,}000$ au). This result is mostly independent from the inferred grain size distribution of the outer dust structures, which do not contribute significantly to the near- to mid-infrared emissions. 
We can estimate the collisional timescale for these grains to understand if they can be responsible for a power law grain size distribution:
\begin{equation}
    \tau_{\text{col}} = \frac{1}{4\pi a_{\rm col}^2 n_{\rm col} v_{\text{rel}}},
\end{equation}
where $a$ is the radius of the grains, $n_{\rm col}$ is the number density of the colliding grains and $v_{\text{rel}}$ is the average relative velocity between dust grains. We can take $a_{\rm col}=$ \SI{0.192}{\micro\meter} from model C in Section \ref{sec:dust_stellar_SED}. 

While it is difficult to know precisely the number density of the dust, we can give an order of magnitude estimate by considering a spherical shell with width $\Delta r$ where the dust resides in shortly after formation. The number density of dust in this spherical shell is given by:
\begin{equation}
    n_{\rm col} \approx \frac{N_{\rm col}}{\frac{4\pi}{3}((r_{in}+\Delta r)^3 - r_{in}^3)} 
\end{equation}
where $N_{\rm col}$ is the total number of dust in the spherical shell, $r_{in} = 228$ au is where the dust is formed. 

One way to estimate $N_{\rm col}$ is to use $n_3$, the radial number density of dust derived in Section \ref{sec:dust_stellar_SED}, where $N_{\rm col} = n_3 \Delta r$. Depending on the value of $\Delta r$ taken (between 100 au and $1000$ au), $n_{\rm col}$ can vary between $\sim 10^{-8}-10^{-10} \mathrm{~cm}^{-3}$. Another way to estimate $N_{\rm col}$ is based on the dust production rate $\dot{M}_d$. If we assume that the total mass of dust is conserved as dust propagates out, 
\begin{equation}
\label{eq:Ncol}
    N_{\rm col} \approx \frac{\dot{M}_d \Delta r}{\frac{4\pi}{3} \rho_d a_{\rm col}^3 v_\infty}
\end{equation}
where $\rho_d \approx 1.81 \mathrm{~g~cm}^{-3}$ is the bulk density of the dust. This estimates $n_{\rm col}$ to be $\sim 10^{-9} - \sim 10^{-10} \mathrm{~cm}^{-3}$. Taking an upper limit between the two estimates ($n_{\rm col}\sim 10^{-8} \mathrm{~cm}^{-3}$), the collision timescale of dust in this spherical shell is given by: 
\begin{equation}
\begin{aligned}
    \tau_{\text{col}} \approx & 20.1 \left(\frac{a_{\rm col}}{192 \mathrm{~nm} }\right)^{-2} \left(\frac{v_{\text{rel}}}{340 \mathrm{~km ~s}^{-1}}\right)^{-1}\\ & \qquad\quad \times \left(\frac{n_{\rm col}}{10^{-8} \mathrm{~cm}^{-3}}\right)^{-1}   \mathrm{yr}.
\end{aligned}
\end{equation}
If the power law grain size distribution in the outer dust structures is indeed the result of collisions of these dust grains in the center, the collision timescale has to be much less than the dynamical timescale $t_{\text{dyn}}$. The relative velocity between dust grains has to be greater than:
\begin{equation}
\begin{aligned}
    v_{\text{rel}} & \gg 340 \left(\frac{a_{\rm col}}{192 \mathrm{~nm} }\right)^{-2} \left(\frac{t_{\text{dyn}}}{20.1 \mathrm{~yr}}\right)^{-1} \\ & \qquad\qquad\times \left(\frac{n_{\rm col}}{10^{-8} \mathrm{~cm}^{-3}}\right)^{-1}   \mathrm{~km ~s}^{-1}.  
\end{aligned}
\end{equation}
If we take the dynamical timescale of the dust to be the orbital period of WR~112 and its companion, then, the relative velocity between dust has to be greater than \SI{340}{\kilo\meter\per\second}, which is roughly $28\%$ of the terminal wind speed of WR~112. 
The dynamical timescale could be even shorter than the orbital period. 
For example, it could be the time the dust grains stay in this denser spherical shell before being accelerated to terminal speed and enter regions with even lower density of dust. This could be much shorter than the orbital period of WR~112, requiring an even faster relative speed between the dust grains. 

To further assess the plausibility of collisions in the system, we estimate the stopping time, which characterizes the timescale on which the velocity of the dust is reduced significantly by the drag from the gas, for the dust.
For the dust to form, the gas must cool to temperatures below the dust nucleation temperature ($\lesssim 2000$ K). The sound speed at $2000$ K is roughly $ c_s \simeq \sqrt{kT/m_\text{H}} \approx 3.52 \mathrm{~km ~s}^{-1}$. 
If the dust grains are traveling at velocities $v_{\text{rel}} \gg$ \SI{340}{\kilo\meter\per\second}, they should be very supersonic and experience significant drag. The stopping time for a dust with radius $a_{\text{grain}}$ is \citep{Draine1979, Commeron2023}:
\begin{equation}
    \tau_{\text{stop}} = \sqrt{\frac{\pi \gamma_h}{8}} \frac{\rho_d}{\rho_g} \frac{a_{\text{grain}}}{c_s} \bigg[ 1 + \frac{9\pi}{128} \bigg( \frac{v_{\text{rel}}}{c_s} \bigg)^2 \bigg]^{-1/2},
\end{equation}
where $\gamma_h$ is the specific heat ratio of the gas, which equals $5/3$ for monoatomic gas, and $\rho_g$ is the gas density. Since the speed of the dust is much faster than the thermal speed ($v_{\text{rel}} \gg c_s$), the stopping time can be approximated as:
\begin{equation}
    \tau_{\text{stop}} \approx \frac{4\sqrt{\gamma_h}}{3} \frac{\rho_d}{\rho_g} \frac{a_{\text{grain}}}{v_{\text{rel}}}.
\end{equation}
It is very difficult to know the gas density, as the gas is not observed. Similarly to the dust density, we can estimate the average gas density in the uniformly expanding spherical shell with width $\Delta r$ using the mass loss rate of WR~112:
\begin{equation}
\label{eq:rhog}
    \rho_g = \dot{M}_{\text{WR}} \frac{\Delta r}{v_{\infty}} \frac{1}{\frac{4\pi}{3}((r_{in} + \Delta r)^3 - r_{in}^3)},
\end{equation}
where $\dot{M}_{\text{WR}}$ is the mass loss rate of WR~112 and $v_{\infty}$ is the terminal speed of the stellar wind. Using Equation \ref{eq:Ncol}, the ratio of the stopping time to the collision time can be written as:
\begin{equation}
    \frac{\tau_{\rm stop}}{\tau_{\rm col}} \approx 4 \sqrt{\gamma_h} \frac{\dot{M}_d}{\dot{M}_{\text{WR}}},
\end{equation}
which only depends on the mass loss rate of WR~112 and the dust production rate. With a mass loss rate of $2.7 \times 10^{-5}$ $M_{\odot}~\mathrm{yr^{-1}}$ estimated in Section \ref{sec:stellarsed} and dust production rate of $\sim (1$--$2) \times 10^{-7} ~M_{\odot}~\mathrm{yr^{-1}}$, $\tau_{\rm stop} \approx 0.02$--$0.04~ \tau_{\rm col}$.
This suggests that the stopping time is always less than the collision time for supersonic dust, independent of the relative speed between the dust grains. 

Our estimates of collision time suggests that the relative velocity between \SI{0.2}{\micro\meter} dust grains needs to be higher than \SI{340}{\kilo\meter\per\second} for there to be enough collisions. Nevertheless, there cannot be dust traveling at such supersonic speed relative to the gas, as it would be quickly coupled to the gas by drag, considering the relatively short stopping time. For there to be collisions that drive the system to a steady-state power-law grain size distribution, the gas has to be turbulent and carry the dust to over \SI{340}{\kilo\meter\per\second} ($28\%$ of the gas terminal speed). Such a level of turbulence is still possible given that the dust forms at the shock of the colliding wind from WR~112 and its companion. 
However, we do emphasize that these estimates of timescales are order-of-magnitude estimates and there can be large uncertainties, such as our approximation for the number density of dust or gas in Eqs.~\ref{eq:Ncol} and \ref{eq:rhog}. 

\subsubsection{Time-varying power law distribution}
\label{sec:power_vary}
\begin{figure}
    \centering
    \includegraphics[width=0.9\linewidth]{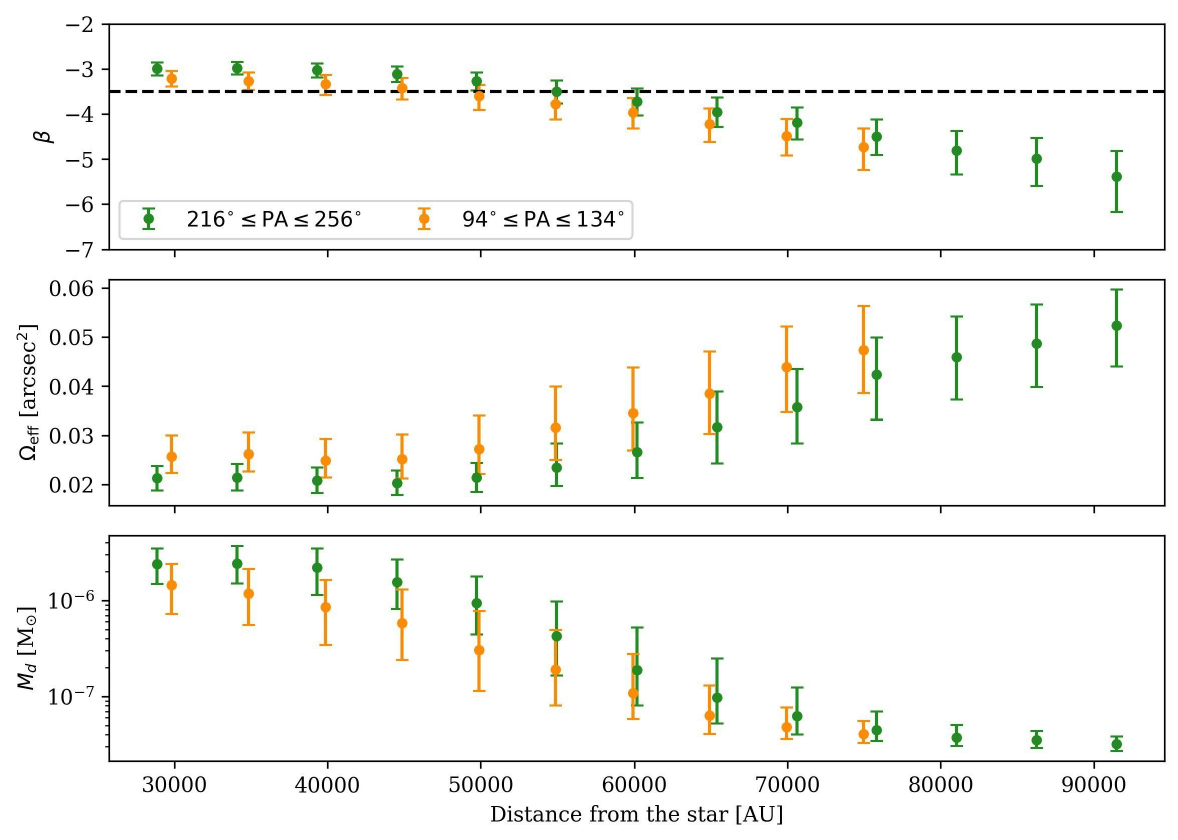}
    \caption{Power law index (top), effective solid angle (middle) and total mass of the dust in the aperture (bottom) for dust in apertures in the first (green) and second sectors at different distance away from the star. The black dashed line in the top panel indicates the power index for a collisional steady-state $\beta=-3.5$. }
    \label{fig:alter_powerlaw}
\end{figure}

Importantly, adopting a power law grain size distribution only provides an acceptable, but not fully satisfactory, reproduction of the observed spatially resolved SED ($\chi^2 = 144$). The model underestimates the emission at \SI{21}{\micro\meter} for the inner parts of the dust structure and at \SI{7.7}{\micro\meter} for the outer parts. To understand this discrepancy quantitatively, we fit the SED of each aperture individually, with fixed lower bounds and upper bounds for the grain size distribution but free normalizations and power law index, different for each aperture. The results are summarized in Figure~\ref{fig:alter_powerlaw}.

We found that the power law index decreases as the dust travels further away from the star. This suggests that there are relatively more larger grains closer to the star and more smaller grains further away from the star. While the effective solid angle of dust increases with distance from the star, the total mass of dust decreases when most of the mass is contained in the larger grains. This hints at the continual destruction of larger grains at $r>25{,}000$ au, and only a fraction of the mass gets converted into smaller dust grains. 

Nevertheless, there lacks physical mechanisms that can explain such variations in the grain size distribution. 
Collisions are unlikely the cause as the number density of dust at $r>25{,}000$ au is very low and there should be a limited amount of turbulence. 
Since radiative torques only disrupt grains above a certain critical grain size and the critical grain size increases with distance from the star, it cannot account for the decrease in number of larger grains with increasing distance from the star (\citealt{Hoang2019}, see more in Section \ref{sec:disc_size_two}). Our previous estimates of stopping time suggests that the dust should be coupled to the gas. Taken together with the lack of any misalignment among the three JWST images at different wavelengths, this indicates that grains of different sizes share the same velocity, making differential motion between them an unlikely explanation.

We also caution that the parameters of this model have considerable uncertainties. A static power index of $-3.5$ is consistent with the data in 2--3$\sigma$. The variable–slope model achieves $\chi^2 = 36$ with 46 parameters, compared to $\chi^2 = 147$ with only 5 parameters in the fixed-slope case. 
Yet, it is difficult to determine if variations in the power law is statistically significant. Information criteria such as the Bayesian information criterion cannot be applied, since the uncertainties in the SED are not precisely known and assumed at $20\%$.

By contrast, the two-grain-radii model has only 5 free parameters and yields $\chi^2$ of $30.1$. It performs considerably better than both cases of the power law size distribution, no matter what the uncertainties in the SED are. For this reason, we favor the two-grain-radii grain size distribution over the power law.

\subsection{Bimodal grain size populations}
\label{sec:disc_size_two}
The emission from two grain populations, each characterized by a single grain radius and constant in number as the dust expands outward, is in excellent agreement with the observed spatially resolved SEDs.
The population of smaller grains has radius around $1.25^{+0.25}_{-0.16}$ \SI{}{\nano\meter}, while the population of larger grains has radius around $134^{+84}_{-26}$ \SI{}{\nano\meter} and is $2.85^{+1.27}_{-0.82} \times 10^{-5}$ less abundant in number. The relatively small uncertainties for the grain radii suggests that the number of dust grains with intermediate sizes is small, which also explains why a power law grain size distribution does not reproduced the observed SED well.
The presence of a population of slightly larger grains with radius of $100-200$\SI{}{\nano\meter}, whose abundance is lower by $\sim 10^{-5}$, is also independently supported by the unresolved infrared observations. This evidence further favors the bimodal grain size distribution over alternative distributions.  

Interestingly, the SED of WR 140 can also be reproduced with two populations of grains, including one with radius $\sim$ \SI{1}{\nano\meter} \citep{Lau2023}. The larger grains in WR~112 are estimated to have a radius of 30--50 \SI{}{\nano\meter}, which is $3-5$ times smaller than our estimate for the larger grains in WR~112.

Importantly, the bimodal distribution partially resolves the conflict between the grain size distribution derived by previous works. In WR~112 and potentially other WC systems, both small grains of a few nanometers and sub-micron grains are present. As discussed in Section \ref{sec:dust_stellar_SED}, the emission is dominated by the larger grains closer to the star, leading to a large grain size inferred from the observations (e.g. in \citealt{Marchenko2002}). In other cases, the stochastically heated small grains that are much more abundant in number could contribute more significantly to the emission. 


The excellent reproduction of the observed SEDs, without requiring the grain size distribution to evolve in any way over time, is noteworthy. 
This is consistent with our discussion in Section \ref{sec:disc_size_power} that there lacks physical mechanisms that can alter the grain size distribution of the outer structures. Collisions happen at very low frequency at $r>25{,}000$ au, where the dust number density and turbulence level in the gas should be low. Because the grain size above which radiative torques can cause disruption becomes larger farther from the star \citep{Hoang2019}, any large grains that already are able to survive near the star should survive without being disrupted as they move outward. 

\subsubsection{Theoretical challenge: origin of bimodality}

The challenge with two grain populations with different radii is positing the mechanisms by which dust evolves into these two populations shortly after formation. These mechanisms must exhibit bimodality: they allow grains with quite different radii, $\sim$ \SI{134}{\nano\meter} and $\sim$ \SI{1}{\nano\meter}, to exist, but also create a ``grain size valley'' in between, where intermediate-sized grains are scarce. 

One possible explanation is that the bimodal size distribution is already established at the dust formation process: only grains with radius of $\sim$\SI{1}{\nano\meter} and $\sim$\SI{0.134}{\micro\meter} are allowed to condense. \citet{Rotundi1998} found that amorphous carbon dust can be in a lot of different structures, including chain-like aggregates and more bulky structures (see more in Section \ref{sec:disc_chem}). The chain-like aggregates can be $\sim 100$ nm in length and consist of more fine-grained clumps with size of a few nanometers, while the bulky structures can be sub-micrometer to micrometer in size. The two populations of grains inferred could correspond to the different structures formed: individual clumps in the chain-like aggregates might behave as nanometer-sized grains, especially if the chains can be broken down in a violent environment, and the bulky structures can be the grains with radius $\sim$ \SI{100}{\nano\meter}. However, the condensation process of dust in WR~112 and other WC systems is still poorly understood. 

Other explanations of the bimodal size distribution require dust destruction mechanisms, in which case the grain size distribution is altered shortly after the initial condensation of dust.
Erosion is a possible mechanism that exhibits bimodality. When larger grains are eroded by an initial small population of nanometer-sized grains at low velocity ($\sim10$\,m/s), fragments that have mass comparable or slightly larger than the smaller grains are ejected from the larger grains \citep{Seizinger2013}. This would increase the abundance of small grains without producing intermediate-sized grains. However, given the low number density of dust, it is uncertain if the erosion rate is high enough to produce a large population of small grains. 

Grain destruction mechanisms that only destroy grains with a certain range of sizes is the most likely explanation for the bimodality. If dust nucleation is rapid, small grains can condense and coagulate into large grains on very short timescales. Destruction mechanisms then subsequently destroy grains of intermediate sizes, leading to a bimodal size distribution.  

A fragmentation barrier due to collisions has been invoked to explain the bimodality in sizes in the formation of planetesimals. There is only a population of small bodies constantly replenished by the destruction of intermediate bodies and a population of large bodies that somehow pass the barrier and can grow in size \citep{Windmark2012a, Windmark2012b}. In our case, while collisions can happen, it is unknown how it can lead to a fragmentation barrier of dust at \SI{0.1}{\micro\meter}. The critical velocity of fragmentation for micron and sub-micron grains should be proportional to $a^{-5/6}$ \citep{Dominik1997}. Since dust should be coupled to the gas, grains of different sizes should be traveling at the same velocity (unless the gas is turbulent). That would suggest larger grains are more likely fragmented than smaller grains, which is contrary to the idea of a barrier at \SI{0.1}{\micro\meter}.
Similarly, thermal and non-thermal sputtering can be ruled out as sputtering leads to the destruction of grains of all sizes and quicker destruction of the nanometer-sized dust \citep{Tielens1994}. Shock processing of dust, which include the collective effects of sputtering, shattering and vaporization, also destroy larger grains more quickly and thus cannot establish such kind of barrier \citep{Jones1996}. 

\subsubsection{Radiative torque disruption}
One plausible grain destruction mechanism is the radiative torque disruption (RATD, \citealt{Hoang2019}). It happens when the torque exerted by radiation on grains with irregular shape cause them to spin so fast that they fragment into smaller pieces. Since most structures found in amorphous carbon dust in the lab are not spherical \citep{Rotundi1998}, it is likely that RATD plays a role in the system.

Quantitatively, destruction happens when the rotation speed of the grain $\omega_{\text{gr}}$ exceeds the the maximum rotation speed at which the grain can sustain the tensile stress due to rotation $\omega_{\text{disr}}$. Whether or not this happen is determined by a complex interplay between radiation, the dust grains and the gas in the environment. We can obtain order-of-magnitude estimates to understand the possible effects of RATD in the system. 

Under a constant radiation source, the angular velocity of grains increases until it reaches a steady state \citep{Hoang2019, Hoang2020}:
\begin{equation}
    \omega_{\text{gr}}(t) = \omega_{\text{RAT}} \bigg[ 1 - \exp\big(- \frac{t}{\tau_{\text{damp}}} \big) \bigg],
\end{equation}
where $\omega_{\text{RAT}}$ is the maximum rotation speed and $\tau_{\text{damp}}$ is the damping time for the angular velocity of the grains. The angular velocity of the grains is mainly damped by two mechanisms: their collisions with gas in the environment and their infrared emissions. The damping time can be written as:
\begin{equation}
    \tau_{\text{damp}}^{-1} = \tau_{\text{gas}}^{-1} + \tau_{\text{IR}}^{-1},
\end{equation}
where $\tau_{\text{gas}}$ and $\tau_{\text{IR}}$ are the rotational damping timescales due to the gas and the thermal emissions, respectively. \citet{Draine1996} have shown that the rotational damping timescales due to gas for a grain of radius $a$ is roughly:
\begin{equation}
    \tau_{\text{gas}} \simeq 3.14 \times 10^{5} \mu^{-1} \hat{a}_{-5} \hat{\rho}_3 \hat{n}_{1} \hat{T}_2^{-1/2} ~\mathrm{yr},
\end{equation}
where $\hat{a}_{-5} = a /0.1$ \SI{}{\micro\meter}, $\hat{\rho}_3 = \rho_d/3 ~\mathrm{g~cm^{-3}}$, and $\mu$ is the mean molecular weight of the gas ($2$ for pure helium). $\hat{n}_{1} = n_{\rm gas}/10~\mathrm{cm}^{-3}$, where $n_{\rm gas}$ is the number density of gas in the environment, and $\hat{T}_2 =T_{\rm gas}/100 \mathrm{~K}$, where $T_{\rm gas}$ is the gas temperature. Assuming the gas is $100\%$ helium (which does not change our order-of-magnitude estimation), we can estimate the gas number density of WR~112 using the mass density: $n_{\rm gas} = \rho_{\rm gas}/m_{\rm He}$, where $m_{\rm He}$ is the mass of a single helium atom.
As before, we assumed the gas temperature to be $\sim 2000$\,K. 

The rotational damping timescale due to dust thermal emissions for grains at thermal equilibrium is roughly \citep{Draine1996, Draine1998}:
\begin{equation}
\label{eq:tau_IR}
    \tau_{\text{IR}} \simeq 1.60 \times 10^{5} \frac{\hat{\rho}_3 \hat{a}_{-5}^3}{\langle Q_{\rm abs}\rangle} \left(\frac{T_{\rm d}}{18~\mathrm{K}}\right)^{2} U_{\rm rad}^{-1} ~\mathrm{yr},
\end{equation}
where $T_{\rm d}$ is the dust temperature. $U_{\rm rad}$ is the ratio of the energy density of radiation from the stars to the energy density of interstellar radiation:
\begin{equation}
    U_{\rm rad} = \frac{1}{u_{\rm isrf}} \frac{1}{4\pi r^2 c} \int S_{\ast} \, d\lambda,
\end{equation}
where $u_{\rm isrf}=8.64\times 10^{-13} \mathrm{~erg~cm}^{-3}$, and $c$ is the speed of light. 
$\langle Q_{\rm abs}\rangle$ is the Planck mean absorption efficiency, which is given by:
\begin{equation}
    \langle Q_{\rm abs}\rangle = \frac{\int Q_{\rm abs}(a,\lambda) S_{\ast}(\lambda) \, d\lambda}{\int S_{\ast}(\lambda) \, d\lambda}.
\end{equation}
We emphasize that to obtain more accurate results under the strong radiation from the WR star and for carbonaceous dust, Eq.~\ref{eq:tau_IR} is preferred over approximations in \citet{Draine1998} and \citet{Hoang2019}. Both $Q_{\rm abs}(a,\lambda)$ and $T_{\rm d}$ that are specific to WR~112 have been obtained in Section \ref{sec:tempgrid}.

Radiative torque has been calculated numerically using discrete dipole scattering \citep{Draine1994}. \citet{Lazarian2007} and \citet{Hoang2014} found that the dependence of torque on grain sizes can be captured analytically using the average torque efficiency $\bar{Q}_{\Gamma}$. The radiative torque is proportional to $a^2 \bar{Q}_{\Gamma}(a)$, and $\bar{Q}_{\Gamma}(a)$ is given by:
\begin{equation}
\label{eq:torqueeff}
    \bar{Q}_{\Gamma}(a) \simeq
    \begin{cases}
        2 \left( \frac{\bar{\lambda}}{a} \right)^{-2.7}   & a \leq a_{\rm trans}, \\[1pt]
        0.4 & a > a_{\rm trans},  
    \end{cases}
\end{equation}
where $\bar{\lambda}$ is the mean wavelength of radiation received by the grain and $a_{\rm trans}$ is the radius at which there is a regime change in the behavior of torque efficiency. This transition radius $a_{\rm trans}$ is estimated empirically to be roughly $\sim \bar{\lambda}/1.8$ \citep{Lazarian2007}, though there are relatively large uncertainties.

Using the average torque efficiency, an analytical form can be derived for the maximum rotation speed of the grain the grain can reach under radiative torque. After correcting for the fact that the gas is mostly helium rather than hydrogen, this maximum rotation speed is \citep{Hoang2019, Hoang2020}:
\begin{equation}
\begin{aligned}
&\left(\frac{\omega_{\text{RAT}}} {\mathrm{rad~s}^{-1}}\right) 
= \gamma_{\rm iso} \, U_{\rm rad} \, \hat{n}_1^{-1} \, \hat{T}_2^{-1/2} 
  \frac{1}{1+\mathrm{FIR}} \\ &\qquad\quad\times
\begin{cases}
  5.76 \times 10^{8} \,
  \hat{a}_{-5}^{0.7} \,
  \bar{\lambda}_{0.5}^{-1.7}, 
  & a \leq a_{\rm trans}, \\[6pt]
  1.07 \times 10^{10} \,
  \hat{a}_{-5}^{-2} \,
  \bar{\lambda}_{0.5}, 
  & a > a_{\rm trans},
\end{cases}
\end{aligned}
\end{equation}
where $\gamma_{\rm iso}$ is the degree of anisotropy of the radiation source ($\gamma_{\rm iso}=1$ for unidirectional radiation), $\mathrm{FIR} = \tau_{\text{gas}} / \tau_{\text{IR}}$ is the ratio of the damping timescales, and $\bar{\lambda}_{0.5} = \bar{\lambda}/0.5$ \SI{}{\micro\meter}. 

By analyzing the tensile stress due to rotation, the maximum rotation speed the grain can endure is given by \citep{Hoang2019, Hoang2020}:
\begin{equation}
    \omega_{\text{disr}} \simeq 1.09 \times 10^9 \hat{\rho}_3^{-1/2} \hat{a}_{-5}^{-1} \hat{S}_{\rm max, 7}^{1/2}~\mathrm{rad s}^{-1},
\end{equation}
where $\hat{S}_{\rm max, 7} = S_{\rm max}/10^7 \mathrm{~erg~cm}^{-3}$, and $S_{\rm max}$ is the maximum tensile strength of the carbon grains. 
The time it takes the grain to reach the critical rotation speed is then given by:
\begin{equation}
\label{eq:tau_RAT}
    \tau_{\rm RAT} \simeq = - \tau_{\rm damp} \ln \left( 1 - \frac{\omega_{\text{disr}}}{\omega_{\text{RAT}}} \right).
\end{equation}

\begin{figure*}
    \centering
    \includegraphics[width=0.58\linewidth]{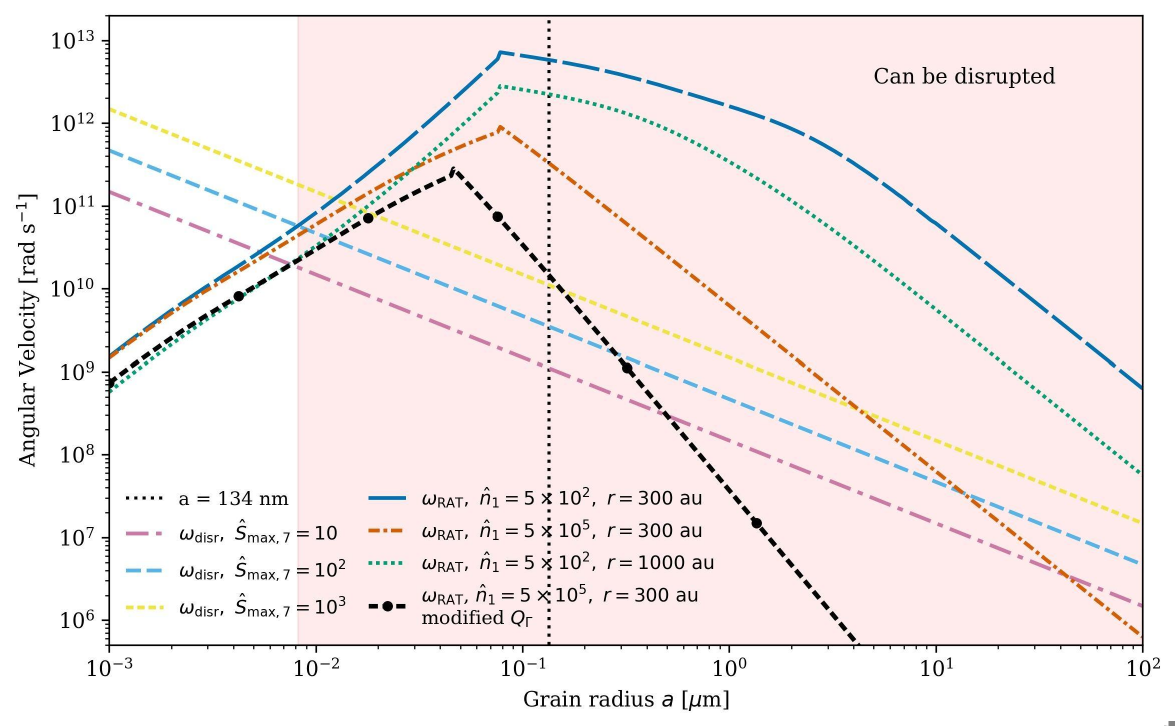}
    \caption{The maximum angular velocity of grains due to radiative torque $\omega_{\text{RAT}}$ and the critical angular velocity of grains under tensile stress $\omega_{\text{disr}}$, with different number density of gas $\hat{n}_1$, maximum tensile stress the grain can endure $\hat{S}_{\rm max, 7}$, and distance from the star $r$. The black vertical dotted line indicates the radius of the larger grains inferred from the observed SED. The red shaded regions are grains that in theory can be disrupted at $r=300$ au with $S_{\rm max} = 10^9 \mathrm{~erg~cm}^{-3}$. The black dashed line with circular markers indicate the $\omega_{\text{RAT}}$ if the torque efficiency is modified into Equation \ref{eq:torqueeff_mod} and there is enhanced damping.}
    \label{fig:RATD_omega}
\end{figure*}

Figure \ref{fig:RATD_omega} shows our estimates for the maximum angular velocity of grains due to radiative torque $\omega_{\text{RAT}}$ and the critical angular velocity of grains under tensile stress $\omega_{\text{disr}}$ in WR~112. When $\omega_{\text{RAT}}>\omega_{\text{disr}}$, the grains can be disrupted. We see that change in the gas number density or maximum tensile stress does not affect the critical radius (the radius above which grains are disrupted) significantly. This is because damping due to thermal emissions is dominant for these small grains. We are thus able to deduce a robust estimate for the smallest grains that can be disrupted by RATD. Our analysis in Section \ref{sec:dust_stellar_SED} suggests that the dust most close to WR~112 is at $r=200-400$ au and our calculation in Section \ref{sec:disc_size_power} estimates the gas number density to be given by $\hat{n}_{1} \sim 100$--$500$. Under such conditions, all grains with radius larger than $\sim 8.2~\mathrm{nm}$ can reach $\omega_{\text{RAT}} > \omega_{\text{disr}}$ and be disrupted. 

Nevertheless, unlike in the interstellar medium, the dust is not static in WR~112 and is driven away from the star. The dust has to reach the critical rotation speed fast enough before the radiation level and the exerted torque decrease significantly. The time it takes for grains smaller than $a_{\rm trans}$ to reach the critical rotation speed ($\tau_{\rm RAT}$) is estimated to be less than a day using Equation \ref{eq:tau_RAT}. At roughly \SI{1200}{\kilo\meter\per\second}, the dust travels by less than $0.6$ au in such short period of time, allowing our estimates to be applicable. This estimation explains the abundance of nanometer-sized grains around WR~112: slightly larger grains are all disrupted rapidly and fragmented into nanometer-sized grains.

However, this also poses a problem to the existence of \SI{0.134}{\micro\meter} in the outer dust structures around WR~112 inferred from the SED. Our estimates of $\tau_{\rm RAT}$ suggest that grains smaller than $\sim$\SI{10}{\micro\meter} can reach the critical angular velocity fast enough to be destroyed, even if there is stronger damping due to higher densities of the gas or if $S_{\rm max}$ is larger. The \SI{0.134}{\micro\meter} grains couldn't survive to reach the outer dust structures.

While our results point to a potential problem, the large uncertainties in the estimates for larger grains mean that this issue cannot be considered definitive.
In particular, the torque efficiency $Q_{\Gamma}$ is subject to very large uncertainties. Equation \ref{eq:torqueeff} is often adopted just for convenience. Simulations of \citet{Herranen2019} found that the actual torque efficiency deviates from the analytical form in Equation \ref{eq:torqueeff}. It has been observed that the torque efficiency flattens out starting at $a \gtrsim \bar{\lambda}/5 - \bar{\lambda}/3$, earlier than the empirically adopted transition radius $a_{\rm trans} =\bar{\lambda}/1.8$. For $a \gtrsim \bar{\lambda}/3$, the uncertainties in $Q_{\Gamma}$ are so large ($\sim0.5$ order of magnitude) that it is difficult to determine if $Q_{\Gamma}$ remains constant for grains with $a > a_{\rm trans}$. 
For grains that are more spherical, the torque efficiency in Equation \ref{eq:torqueeff} is larger than the simulation by over a factor of $2$, and $Q_{\Gamma}$ can even decrease with increasing grain size for $a > a_{\rm trans}$. 

Besides uncertainties in the torque efficiency law, the significance of damping could be inaccurate as well. Our estimates of the gas number density do not take into account of the collisions of the stellar winds. The collisions likely compress the gas to much higher number density, which allows the dust to nucleate. 
There could also be additional sources of damping besides gas collisions and thermal emissions, such as electromagnetic interactions of the grain molecules with ions from the stellar winds. 

Given the uncertainties, we consider modifications to Equation \ref{eq:torqueeff} and demonstrate that, in principle, such changes can resolve the problem. We propose the modified torque efficiency to be given by:
\begin{equation}
\label{eq:torqueeff_mod}
    \bar{Q}_{\Gamma}(a) \simeq
    \begin{cases}
        0.60 \left( \frac{\bar{\lambda}}{a} \right)^{-2.7}   & a \leq \bar{\lambda} /3, \\
        0.010 \left( \frac{\bar{\lambda}}{a} \right) & a > \bar{\lambda} /3.
    \end{cases}    
\end{equation}
The regime transition in torque efficiency is shifted to $\bar{\lambda}/3$.
The modified law also allows $Q_{\Gamma}$ to decrease with increasing grain size at $a > \bar{\lambda} /3$. Despite these differences, this modified torque efficiency remains consistent with 1--2$\sigma$ uncertainties of the values computed from discrete dipole scattering codes for carbon spherical grains in \citet{Herranen2019}.  In addition to modification in the torque efficiency law, we consider the case where the gas number density is increased by a factor of $10^3$ to $5 \times 10^{6} \mathrm{~cm}^{-3}$ due to compression of the gas by the collisions of the stellar winds. 

The maximum rotation speed with the modified torque efficiency law and denser gas is shown as the black dashed line with circular markers in Figure \ref{fig:RATD_omega}. 
The figure shows that, while the minimum grain size disrupted remains mostly unaffected, $\omega_{\rm RAT}$ drops with a steeper slope starting at $a \sim 50~\mathrm{nm}$. $\omega_{\rm RAT}$ intersects with $\omega_{\rm disr}$ at $a_{\rm disr, max}$= $0.1-0.2$ \SI{}{\micro\meter} if we assume $S_{\rm max} = 10^{10} \mathrm{~erg~cm}^{-3}$ and $\sim$\SI{0.3}{\micro\meter} if $S_{\rm max} = 10^{9} \mathrm{~erg~cm}^{-3}$ is adopted. Under this scenario, it is possible for grains with radius \SI{0.134}{\micro\meter} to survive in the system. 

We emphasize that these estimates only show that it is possible, in principle, for RATD to only disrupt grains between \SI{10}{\nano\meter} and \SI{0.1}{\micro\meter}, given the large uncertainties in the torque efficiencies and the circumstellar environments.
More comprehensive models of RATD and simulations that include time-varying radiation, more accurate estimate of the torque efficiency for larger grains, and accurate estimates for the circumstellar gas density are required to confirm if our speculation is really plausible. However, our result highlights the potential importance of RATD to account for grain size distribution in WR binaries where the radiation is so strong that radiative torques could destroy large grains. It also suggests how such a mechanism is able to disrupt grains with only intermediate sizes between some minimum and maximum grain size cutoffs and leads to a bimodal grain size distribution, such as the one inferred for WR~112. 

\subsubsection{Radiation-driven sublimation}
Finally, we propose that UV radiation, in addition to exerting torques, can heat grains to the point of sublimation, which may also create a ``grain size valley'', leading to a bimodal grain size distribution. 
Large grains remain at relatively low temperatures, because they have large heat capacities and emit effectively to cool down. Very small, nanometer-sized grains have very small absorption cross section and heat capacities. They absorb photons almost one at a time and have time to cool down before the next absorption. This results in stochastic heating, where the temperature experiences occasional short spikes. 
Conceivably, a fraction of these nanometer-sized grains are ``lucky'' enough to survive if they do not absorb high-energy photons in a row and remain at relatively low temperatures. On the other hand, the intermediate-sized grains, with larger absorption cross sections than the small grains but smaller heat capacities than the large grains, are heated relatively continuously, remain at high equilibrium temperatures, and eventually sublimate.

\begin{figure}
    \centering
    \includegraphics[width=0.99\linewidth]{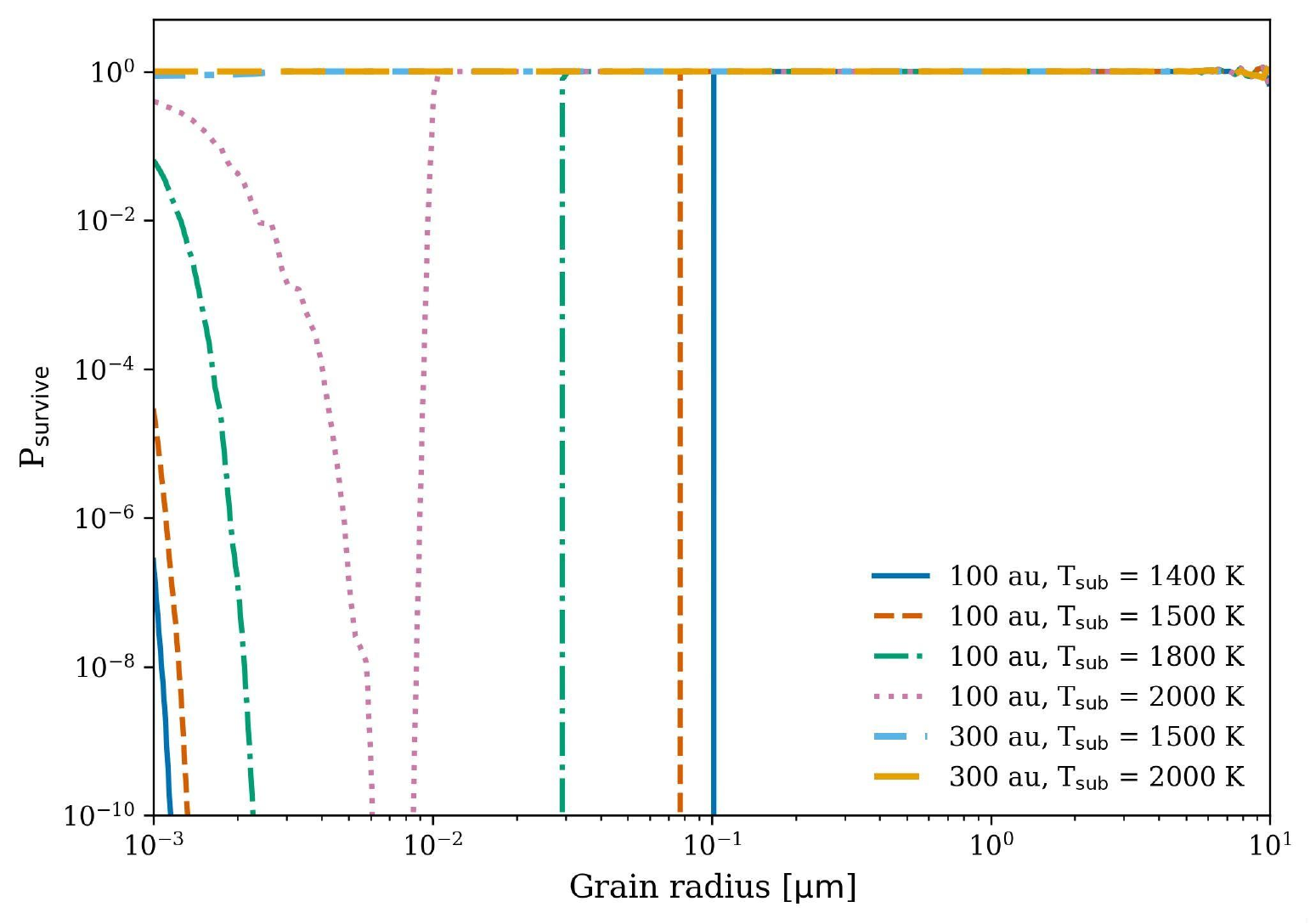}
    \caption{The probability of the dust to be below the sublimation temperature using the steady-state temperature distribution computed by DustEM, for different sublimation temperatures and at different distance from the star. }
    \label{fig:Psurvive}
\end{figure}

To understand this quantitatively, we can compute the temperature distributions of grains with different sizes using DustEM, just like in Section \ref{sec:stochastic}. We can calculate the probability of the dust to be below the sublimation temperature $T_{\rm sub}$, which we refer to as the survival probability $P_{\rm survive}$. Figure \ref{fig:Psurvive} shows the survival probability for different sublimation temperatures and at different distance from the star. 
At $r=100$ au, there is a ``grain size valley'' between a few nanometer and $0.01$--\SI{0.1}{\micro\meter}, where grains with these sizes are always hotter than the sublimation temperature. Grains larger than $\sim$\SI{0.1}{\micro\meter} are always cool enough to survive. 
Nanometer-sized grains (radii $<$\SI{2}{\nano\meter} for a sublimation temperature of $1800$ K) have nonzero but very small survival probability. It is interesting that this radius cutoff is even smaller than that predicted by RATD ($\sim$\SI{8}{\nano\meter}) and agrees well with the inferred grain radius of the population of small grains in WR~112. 
The nonzero but small survival probability suggests that even these very small grains absorb photons frequently enough under the intense radiation that they do not have time to cool sufficiently before the next absorption. As a result, they are mostly above the sublimation temperature. 

Whether or not these nanometer-sized grains can eventually survive depends on the timescale at which sublimation happens and the time for which the grain is under the intense radiation. 
As shown in Figure \ref{fig:Psurvive}, while the radiation is intense enough to sublimate grains at $r=100$ au, it is no longer strong enough at $r=300$ au. If the dust travels at the terminal wind speed of WR~112, this happens in $\sim 0.8$\,yr. 
It is likely that the grains sublimate in time much shorter than $0.8$\,yr at temperatures above the sublimation temperature. This would suggest that these nanometer-sized grains should sublimate as well. 

However, our calculation is subject to a significant caveat: the temperature distribution computed by DustEM is the steady-state distribution under a constant radiation field when no sublimation is considered \citep{Desert1986, Compigne2011}. In this steady state, each grain performs random walks in the temperature, and the probability $P(T)$ is the probability of finding a grain at temperature $T$ during this random walk. In reality, if sublimation happens fast enough, sublimation would introduce an absorbing boundary at the sublimation temperature for the random walk. 
Moreover, the radiation becomes weaker over time as the grain travels further away from the star. 
As a result, the survival probability cannot be calculated as a steady state solution. It would be the probability that the grain never reaches the sublimation boundary given a set of initial conditions under the time-varying radiation field. Our simple but unrealistic steady state solution could underestimate the actual survival probability, even when the sublimation timescale is relatively short. 

In addition, the properties of the nanometer-sized grains have large uncertainties. We derived the absorption cross sections of grains from the bulk optical constants measured in laboratory. This may overestimate the absorption cross sections of the nanometer-sized grains, as their energy states are quite discrete and cannot be well approximated by a density of states. This leads to an underestimation of the time between photon absorptions. If absorption is less frequent, the very small grains would have sufficient time to cool before absorptions and remain at lower temperatures between temperature spikes. The steady-state temperature distribution also assumes the bulk heat capacities. As the energy states become more discrete, the bulk heat capacities may not be good approximations. 

Given the oversimplified steady-state temperature distribution, which does not consider sublimation, and the uncertainties in nanometer-sized grain properties, our results are inconclusive. 
Nevertheless, the probability for grains to be lower than the sublimation temperature being zero for only intermediate-sized grains clearly indicates that radiation-driven sublimation exhibits bimodality. 
The bimodality suggests that it is, in principle, possible for radiation-driven sublimation to create a ``grain size valley'' and drive the system to bimodal size distribution, under the right conditions.
More careful modeling that accounts for the time-varying radiation, the sublimation during stochastic heating and more accurate absorption cross sections or heat capacities is required to determine if this bimodality really allows nanometer-sized grains to survive. 

To summarize this section, we discussed how a bimodal grain size distribution with a ``grain size valley'' for intermediate-sized grains is able to reproduce the observed SED excellently and partially resolve the previous debates over inferred grain sizes in WR~112. We demonstrated, in principle, that radiative processes, specifically radiative torque disruption and radiation-driven sublimation, may only destroy grains between a minimum and maximum radius cutoff, driving the system to a bimodal size distribution. Current models face challenges where radiative torques would destroy the inferred population of larger grains and nanometer-sized grains would sublimate under intense radiation. These challenges likely arise from the highly simplistic assumptions and large uncertainties of these models. 
To more accurately assess whether these processes can actually drive the system into a bimodal grain size distribution, we need to develop more comprehensive models that account for the time-varying radiation field, the detailed response of dust to radiative torques, and the exact sublimation processes.

\subsection{Types of carbon grains}
\label{sec:disc_chem}
We applied our emission model to three different types of amorphous carbon grain prepared by \citet{Colangeli1995} and measured by \citet{Zubko1996}, and the results are presented in Section \ref{sec:stochastic} and Appendix \ref{Appendix:results_types}. These three types of grains correspond to different formation pathways in the laboratory: the ACAR-type grain is produced by arc discharge of amorphous carbon electrodes in inert Argon atmosphere; the ACH2-type grain is produced by arc discharge in H$_2$ atmosphere; and the BE-type grain is produced by burning of benzene in Earth air. 

The varying formation pathways leads to different chemical compositions and structures. In the laboratory, \citet{Rotundi1998} found that the ACAR-type and the ACH2-type grains are dominated by chain-like aggregates. Small fine-grained particles with diameter of $7-15$ \SI{}{\nano\meter} are organized into agglomerates, which are then arranged into chains. In addition to chain-like aggregates, other structures were also observed such as ``poorly graphitized carbon'' (loops or rings on smooth sheet) or ``bucky carbon'' (concentric onion-like structures and tubes). Some of these structures can be aggregated into patches up to micrometer size. 

Despite the shared structures, the ACAR-type and the ACH2-type grains differ due to different availability of hydrogen in the formation environment. The ACAR-type grains are hydrogen-poor, while the ACH2-type grains are more hydrogenated. The C/H ratio also affects the structures of the chain-like aggregates, which lead to differences in the optical properties. On the other hand, the BE-type grains produced by combustion of benzene are also hydrogenated.

We expect the formation environment of circumstellar dust around WR~112 to be hydrogen-poor, since WR stars and their stellar winds lack hydrogen \citep{Sander2012}. However, the young O-star companion of WR~112 may have a stellar wind dominated by hydrogen \citep{HainichR2019}. Since the dust forms at the collision front of WR~112 and the O-star companion, it is still possible for there to be some hydrogen present at the formation site of dust. The mass loss rates of the WR~112 and its companion we presented in Section \ref{sec:stellarsed} is subject to large uncertainties. Nonetheless, estimates of the ratio of wind momentum suggest that the mass loss rate of the companion has to be at least one order of magnitude lower than that of WR~112 \citep{Lau2020b}. As a result, the C/H ratio should be $\gtrsim 10$, making the formation of hydrocarbons unlikely. One possible concern is that if a large amount of carbon is locked into CO gas, the C/H ratio would be smaller, allowing more hydrogenated carbon grains to form. Nevertheless, our ALMA observations do not detect any significant amount of CO around WR~112. 

In addition, UV irradiation leads to dehydrogenation. When hydrogenated amorphous carbon grains were irradiated with a UV flux of $\sim 10^{16}$--$10^{17}$ \SI{}{\eV\per\second\per\centi \meter\squared} for a few hundred hours, the grains lose much of the hydrogen \citep{Mennella1996, Gadallah2011}. The UV flux for $\lambda = 0.1-0.3$ \SI{}{\nano\meter} at $\sim 1000$ au away from WR~112 is on the same orders of magnitude. The irradiation time is presumably longer, considering the dynamical timescale of the system. Even if grains formed initially are hydrogenated, they would soon be dehydrogenated by the UV photons from WR~112. 
Considering all the arguments discussed, we believe that the dust grains around WR~112 should be mostly hydrogen-poor carbon grains. 

Although the chemical compositions of dust does affect their optical properties and emissions, we are not able to identify the type of carbon grain using the spatially resolved mid-infrared dust SED. With a two-radii grain size distribution, the three types of grains produce very similar SED, with similar $\chi^2$ values (see Appendix \ref{Appendix:results_types}). The underlying reason is the degeneracy between the choice of grain type and the grain size, both of which affects the value of $Q_{\text{abs}}$. With the grain size distribution largely unknown, models of the continuum emissions from dust cannot place constraints on the hydrogen content of the grains. 

On the other hand, previous spectral observations provide some evidence that the grains are hydrogen-poor. 
One important interstellar feature related to hydrogenated carbonaceous dust is the absorption feature at \SI{3.4}{\micro\meter} first discovered by \citet{Soifer1976}. \citet{Sandford1991} analyzed this absorption feature and attributed it to C-H stretching in -CH$_2$ and -CH$_3$ of aliphatic hydrocarbons. 
No strong absorption feature at \SI{3.4}{\micro\meter} is observed in the ISO-SWS spectrum of WR~112 \citep{Schutte1998}. \citet{Chiar2001} measured the optical depth of WR~112 at \SI{3.4}{\micro\meter} and deduced that the weak absorption feature present is mostly of interstellar rather than circumstellar origin.  
These observations make the presence of hydrogenated amorphous carbon grains unlikely. 
Polycyclic aromatic hydrocarbons (PAH) molecules, on the other hand, have been proposed to cause an emission feature at \SI{3.3}{\micro\meter} \citep{Leger1984, Allamandola1985}. 
Such an emission features is missing from the ISO-SWS spectrum of WR~112 \citep{Schutte1998}. In addition, high flux densities observed at both \SI{3.4}{\micro\meter} and \SI{4.6}{\micro\meter} are better explained by the continuum emission from larger grains. 
More recently, \citet{Taniguchi2025} found that the mid-infrared features of dust around WR~140 are consistent with hydrogen-poor R Coronae Borealis stars. This supports the idea that the dust formed in WC binaries are hydrogen-poor. 

However, there have been studies suggesting that PAH molecules should be present in the circumstellar dust of WC stars.
Besides detecting features consistent with hydrogen-poor conditions, \citet{Taniguchi2025} also tentatively detected \SI{11.2}{\micro\meter} aromatic infrared band, which is associated with the C-H out-of-plane bending mode, in one of the dust shell in WR~140. They suggested that the hydrogenated carbon dust may acquire hydrogen after formation by the hydrogen-rich companion star of WR~140, while also cautioning that the feature could be contaminated by nearby He emission lines.
\citet{Marchenko2017} re-examined the ISO-SWS spectrum of WR~112 and other WC stars. They proposed that broad emission features at $\sim 6.5, 8.0, 8.8$ and \SI{10.0}{\micro\meter} are from PAH molecules or cations. Nevertheless, besides PAH molecules, hydrogen-poor amorphous carbon grains can also produce strong peaks at \SI{6.3}{\micro\meter} that can be redshifted to $\sim$\SI{6.5}{\micro\meter} \citep{Borghesi1987}. In labs, they were also observed to develop emission features at $7.8-$\SI{8.0}{\micro\meter} that can explain the \SI{8.0}{\micro\meter} peak observed \citep{Colangeli1995, Chiar2002, Garca-Hernndez2013}. The \SI{8.8}{\micro\meter} and \SI{10.0}{\micro\meter} features are not obvious in the spectrum of WR~112 and other WC stars (except WR 48a). 

Another potential evidence for PAH molecules is the absorption feature observed in the spectra of WR~112 and other WC stars at \SI{6.2}{\micro\meter}, which has been associated with C-C stretching of PAH molecules \citep{Schutte1998, Chiar2001}. While \citet{Schutte1998} and \citet{Marchenko2017} considered this absorption feature to be interstellar, \citet{Chiar2001} argued that it might be from C-C stretching of circumstellar aromatic carbon materials. 
A perhaps more convincing argument comes from the high-resolution high-signal-to-noise-ratio spectrum of WR~140 obtained by \citet{Lau2022}. They observed obvious spectral emission feature at \SI{7.7}{\micro\meter}, which is typically attributed to C-C stretching in PAH molecules. Nevertheless, the absence of emission feature at \SI{8.6}{\micro\meter} which is associated with C-H bending modes and the presence of emission feature at \SI{6.4}{\micro\meter}, commonly attributed to hydrogen-poor amorphous carbon grains, suggest that the environment is hydrogen-poor. 

There seems to be contradicting evidence regarding the presence of PAH molecules in WC systems. However, evidence supporting the presence of PAH molecules are not direct: the spectral features are usually not directly associated with C-H modes, with the exception of tentative detection of \citet{Taniguchi2025} for WR~140. The C-C modes contributing to \SI{6.2}{\micro\meter} and \SI{7.7}{\micro\meter} features in theory could also be from ``dehydrogenated'' variants of the PAH molecules (e.g., fully substituted hydrogen-poor aromatic compounds) or somewhat graphitized carbon grains. Therefore, we continue to support the idea that grains formed in WR~112 (and other WC systems) are mostly hydrogen-poor, even though we cannot rule out the presence of PAH molecules with certainty. 

The grain composition has implications on the interstellar dust budget. WC systems produce a significant amount of carbonaceous dust. Some WC systems such as WR~112 can be significant dust producers. \citet{Lau2020a} found, using Binary Population and Spectra Synthesis, that WC binary systems may dominate the dust production in the first few Myrs in systems with metallicities $Z \gtrsim 0.008$.
Despite their high dust production rates, if the dust formed is hydrogen-poor, WC systems do not contribute significantly to the UV bump at \SI{217.5}{\nano\meter} in extinction curves observed for interstellar medium, which is more significantly produced by hydrogenated amorphous carbon grains \citep{Mennella1996, Rotundi1998}. The detection of the UV bump in high-redshift ($z\sim 7$) galaxies thus cannot be explained by WC binary systems \citep{Witstok2023}. This may place constraints on the theoretical models of other dust producers in the early universe, such as supernovae. However, it is possible that once the dust from the WC star enters the hydrogen-rich interstellar medium, hydrogenation could occur \citep{Taniguchi2025}, although the timescale associated with this process requires further investigation.

\subsection{Robustness and sources of errors}
\label{sec:robust}
In this section, we discuss potential sources of errors that may affect the robustness of the inferred dust properties from the spatially resolved JWST and ALMA observations. 

Since the properties of dust are derived from fits to the observed SED, uncertainties and errors of the photometric data points must be carefully considered. We adopted a conservative $20\%$ uncertainty for photometry, as discussed in Section \ref{sec:dust_sed_aperture}, which should account for estimates for various uncertainties. Nevertheless, the uncertainties due to PSF subtraction are difficult to characterize given WR~112 is very bright. Since the PSF is the most prominent for the \SI{7.7}{\micro\meter} images, one concern would be that the uncertainties for flux at \SI{7.7}{\micro\meter} are underestimated, which introduces biases to the inferred grain size distribution. 

While the uncertainty is not precisely known, we found that the radial brightness profile of the \SI{7.7}{\micro\meter} image exhibits troughs between the outer apertures, which can only be attributed to a genuine decrease in the dust column density rather than to PSF effects. Thus, the emission at \SI{7.7}{\micro\meter} in outer apertures should still be largely attributed to dust. The properties of dust that rely particularly on the \SI{7.7}{\micro\meter} observations (e.g., importance of stochastic heating) should still be reliable. 

Another concern with constraining the grain size distribution from the SED would be that our models only include the continuum emission. As discussed in the previous section, the spectra of WR stars such as WR~112 can show emission features (e.g., \citealt{Schutte1998, Marchenko2017}). Prominent emission features can increase the integrated fluxes measured within a filter, causing the total flux to be overestimated relative to the actual continuum level. \citet{Endo2022} observed broad \SI{8}{\micro\meter} emission features in dust around several WR stars including WR~112. \citet{Taniguchi2025} found narrower emission features around \SI{7.7}{\micro\meter} for the resolved dust shells in WR~140. 

However, in the ISO spectrum of WR~112 examined by \citet{Taniguchi2025}, the emission bands are brighter than the continuum level by less than $30\%$. In the resolved spectra of WR~140, the emission features are even lower in magnitude. The integrated flux in the relatively broad MIRI \SI{7.7}{\micro\meter} filter should be affected by even less. The net effects are likely within the $20\%$ uncertainties we assumed. More importantly, the observed flux at \SI{7.7}{\micro\meter} is several orders of magnitudes larger than the predicted emissions if no nanometer-sized grains are present. Our inference that there are nanometer-sized grains should be robust against uncertainties due to emission bands. The bimodal grain size distribution provides a better fit to the observed SED than the other distributions because it more accurately reproduces the slope between \SI{15}{\micro\meter} and \SI{21}{\micro\meter}, which should not be affected by emission features, at least the \SI{7.7}{\micro\meter} one. As a result, the inferred grain size distribution is unlikely to be biased by the presence of the emission features.

We also note that the constraints on the grain size distribution placed by joint fitting of the SEDs in all apertures rely on the assumption that the grain properties, including grain sizes, do not vary from one aperture to another. Grain properties can evolve close to the stars, as indicated by the evolution of PAH features in WR~140 at $r\lesssim5000$ au \citep{Taniguchi2025}. However, we modeled the emission of dust structures at $r>25{,}000$ au, where evolution in grain properties should not be significant. Collisions, radiative torque disruption, and radiation-driven sublimation are insignificant at $r>25{,}000$ au away from the star, as suggested by our estimates in Section \ref{sec:disc_size_power} and \ref{sec:disc_size_two}. 
In Section \ref{sec:power_vary}, we also relaxed the assumption for the power law grain size distribution by allowing the power law index and normalization to vary as dust travels. However, even with the assumption relaxed, the power law size distribution yields poorer reproduction of the SEDs than the bimodal size distribution under the original assumption.
For these reasons, we believe that this assumption should be valid to first order and does not significantly affect the robustness of the inferred grain size distribution.

In Section \ref{sec:stellarsed}, we discussed how we are not able to constrain the stellar spectra of WR~112 and its companion most precisely with limited number of ultraviolet-to-optical observations. We adopted the median stellar spectra among acceptable models when we constrain the grain size distribution in Section \ref{sec:dust_sed}. To assess the robustness of the inferred grain size distribution against variations in the stellar spectrum, we performed MCMC fits using the 16th- and 84th-percentile stellar spectra among the acceptable models. The results are summarized in Appendix \ref{Appendix:results_varyspectra} in Tables \ref{tbl:grain_stoch_p84} and \ref{tbl:grain_stoch_p16}. 

The tables show the inferred grain size distributions when different spectra are adopted and the $\chi^2$ for the median posterior model. Some grain size distributions are able to better reproduce the observed SED when a more or less luminous spectrum is adopted. Greater grain sizes are inferred when a luminous spectrum is adopted, as expected from the degeneracy between temperature of the dust and the grain radius. 

However, the variation in the stellar spectra does not change our main conclusions. Among the four parameterizations of grain size distribution, the two-radii distribution is always the best. The inferred two-radii distribution is also relatively robust against changes in stellar spectra: the model still infers a population of grains with radius of $\sim$\SI{1}{\nano\meter} and another population of $\sim0.1$--\SI{0.2}{\micro\meter} that is $\sim 10^{-5}$ less abundant. The effective solid angle can vary by a factor of 2 as the stellar spectrum changes, but this change mainly results from variations in the grain radii. The dust production rate is $2.05^{+1.14}_{-0.47} \times 10^{-7} ~\mathrm{M}_{\odot} ~\mathrm{yr}^{-1}$ and $1.73^{+1.12}_{-0.33} \times 10^{-7} ~\mathrm{M}_{\odot} ~\mathrm{yr}^{-1}$ for the 16th- and 84th-percentile stellar spectra, respectively. These are consistent with the dust production rate of $1.93^{+1.31}_{-0.47} \times 10^{-7}~\mathrm{M}_{\odot}~\mathrm{yr}^{-1}$ inferred when the median stellar spectrum is adopted. It is thus relatively safe to conclude that our main conclusions are robust against uncertainties in the stellar spectra. 

We also emphasize that uncertainties in the distance to WR~112 should not affect the temperature of the dust. The stellar photon flux the dust receives in Equation \ref{eq:Tgrid} can be written as $\frac{S_{\ast}}{4\pi r^2} = \frac{S_{\ast}}{4\pi d^2 \theta^2}$, where $\theta$ is the distance of dust from the star measured in angles. Since $\frac{S_{\ast}}{4\pi d^2}$ is fixed by the observed stellar SED when we conducted the fits, the stellar photon flux received by the dust should be independent of the distance to WR~112. 

Another source of uncertainty comes from the modeling of the continuum emission. The optical constants of the amorphous carbon dust we adopted are measured in laboratories by \citet{Zubko1996}. We are uncertain on the range of grain size the optical constants apply to. In the lab, the amorphous carbon grains created condense into micron- or sub-micron-sized clusters or chains consisting of small grains with radii of $5-15$\SI{}{\nano\meter}. While it is generally assumed that optical constants depend only on the material, we are not entirely certain if they extend to smaller grains (e.g., \SI{1}{\nano\meter}). 

When considering stochastic heating of small grains, we used the bulk heat capacity to describe the behavior of small grains when absorbing or emitting photons. We followed \citet{Michelsen2008} and assumed that the heat capacity of graphite is a good estimate for the heat capacity of amorphous carbon. The accuracy of this approximation, however, remains uncertain, especially for the nanometer-sized grains. 

 

\section{Conclusions}
We have presented a comprehensive analysis on the dust emissions from the colliding-wind Wolf-Rayet binary WR~112. To begin with, this work presented ALMA Band 6 observations of WR~112, which are the first millimeter observations of a WC binary system capable of spatially resolving its dust structure. The observations do not detect extended dust structures, however, showing one unresolved central source, which could be explained by free-free emissions from the colliding stellar winds. 

While we detected emission from the three CO isotopologues, the close match in radial velocity with gas in the interstellar medium (\citealt{Dame2001} and \citealtaliasy{HI4PI2016}) rather than the stellar wind speed, along with the lack of spatial symmetries in the morphology of the emission, led us to believe that the CO emissions are from interstellar gas rather than stellar winds of WR~112.

The ALMA observations place an upper limit on the brightness of dust at \SI{1.3}{\milli\meter} which places stringent constraints on the grain size distribution of dust produced by WR~112. Combining these limits with JWST images at 7.7, 15, and \SI{21}{\micro\meter}, we constructed SEDs for 23 apertures around WR~112. We considered various emission models to reproduce the observed SEDs, and the main findings are as follows:
\begin{enumerate}
    \item The observed spatially-resolved SEDs and the temperature profile of dust are consistent with behaviors of hydrogen-free amorphous carbon dust.
    \item Most grains in the outer dust structures are smaller than \SI{0.5}{\micro\meter}, likely dominated by very small grains with radius $\lesssim 30~\mathrm{nm}$.
    \item Under the assumption that the grain radius distribution does not evolve in the outer dust structures, a bimodal distribution, with abundant nanometer-sized grains, some 0.1-micron grains and a ``grain size valley'' for the intermediate sizes, reproduces the observed SED excellently, despite our conservative estimates of the photometric uncertainties. A power law distribution with index $\beta \simeq -3.5$ yields a poorer, though broadly consistent, fit to the SED.
    \item It is a challenge to account for how the system is driven into the bimodal radius distribution. Collisions can be caused by turbulence in the gas, but it is uncertain how they can lead to a bimodal distribution. We propose that radiative processes, including radiative torque disruption and radiation-driven sublimation, can in principle only disrupt grains between \SI{2}{\nano\meter} and \SI{0.1}{\micro\meter}, creating a ``grain size valley'' for the intermediate-sized grains and leading to a bimodal size distribution. However, determining whether this scenario is physically realizable in WR~112 requires more comprehensive modeling, including the time-varying radiation field, the detailed dust response to radiative torques and the exact sublimation processes. 
    \item The dust production rate is sensitive to the grain size distribution adopted. Adopting either a bimodal or power-law grain size distribution yields an estimated dust production rate of $\sim10^{-7}~\mathrm{M}_{\odot}~\mathrm{yr}^{-1}$. Estimates on the total contribution of WC binaries to interstellar carbonaceous dust budget should account for this dependence.
    \medskip
\end{enumerate}

While we cannot determine the grain chemistry from SED modeling, partially because it is degenerate with the grain size distribution, we discussed evidence supporting that the dust formed in the system consists of hydrogen-free amorphous carbon grains or dehydrogenated variants of PAH molecules, even though observations of PAH features suggest that the presence of PAH molecules cannot be ruled out. 

We verified that our results are robust against uncertainties in the stellar spectra of WR~112 and its companion. We also discussed how uncertainties associated with emission features (e.g. PAH features) are expected to have only minor effects on the inferred grain size distribution, whereas those arising from the adopted optical constants and heat capacities in our emission models remain less well characterized.

Even though we have aimed for a comprehensive analysis, several limitations remain. Our parameterizations of the grain size distribution are necessarily simplified, and additional photometric data points would allow more complex distributions to be tested. The saturation of JWST at $r < 25{,}000$ au also limits our ability to confirm with certainty whether the grain size distribution evolves shortly after dust formation. Future observations of higher quality will be critical to refining these constraints, and extending our approach to other WC binaries will be essential for developing a broader understanding of dust production in these systems.

\begin{acknowledgments}
D.W. acknowledges support from the Summer Undergraduate Research Fellowships from California Institute of Technology. D.W. thanks Konstantin Batygin for helpful discussions.
Y.H. is funded by a Caltech Barr Fellowship. Y.H. acknowledges funding for JWST program \#5842 provided by NASA through a grant from the Space Telescope Science Institute. 
T.O. acknowledges the support by JSPS KAKENHI grant No. JP24K07087.
J.R.C. acknowledges funding from the European Union via the European Research Council (ERC) grant Epaphus (project number 101166008).

This paper makes use of the following ALMA data: ADS/JAO.ALMA\#2023.1.00999.S, ADS/JAO.ALMA\#2024.1.00803.S. ALMA is a partnership of ESO (representing its member states), NSF (USA) and NINS (Japan), together with NRC (Canada), NSTC and ASIAA (Taiwan), and KASI (Republic of Korea), in cooperation with the Republic of Chile. The Joint ALMA Observatory is operated by ESO, AUI/NRAO and NAOJ. This work is based in part on observations made with the NASA/ESA/CSA James Webb Space Telescope. The data were obtained from the Mikulski Archive for Space Telescopes at the Space Telescope Science Institute (\!\dataset[DOI: 10.17909/hvkk-dj46]{https://doi.org/10.17909/hvkk-dj46}), which is operated by the Association of Universities for Research in Astronomy, Inc., under NASA contract NAS 5-03127 for JWST. These observations are associated with program \#4093.

\end{acknowledgments}

\FloatBarrier
\bibliographystyle{aasjournalv7}
\bibliography{paper}{}

\newpage
\FloatBarrier
\appendix
\section{ALMA Observations of CO lines}

\subsection{Channel maps of CO isotopologues}
\label{Appendix:channel_map}

\begin{figure*}
    \centering
    \includegraphics[width=0.85\linewidth]{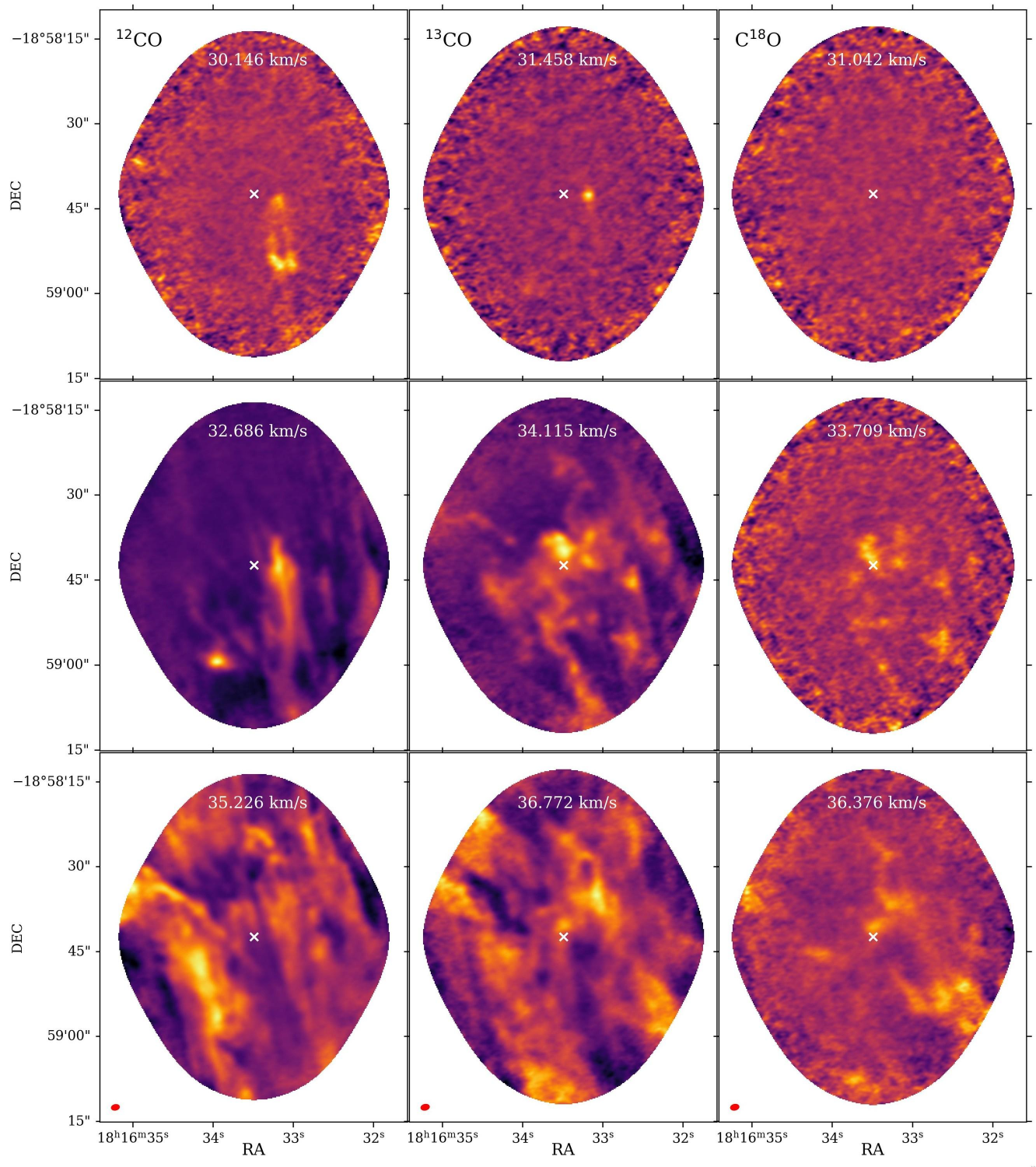}
    \caption{Channel maps of the CO isotopologues, $^{12}$CO (first column), $^{13}$CO (second column), and C$^{18}$O (third column) in velocity channels $30.0 \leq v_{\text{LSR}} \leq 37.0$ \SI{}{\kilo\meter\per\second}. The line-of-sight velocity $v_{\text{LSR}}$ is labeled at the top of each panel. A linear color scale, ranging from the minimum to the maximum emission level in each panel, is used throughout. The white cross indicates the location of WR~112 in each panel. The red ellipse on the bottom left corner of the final row indicates the average beam size for each tracer. }
    \label{fig:channel_3iso_set1}
\end{figure*}

\begin{figure*}
    \centering
    \includegraphics[width=0.9\linewidth]{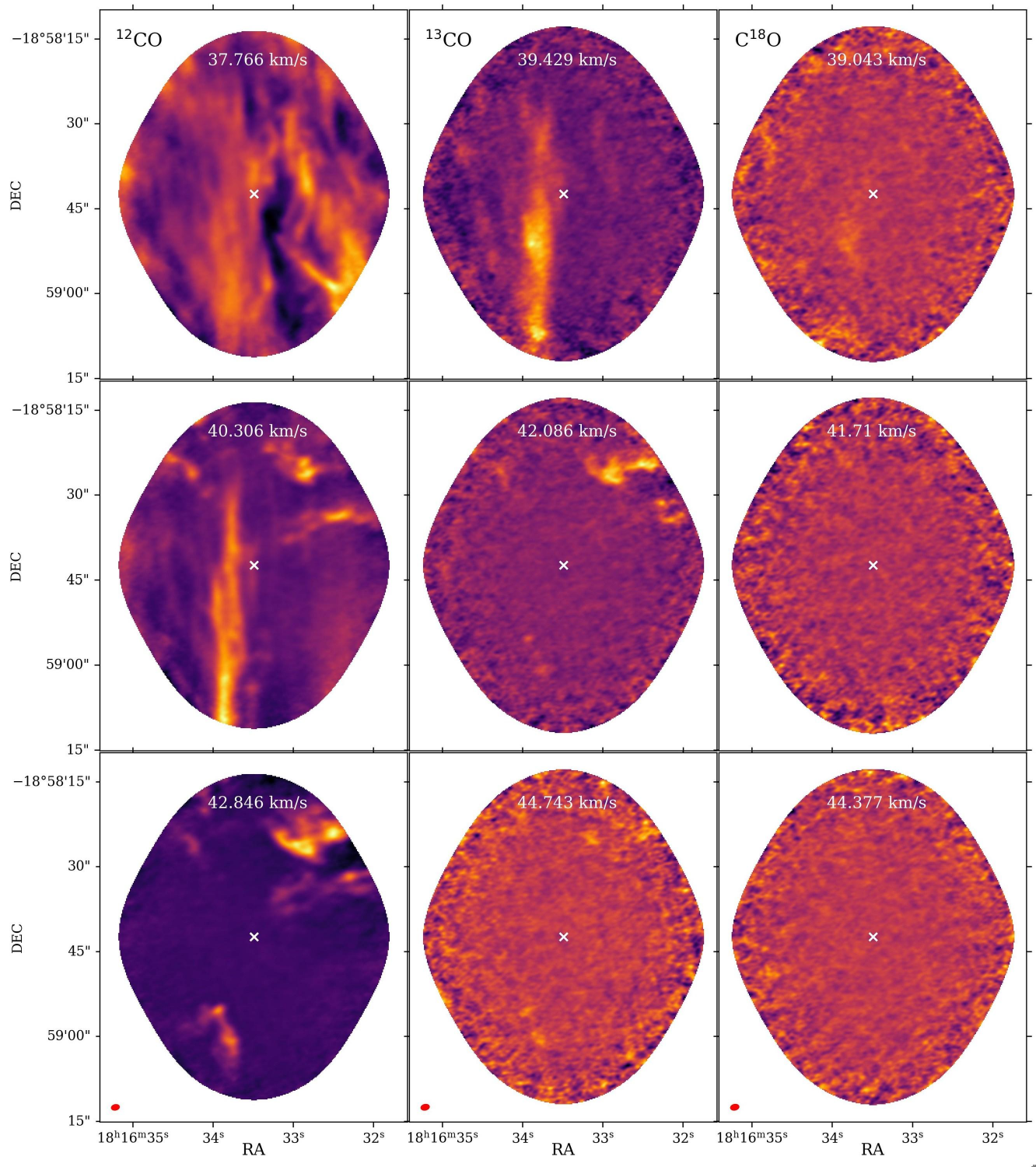}
    \caption{Same as Figure~\ref{fig:channel_3iso_set1}, but for $37.0 \leq v_{\text{LSR}} \leq 45.0$ \SI{}{\kilo\meter\per\second}.}
    \label{fig:channel_3iso_set2}
\end{figure*}

\begin{figure*}
    \centering
    \includegraphics[width=0.9\linewidth]{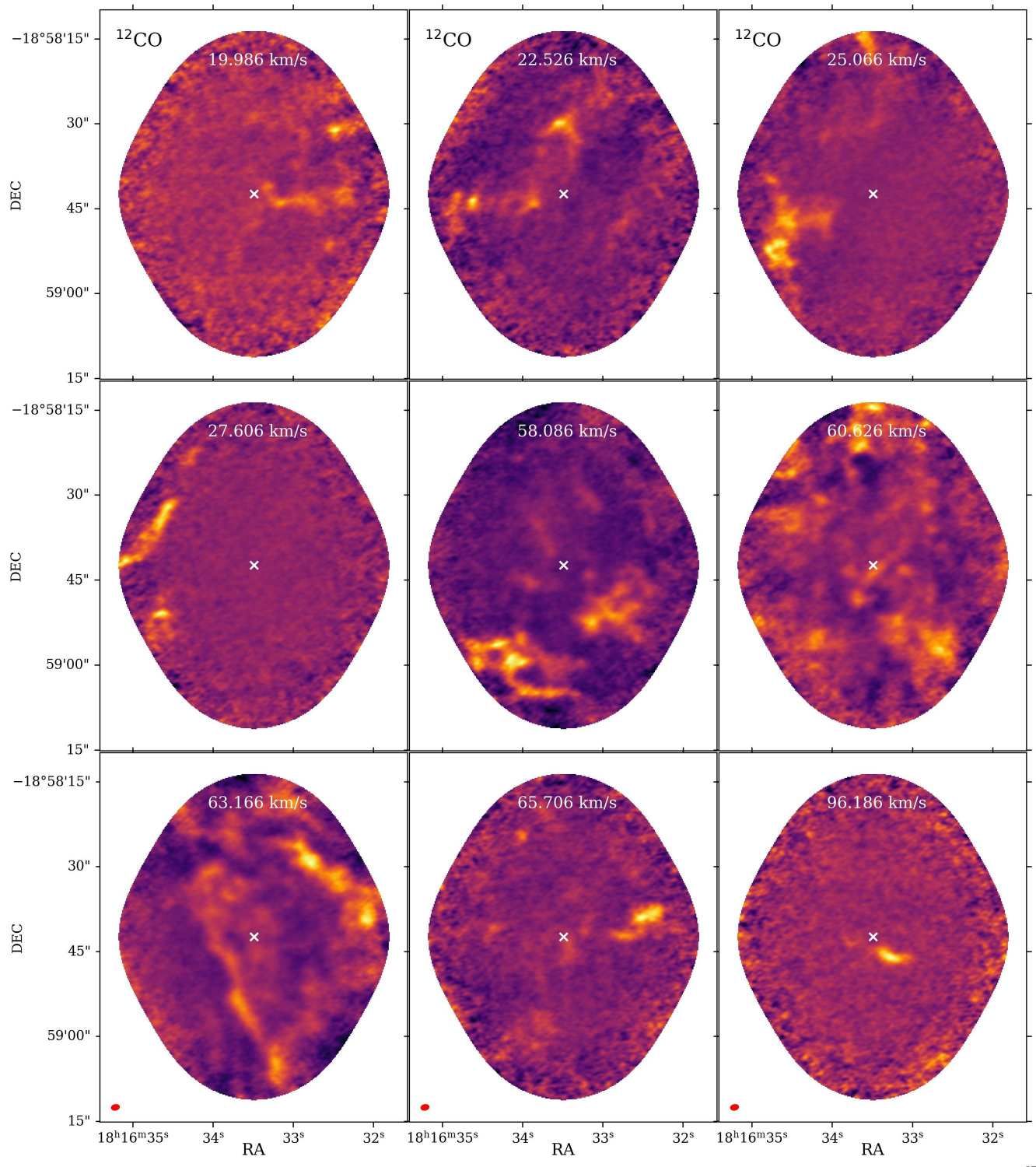}
    \caption{Channel maps of $^{12}$CO, for velocity channels where the two other isotopologues, $^{13}$CO and C$^{18}$O, do not show significant emissions. The rest is the same as Figure~\ref{fig:channel_3iso_set1}. }
    \label{fig:channel_12CO}
\end{figure*}

\FloatBarrier
\subsection{Spectra at the location of WR~112}
\label{Appendix:COspectra}
\begin{figure*}[h]
    \centering
    \includegraphics[width=0.6\linewidth]{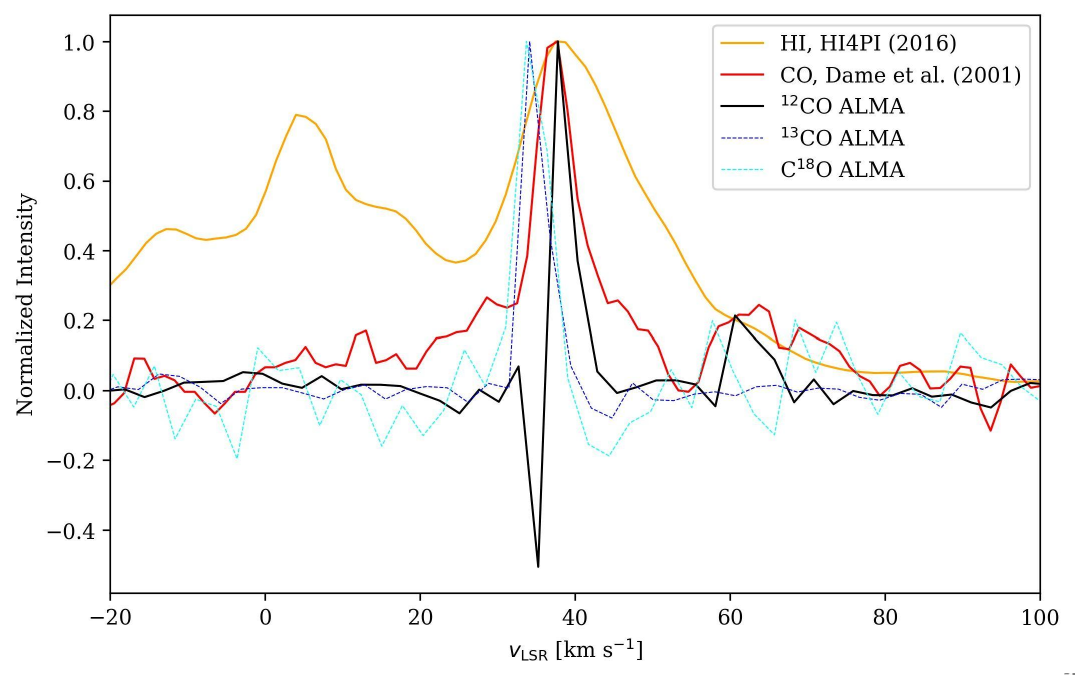}
    \caption{Normalized spectrum of different gas tracers (of the pixel) at the location of WR~112. The yellow line is the spectrum of HI gas from large-scale survey of \citet{HI4PI2016}, while the red line is the spectrum of CO from \citet{Dame2001}. The black, blue and cyan lines correspond to the spectrum of $^{12}$CO, $^{13}$CO and C$^{18}$O from ALMA observations of this work.}
    \label{fig:spectrum_CO}
\end{figure*}

\FloatBarrier
\section{Grain size distribution for the ACH2- and BE-type carbon grains}
\label{Appendix:results_types}

Tables \ref{tbl:grain_stoch_ACH2} and \ref{tbl:grain_stoch_BE} show the MCMC results for the ACH2-type and BE-type carbon grains. 

The results for the BE-type grains are similar to the results for ACAR-type grains that are shown in table \ref{tbl:grain_stoch_ACAR}. On the other hand, for the ACH2-type, besides a considerably good reproduction of the observed SEDs by the two-radii distribution, the other three grain size distributions also achieve an acceptable level of agreement. 

For all three types of carbon grains, the two-grain-radii grain size distribution best reproduces the observed SED. Among models adopting the two-grain-radii distribution, the BE-type grain best reproduces the data, followed by the ACAR-type. However, the differences in $\chi^2$ for the three types of grain are within $8$. The $\chi^2$ distribution has standard deviation $\sqrt{2\nu} \approx 11$, where $\nu$ is the degree of freedom. This suggests that the differences in $\chi^2$ for the three grain types with the two grain radii distribution are not statistically significant ($<1\sigma$). The caveat to this argument is that the uncertainties of the photometric data points in the SED are not precisely known but assumed to be $20\%$, which could cause the $\chi^2$ to be inaccurate. Nevertheless, it is reasonable to conclude that we do not have any evidence to favor any type of carbon grains over the other, from the observed SEDs themselves.

\begin{table*}[h]
\caption{Summary of MCMC fit results for models considering stochastic heating of small grains and the ACH2-type amorphous carbon dust adopted.}
\label{tbl:grain_stoch_ACH2}
\centering
\begin{tabular}{@{}>{\centering\arraybackslash}p{4cm} 
                >{\centering\arraybackslash}p{6.5cm} 
                >{\centering\arraybackslash}p{6.5cm}@{}}
\toprule
\makecell{} 
    & \makecell{One radius} 
    & \makecell{Base-10 log-normal} \\
\midrule
\makecell{Parameter medians} 
    & \makecell{$\Omega_{\text{eff}}^{\text{I}} = 0.093^{+0.003}_{-0.003}~\mathrm{arcsec}^2$ \\[3pt]
    $\Omega_{\text{eff}}^{\text{II}} = 0.109^{+0.004}_{-0.004}~\mathrm{arcsec}^2$ \\[3pt]
    $a_e = 1.87^{+0.030}_{-0.029}~\mathrm{nm}$}
    & \makecell{$\Omega_{\text{eff}}^{\text{I}} = 0.082^{+0.003}_{-0.003}~\mathrm{arcsec}^2$ \\[3pt]
    $\Omega_{\text{eff}}^{\text{II}} = 0.093^{+0.004}_{-0.004}~\mathrm{arcsec}^2$ \\[3pt]
    $a_c = 1.085^{+0.133}_{-0.063}~\mathrm{nm}$ \\[3pt]
    $\sigma_{\log a} = 0.228^{+0.019}_{-0.010}$} \\
\makecell{Chi-squared} 
    & \makecell{$\chi^2 = 91.9$ \\ $\chi^2_\nu = 1.39$} 
    & \makecell{$\chi^2 = 56.4$ \\ $\chi^2_\nu = 0.868$} \\
\midrule
\makecell{} 
    & \makecell{Two radii} 
    & \makecell{Power law} \\
\midrule
\makecell{Parameter medians} 
    & \makecell{$\Omega_{\text{eff}}^{\text{I}} = 0.070^{+0.292}_{-0.061}~\mathrm{arcsec}^2$ \\[3pt]
    $\Omega_{\text{eff}}^{\text{II}} = 0.077^{+0.325}_{-0.068}~\mathrm{arcsec}^2$ \\[3pt]
    $a_1 = 1.57^{+0.26}_{-0.10}~\mathrm{nm}$ \\[3pt]
    $a_2 = 0.078^{+0.097}_{-0.050}$ \SI{}{\micro\meter} \\[3pt]
    $f_2 = (4.30^{+45.3}_{-3.40}) \times 10^{-5}$}
    & \makecell{$\Omega_{\text{eff}}^{\text{I}} = 0.067^{+0.003}_{-0.003}~\mathrm{arcsec}^2$ \\[3pt]
    $\Omega_{\text{eff}}^{\text{II}} = 0.075^{+0.005}_{-0.004}~\mathrm{arcsec}^2$ \\[3pt]
    $a_l = 1.035^{+0.053}_{-0.024}~\mathrm{nm}$ \\[3pt]
    $a_u = 0.707^{+3.32}_{-0.612}$ \SI{}{\micro\meter} \\[3pt]
    $\beta = -4.283^{+0.064}_{-0.122}$} \\
\makecell{Chi-squared} 
    & \makecell{$\chi^2 = 35.1$ \\ $\chi^2_\nu = 0.548$} 
    & \makecell{$\chi^2 = 41.2$ \\ $\chi^2_\nu = 0.644$} \\
\bottomrule
\end{tabular}
\end{table*}

\begin{table*}[h]
\caption{Summary of MCMC fit results for models considering stochastic heating of small grains and the BE-type amorphous carbon dust adopted.}
\label{tbl:grain_stoch_BE}
\centering
\begin{tabular}{@{}>{\centering\arraybackslash}p{4cm} 
                >{\centering\arraybackslash}p{6.5cm} 
                >{\centering\arraybackslash}p{6.5cm}@{}}
\toprule
\makecell{} 
    & \makecell{One radius} 
    & \makecell{Base-10 log-normal} \\
\midrule
\makecell{Parameter medians} 
    & \makecell{$\Omega_{\text{eff}}^{\text{I}} = 0.016^{+0.002}_{-0.001}~\mathrm{arcsec}^2$ \\[3pt]
    $\Omega_{\text{eff}}^{\text{II}} = 0.020^{+0.003}_{-0.001}~\mathrm{arcsec}^2$ \\[3pt]
    $a_e = 8.67^{+1.65}_{-1.59}~\mathrm{nm}$}
    & \makecell{$\Omega_{\text{eff}}^{\text{I}} = 0.0612^{+0.0023}_{-0.0026}~\mathrm{arcsec}^2$ \\[3pt]
    $\Omega_{\text{eff}}^{\text{II}} = 0.0743^{+0.0031}_{-0.0036}~\mathrm{arcsec}^2$ \\[3pt]
    $a_c = 2.780^{+0.0750}_{-0.131}~\mathrm{nm}$ \\[3pt]
    $\sigma_{\log a} = 0.0354^{+0.0237}_{-0.0149}$} \\
\makecell{Chi-squared} 
    & \makecell{$\chi^2 = 256.4$ \\ $\chi^2_\nu = 3.89$} 
    & \makecell{$\chi^2 = 198.7$ \\ $\chi^2_\nu = 3.06$} \\
\midrule
\makecell{} 
    & \makecell{Two radii} 
    & \makecell{Power law} \\
\midrule
\makecell{Parameter medians} 
    & \makecell{$\Omega_{\text{eff}}^{\text{I}} = 0.052^{+0.032}_{-0.018}~\mathrm{arcsec}^2$ \\[3pt]
    $\Omega_{\text{eff}}^{\text{II}} = 0.056^{+0.035}_{-0.019}~\mathrm{arcsec}^2$ \\[3pt]
    $a_1 = 1.373^{+0.195}_{-0.172}~\mathrm{nm}$ \\[3pt]
    $a_2 = 0.182^{+0.0884}_{-0.0427}~\si{\micro\meter}$ \\[3pt]
    $f_2 = 1.33^{+0.65}_{-0.50} \times 10^{-5}$}
    & \makecell{$\Omega_{\text{eff}}^{\text{I}} = 0.034^{+0.003}_{-0.003}~\mathrm{arcsec}^2$ \\[3pt]
    $\Omega_{\text{eff}}^{\text{II}} = 0.040^{+0.004}_{-0.003}~\mathrm{arcsec}^2$ \\[3pt]
    $a_l = 1.027^{+0.0245}_{-0.0191}~\mathrm{nm}$ \\[3pt]
    $a_u = 2.925^{+3.84}_{-1.84}~\si{\micro\meter}$ \\[3pt]
    $\beta = -3.715^{+0.077}_{-0.099}$} \\
\makecell{Chi-squared} 
    & \makecell{$\chi^2 = 27.7$ \\ $\chi^2_\nu = 0.434$} 
    & \makecell{$\chi^2 = 147.2$ \\ $\chi^2_\nu = 2.30$} \\
\bottomrule
\end{tabular}
\end{table*}

\FloatBarrier
\section{Grain size distribution for the 16th and 84th percentile stellar spectra}
\label{Appendix:results_varyspectra}

\begin{table*}[h]
\caption{Summary of MCMC fit results for models with stochastic heating for ACAR-type grains, and the 84th percentile stellar spectra.}
\label{tbl:grain_stoch_p84}
\centering
\begin{tabular}{@{}>{\centering\arraybackslash}p{4cm} 
                >{\centering\arraybackslash}p{6.5cm} 
                >{\centering\arraybackslash}p{6.5cm}@{}}
\toprule
\makecell{} 
    & \makecell{One radius} 
    & \makecell{Base-10 log-normal} \\
\midrule
\makecell{Parameter medians} 
    & \makecell{$\Omega_{\text{eff}}^{\text{I}} = 0.0045^{+0.0002}_{-0.0002}~\mathrm{arcsec}^2$ \\[3pt]
    $\Omega_{\text{eff}}^{\text{II}} = 0.0054^{+0.0003}_{-0.0002}~\mathrm{arcsec}^2$ \\[3pt]
    $a_e = 33.7^{+3.5}_{-3.9}~\mathrm{nm}$}
    & \makecell{$\Omega_{\text{eff}}^{\text{I}} = 0.0060^{+0.0003}_{-0.0003}~\mathrm{arcsec}^2$ \\[3pt]
    $\Omega_{\text{eff}}^{\text{II}} = 0.0069^{+0.0004}_{-0.0003}~\mathrm{arcsec}^2$ \\[3pt]
    $a_c = 1.19^{+0.49}_{-0.15}~\mathrm{nm}$ \\[3pt]
    $\sigma_{\log a} = 0.580^{+0.018}_{-0.031}$} \\
\makecell{Chi-squared} 
    & \makecell{$\chi^2 = 243.0$ \\ $\chi^2_\nu = 3.68$} 
    & \makecell{$\chi^2 = 192.2$ \\ $\chi^2_\nu = 2.96$} \\
\midrule
\makecell{} 
    & \makecell{Two radii} 
    & \makecell{Power law} \\
\midrule
\makecell{Parameter medians} 
    & \makecell{$\Omega_{\text{eff}}^{\text{I}} = 0.0186^{+0.0075}_{-0.0080}~\mathrm{arcsec}^2$ \\[3pt]
    $\Omega_{\text{eff}}^{\text{II}} = 0.0199^{+0.0077}_{-0.0083}~\mathrm{arcsec}^2$ \\[3pt]
    $a_1 = 1.125^{+0.535}_{-0.090}~\mathrm{nm}$ \\[3pt]
    $a_2 = 0.204^{+0.0643}_{-0.0322}$ \SI{}{\micro\meter} \\[3pt]
    $f_2 = 1.32^{+1.41}_{-0.33} \times 10^{-5}$}
    & \makecell{$\Omega_{\text{eff}}^{\text{I}} = 0.0091^{+0.0006}_{-0.0006}~\mathrm{arcsec}^2$ \\[3pt]
    $\Omega_{\text{eff}}^{\text{II}} = 0.0103^{+0.0006}_{-0.0007}~\mathrm{arcsec}^2$ \\[3pt]
    $a_l = 1.087^{+0.167}_{-0.063}~\mathrm{nm}$ \\[3pt]
    $a_u = 5.02^{+2.92}_{-2.03}$ \SI{}{\micro\meter} \\[3pt]
    $\beta = -3.04^{+0.04}_{-0.04}$} \\
\makecell{Chi-squared} 
    & \makecell{$\chi^2 = 34.5$ \\ $\chi^2_\nu = 0.538$} 
    & \makecell{$\chi^2 = 129.1$ \\ $\chi^2_\nu = 2.02$} \\
\bottomrule
\end{tabular}
\end{table*}

\begin{table*}[h]
\caption{Summary of MCMC fit results for models with stochastic heating for ACAR-type grains, and the 16th percentile stellar spectra.}
\label{tbl:grain_stoch_p16}
\centering
\begin{tabular}{@{}>{\centering\arraybackslash}p{4cm} 
                >{\centering\arraybackslash}p{6.5cm} 
                >{\centering\arraybackslash}p{6.5cm}@{}}
\toprule
\makecell{} 
    & \makecell{One radius} 
    & \makecell{Base-10 log-normal} \\
\midrule
\makecell{Parameter medians} 
    & \makecell{$\Omega_{\text{eff}}^{\text{I}} = 0.150^{+0.006}_{-0.006}~\mathrm{arcsec}^2$ \\[3pt]
    $\Omega_{\text{eff}}^{\text{II}} = 0.177^{+0.008}_{-0.009}~\mathrm{arcsec}^2$ \\[3pt]
    $a_e = 2.089^{+0.060}_{-0.068}~\mathrm{nm}$}
    & \makecell{$\Omega_{\text{eff}}^{\text{I}} = 0.131^{+0.006}_{-0.006}~\mathrm{arcsec}^2$ \\[3pt]
    $\Omega_{\text{eff}}^{\text{II}} = 0.153^{+0.008}_{-0.008}~\mathrm{arcsec}^2$ \\[3pt]
    $a_c = 1.140^{+0.208}_{-0.104}~\mathrm{nm}$ \\[3pt]
    $\sigma_{\log a} = 0.233^{+0.0156}_{-0.0282}$} \\
\makecell{Chi-squared} 
    & \makecell{$\chi^2 = 131.0$ \\ $\chi^2_\nu = 1.99$} 
    & \makecell{$\chi^2 = 110.0$ \\ $\chi^2_\nu = 1.70$} \\
\midrule
\makecell{} 
    & \makecell{Two radii} 
    & \makecell{Power law} \\
\midrule
\makecell{Parameter medians} 
    & \makecell{$\Omega_{\text{eff}}^{\text{I}} = 0.1081^{+0.1226}_{-0.0431}~\mathrm{arcsec}^2$ \\[3pt]
    $\Omega_{\text{eff}}^{\text{II}} = 0.1150^{+0.1200}_{-0.0458}~\mathrm{arcsec}^2$ \\[3pt]
    $a_1 = 1.246^{+0.279}_{-0.190}~\mathrm{nm}$ \\[3pt]
    $a_2 = 0.0777^{+0.0450}_{-0.0199}$ \SI{}{\micro\meter} \\[3pt]
    $f_2 = 4.69^{+3.05}_{-2.53} \times 10^{-5}$}
    & \makecell{$\Omega_{\text{eff}}^{\text{I}} = 0.1081^{+0.0054}_{-0.0059}~\mathrm{arcsec}^2$ \\[3pt]
    $\Omega_{\text{eff}}^{\text{II}} = 0.1242^{+0.0071}_{-0.0083}~\mathrm{arcsec}^2$ \\[3pt]
    $a_l = 1.028^{+0.0336}_{-0.0220}~\mathrm{nm}$ \\[3pt]
    $a_u = 1.541^{+4.63}_{-1.23}$ \SI{}{\micro\meter} \\[3pt]
    $\beta = -4.22^{+0.07}_{-0.10}$} \\
\makecell{Chi-squared} 
    & \makecell{$\chi^2 = 33.8$ \\ $\chi^2_\nu = 0.528$} 
    & \makecell{$\chi^2 = 87.2$ \\ $\chi^2_\nu = 1.36$} \\
\bottomrule
\end{tabular}
\end{table*}

\newpage

\section{Model SED}
\subsection{Model SEDs with temperatures are free parameters}
\label{sec:figures_SED_Tfree}
\FloatBarrier
\begin{figure*}[h]
    \centering
    \includegraphics[width=0.8\linewidth]{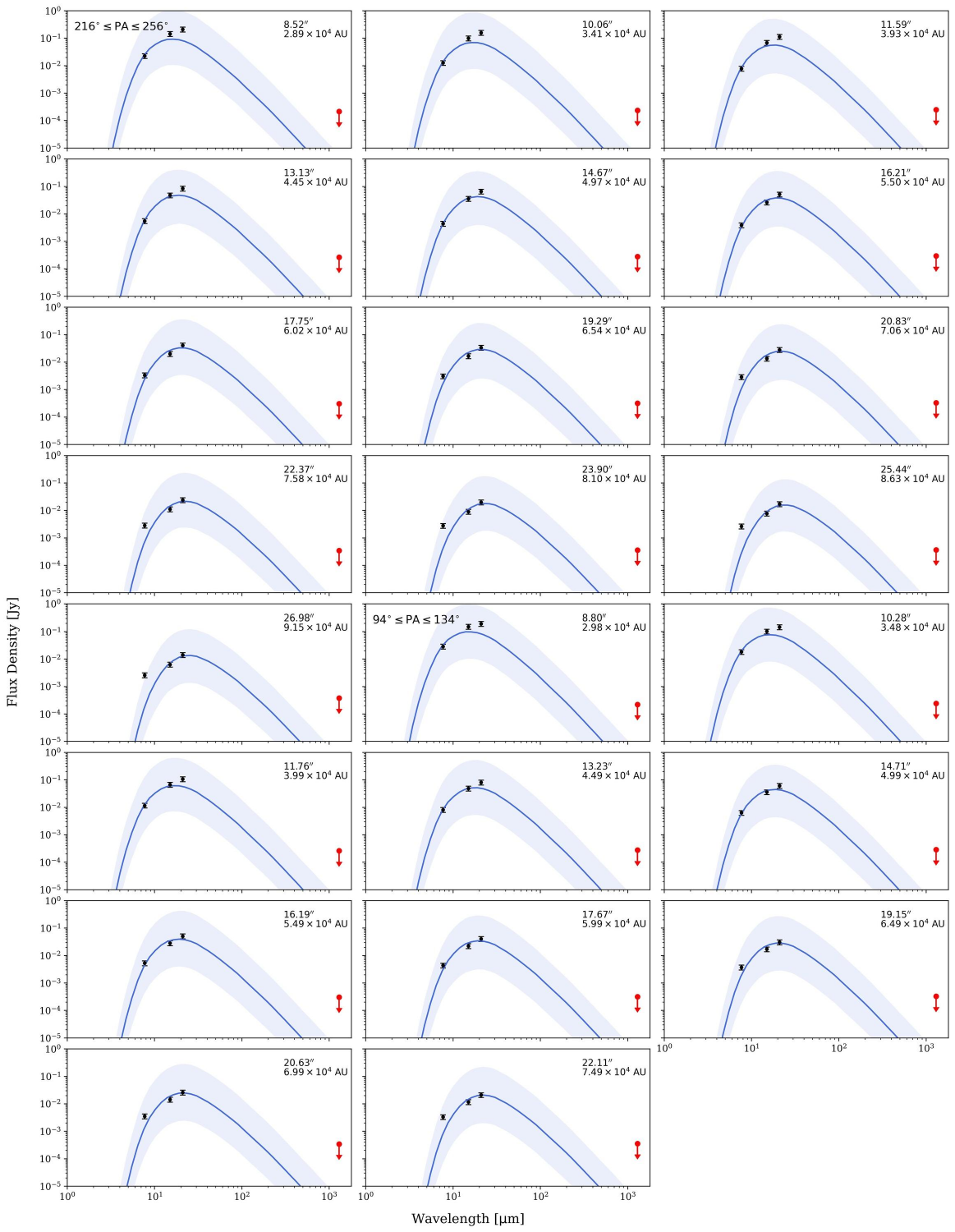}
    \caption{The SEDs of the 23 apertures in the first sector (first 13 panels) and the second sector (last 10 panels). The black data points are from JWST observations, while the red upper limit is from the ALMA continuum image. The blue line shows the model SEDs predicted by the modified blackbody model described in Section \ref{sec:Tfree} with effective solid angle, grain size and temperatures all as free parameters. The light blue region indicates the $68\%$ credible interval. The deprojected distance to each aperture in arcsecond and in au is labeled at the top right corner of each panel. }
    \label{fig:SED_Tfree}
\end{figure*}

\subsection{Model SEDs with equilibrium temperature grids}
\FloatBarrier
\label{sec:figures_SED_Tgrid}
\begin{figure*}[h]
    \centering
    \includegraphics[width=0.85\linewidth]{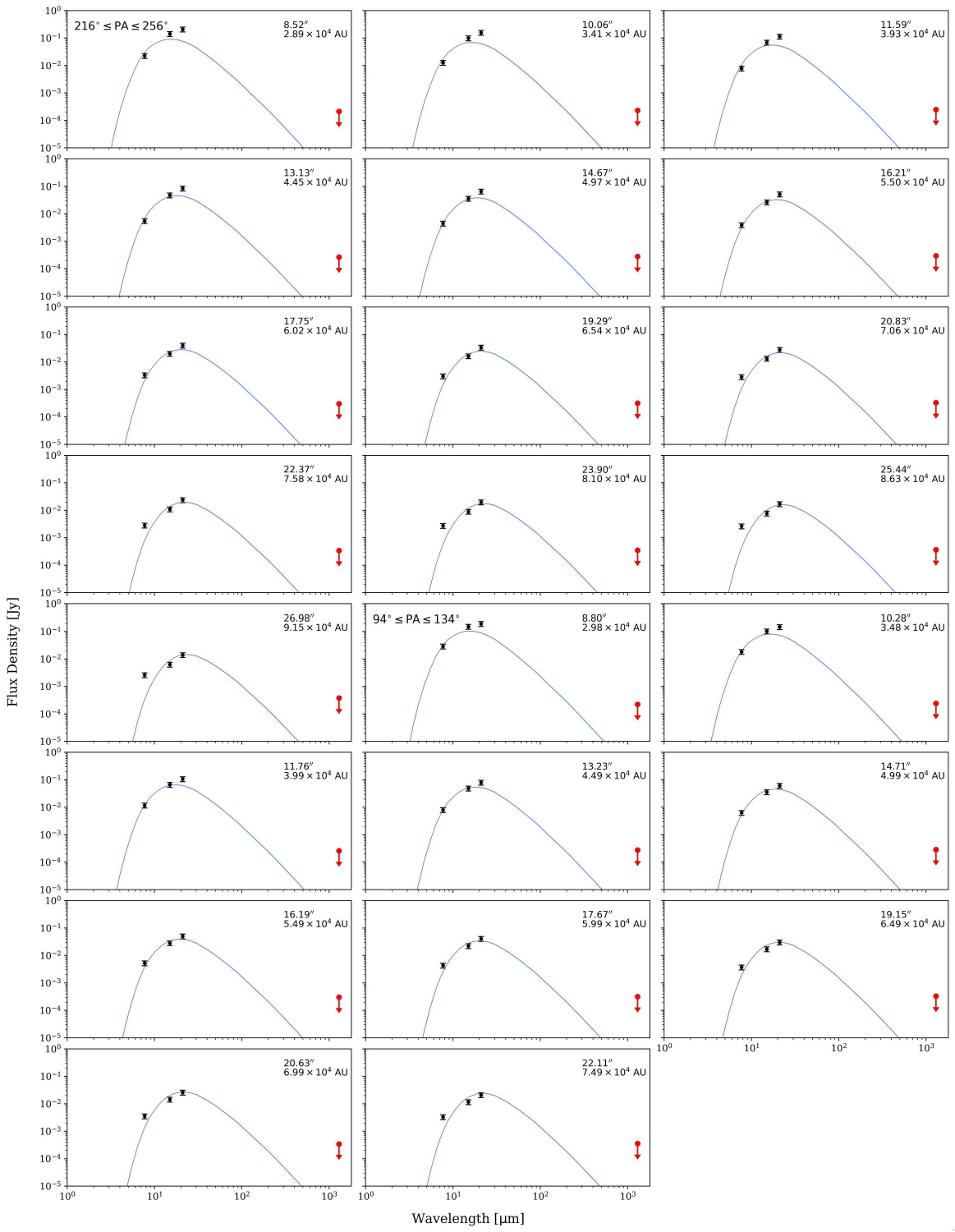}
    \caption{Same as Figure \ref{fig:SED_Tfree}, except for the emission model described in Section \ref{sec:tempgrid} using the one-radius grain size distribution.}
    \label{fig:SED_Tgrid_S}
\end{figure*}

\begin{figure*}
    \centering
    \includegraphics[width=0.9\linewidth]{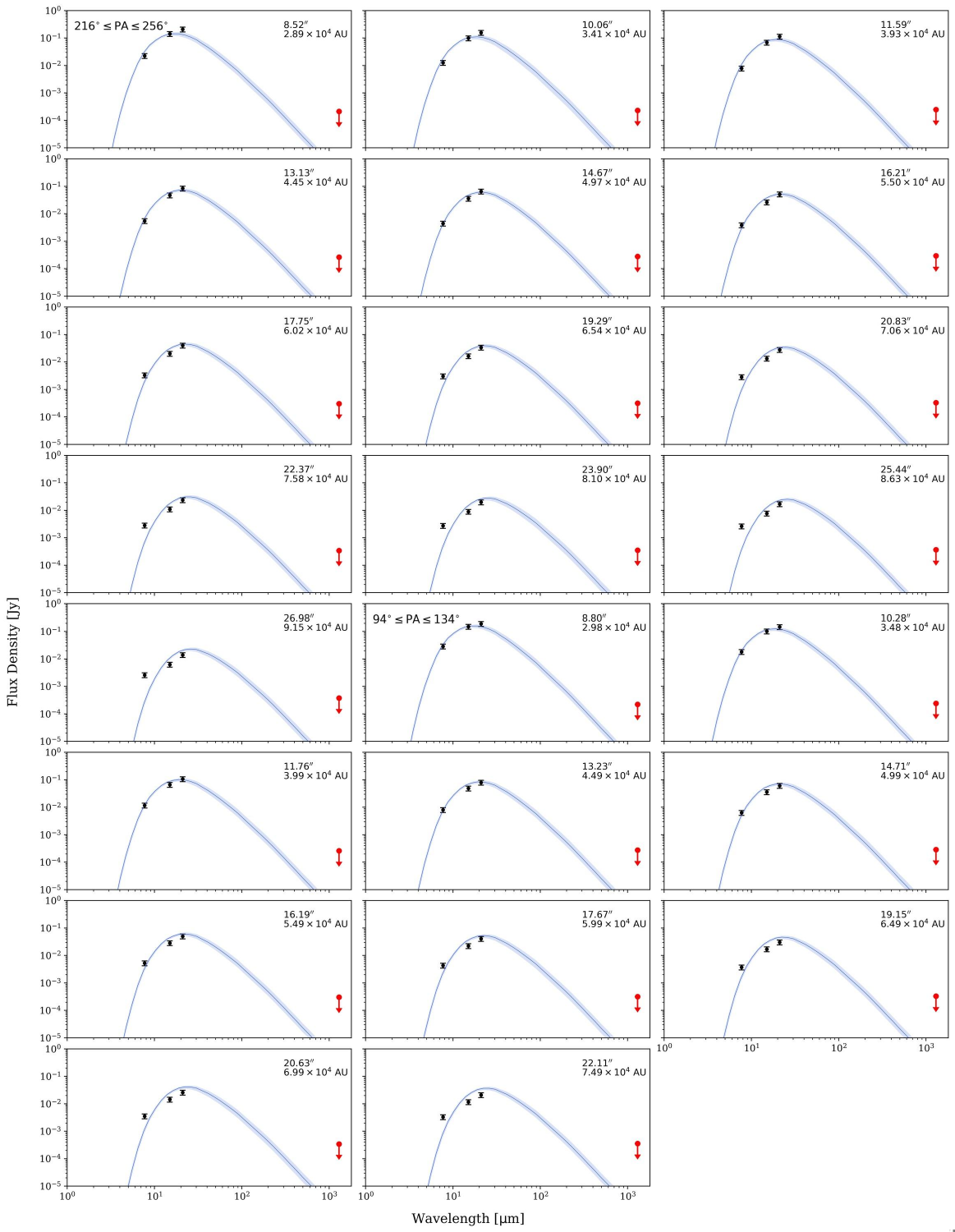}
    \caption{Same as Figure \ref{fig:SED_Tfree}, except for the emission model described in Section \ref{sec:tempgrid} using the base-10 log-normal grain size distribution.}
    \label{fig:SED_Tgrid_G}
\end{figure*}

\begin{figure*}
    \centering
    \includegraphics[width=0.9\linewidth]{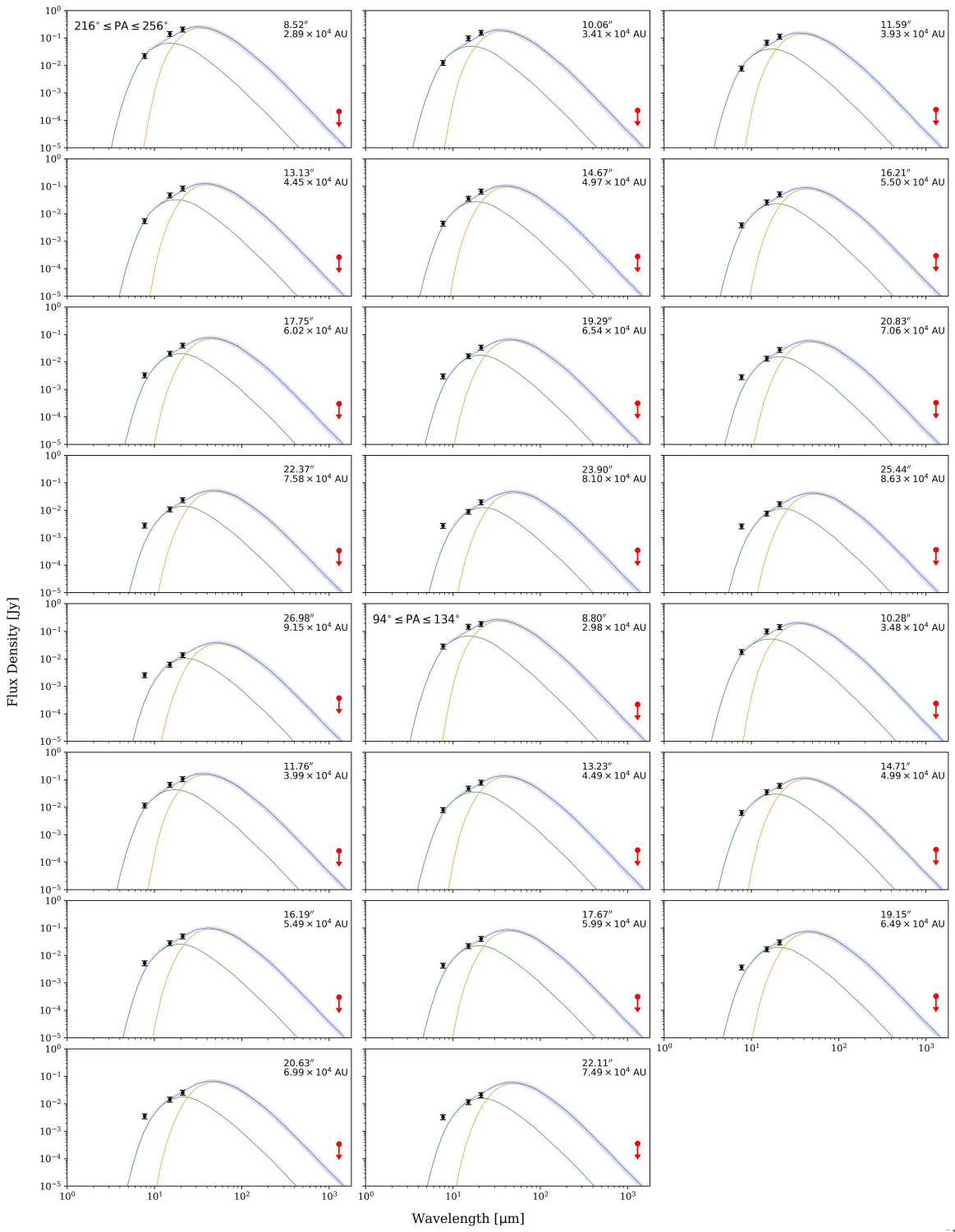}
    \caption{Same as Figure \ref{fig:SED_Tfree}, except for the emission model described in Section \ref{sec:tempgrid} using the two-radii grain size distribution. The emissions from each population of grains are also shown, with smaller grains in green and larger grains in yellow.}
    \label{fig:SED_Tgrid_D}
\end{figure*}

\begin{figure*}
    \centering
    \includegraphics[width=0.9\linewidth]{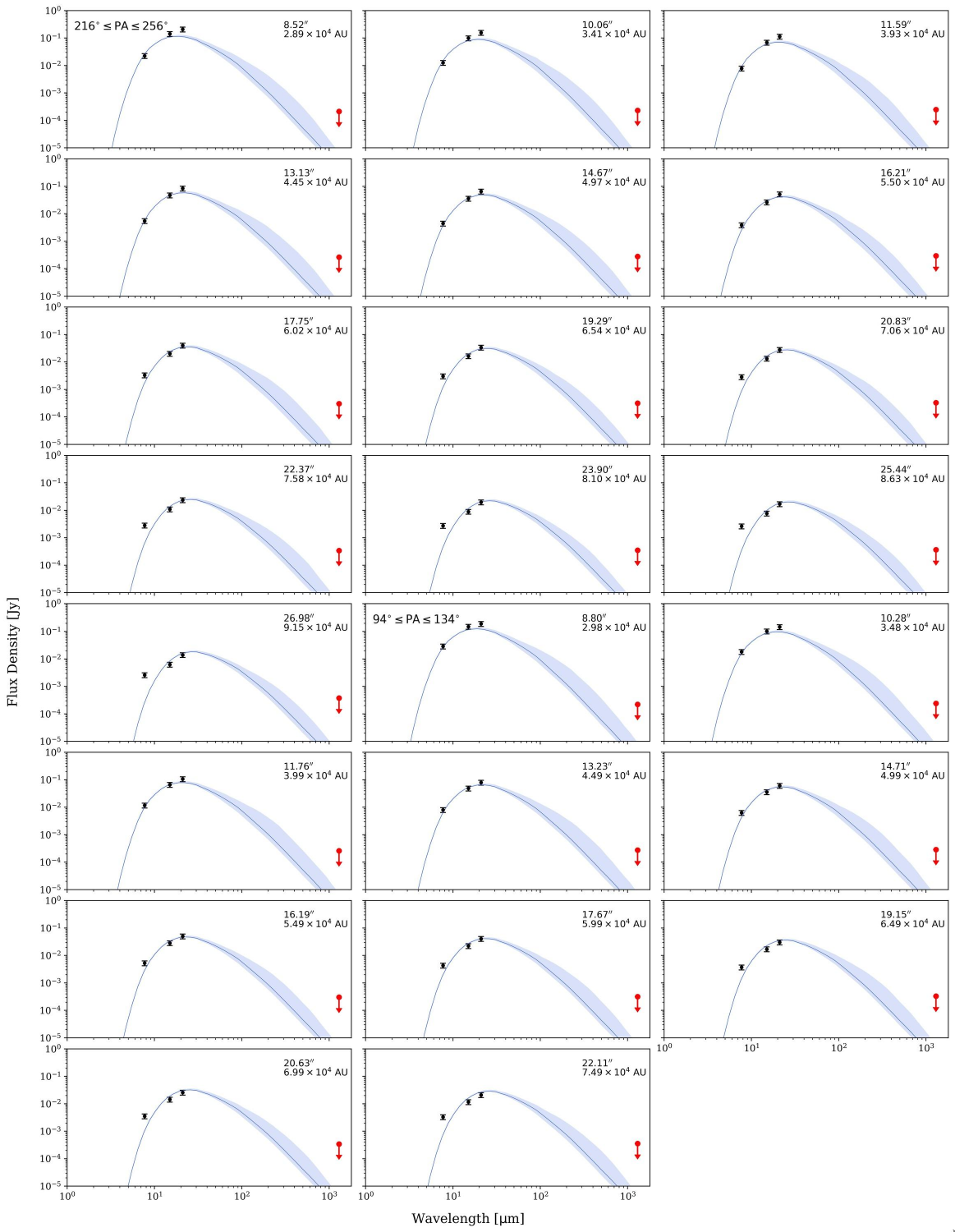}
    \caption{Same as Figure \ref{fig:SED_Tfree}, except for the emission model described in Section \ref{sec:tempgrid} using the power law grain size distribution.}
    \label{fig:SED_Tgrid_P}
\end{figure*}

\FloatBarrier
\subsection{Model SEDs with temperature distributions computed by DustEM for the ACAR-type amorphous carbon dust}
\label{sec:figures_SED_stoch}

\begin{figure*}[h]
    \centering
    \includegraphics[width=0.85\linewidth]{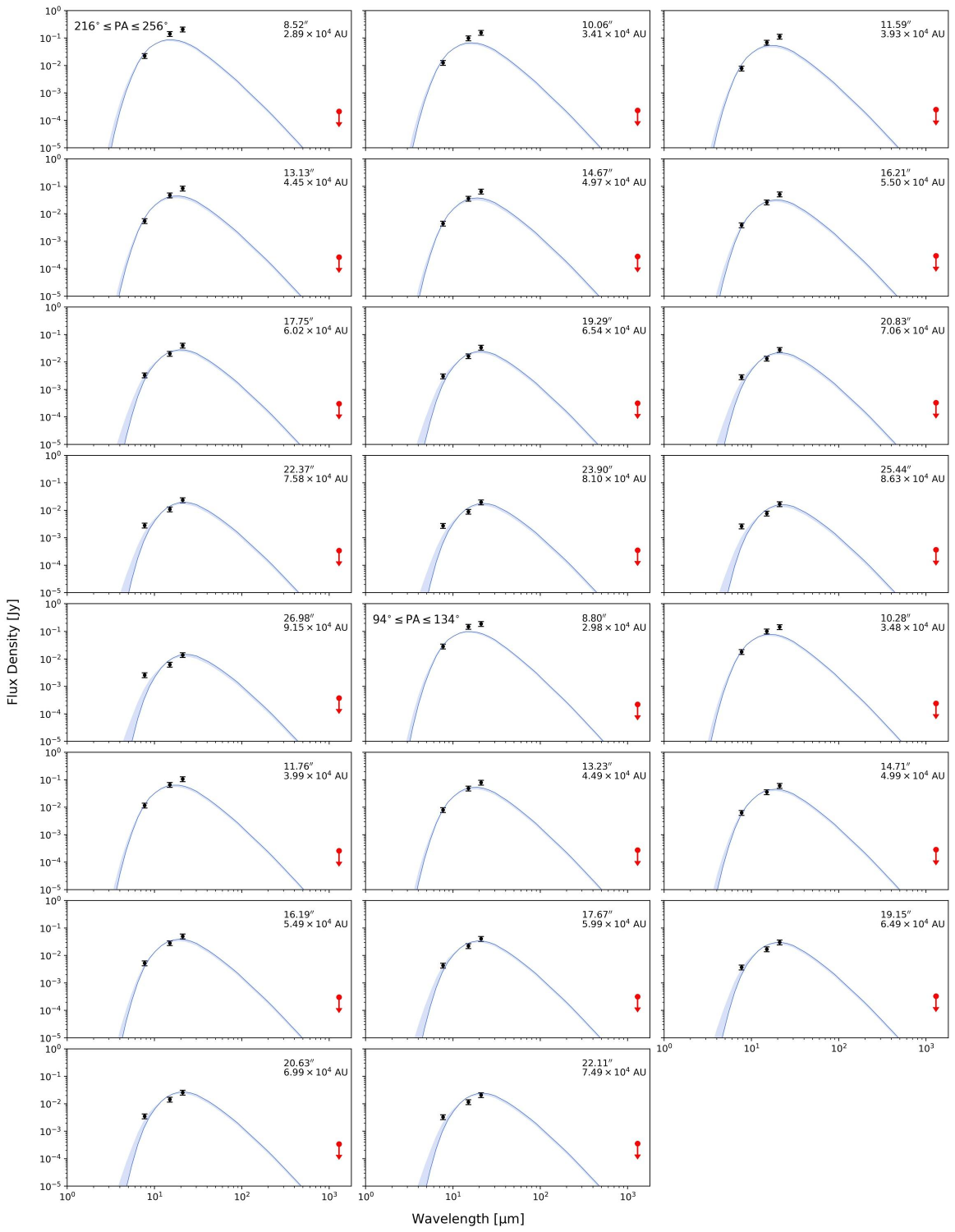}
    \caption{Same as Figure \ref{fig:SED_Tfree}, except for the emission model with stochastic heating described in Section \ref{sec:stochastic} for ACAR-type grains using the single-radius grain size distribution.}
    \label{fig:SED_DM_ACAR_S}
\end{figure*}

\begin{figure*}[h]
    \centering
    \includegraphics[width=0.9\linewidth]{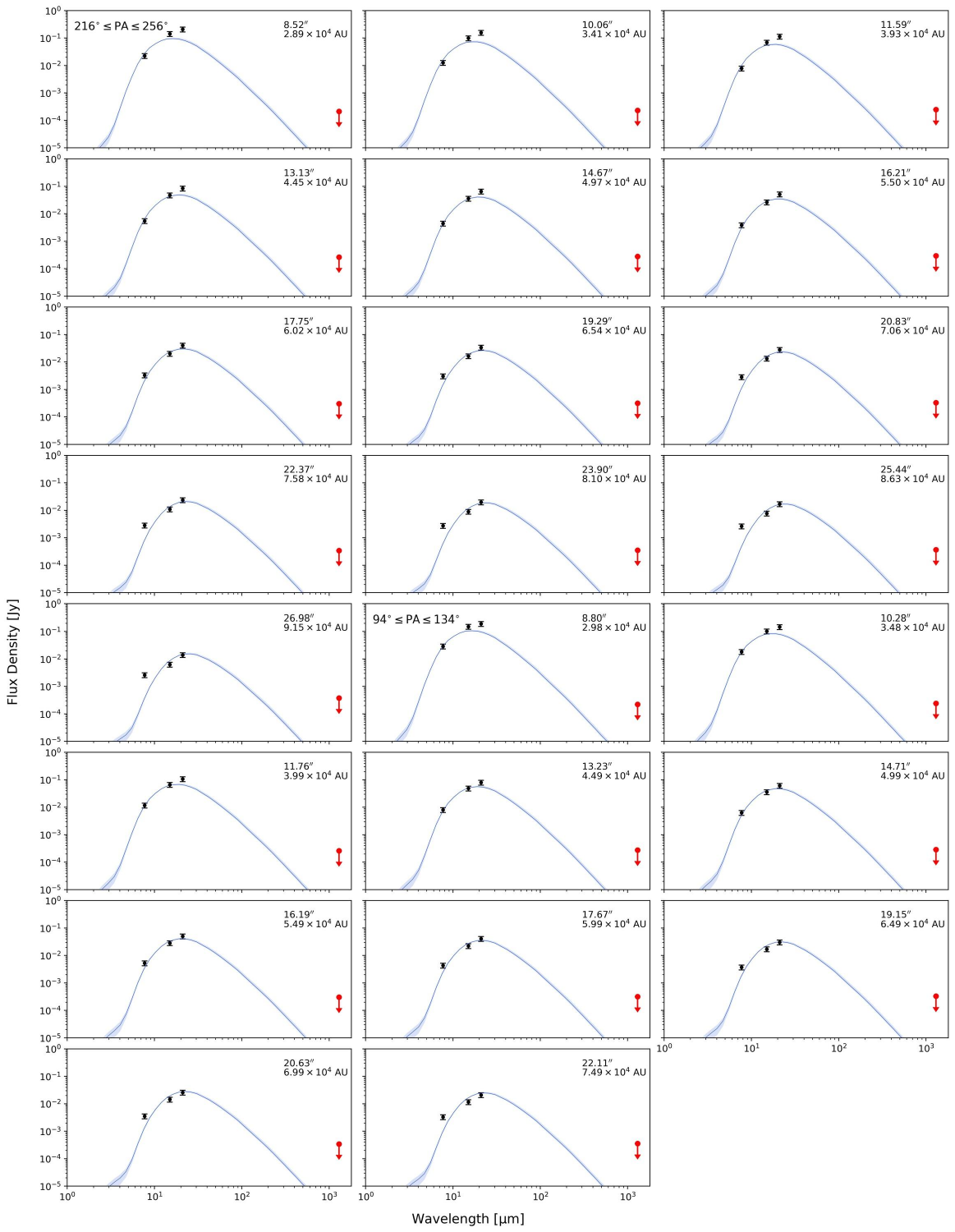}
    \caption{Same as Figure \ref{fig:SED_Tfree}, except for the emission model with stochastic heating described in Section \ref{sec:stochastic} for ACAR-type grains using the base-10 log-normal grain size distribution.}
    \label{fig:SED_DM_ACAR_G}
\end{figure*}

\begin{figure*}[h]
    \centering
    \includegraphics[width=0.9\linewidth]{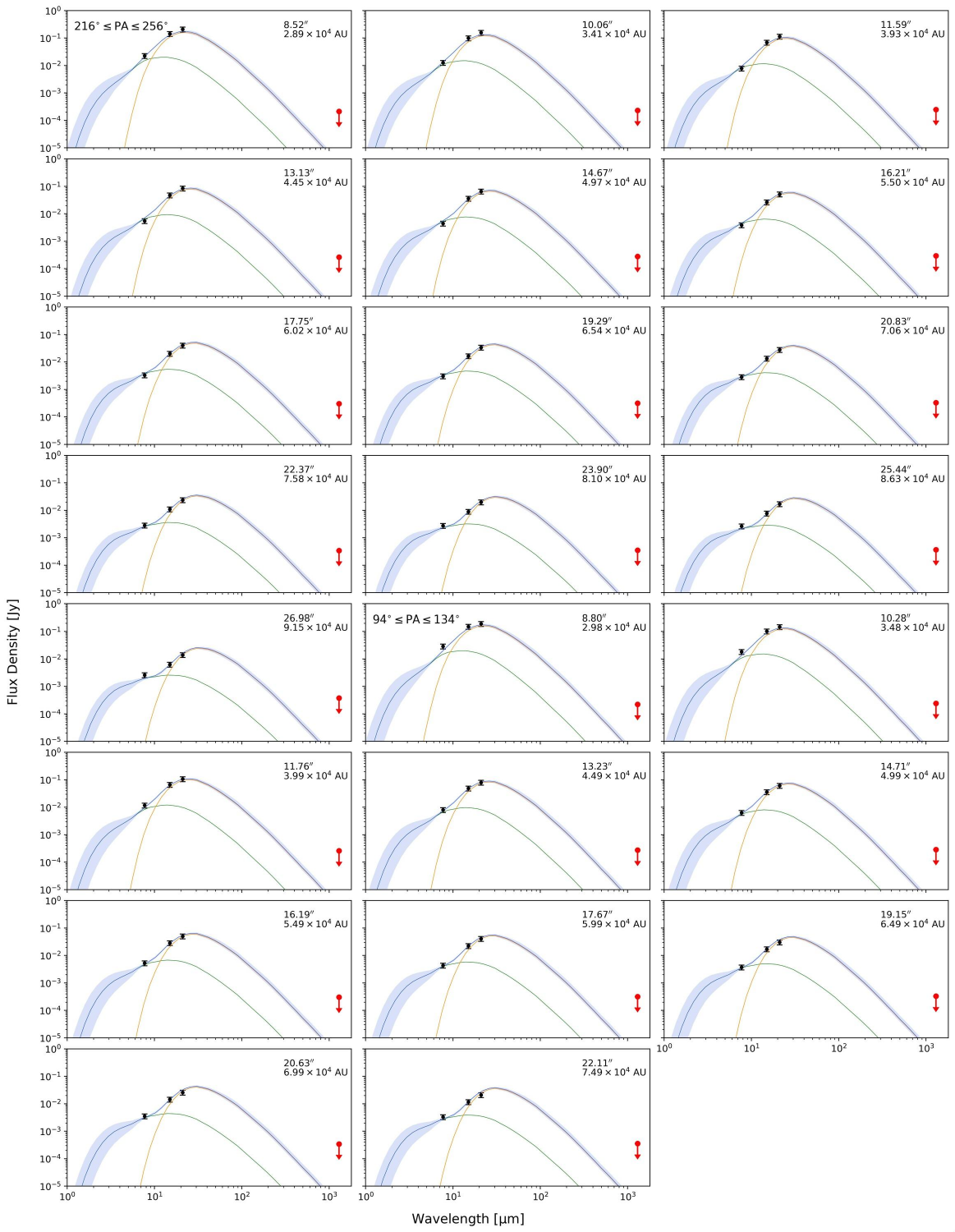}
    \caption{Same as Figure \ref{fig:SED_Tfree}, except for the emission model with stochastic heating described in Section \ref{sec:stochastic} for ACAR-type grains using the two-radii grain size distribution. The emissions from each population of grains are also shown, with smaller grains in green and larger grains in yellow.}
    \label{fig:SED_DM_ACAR_D}
\end{figure*}

\begin{figure*}[h]
    \centering
    \includegraphics[width=0.9\linewidth]{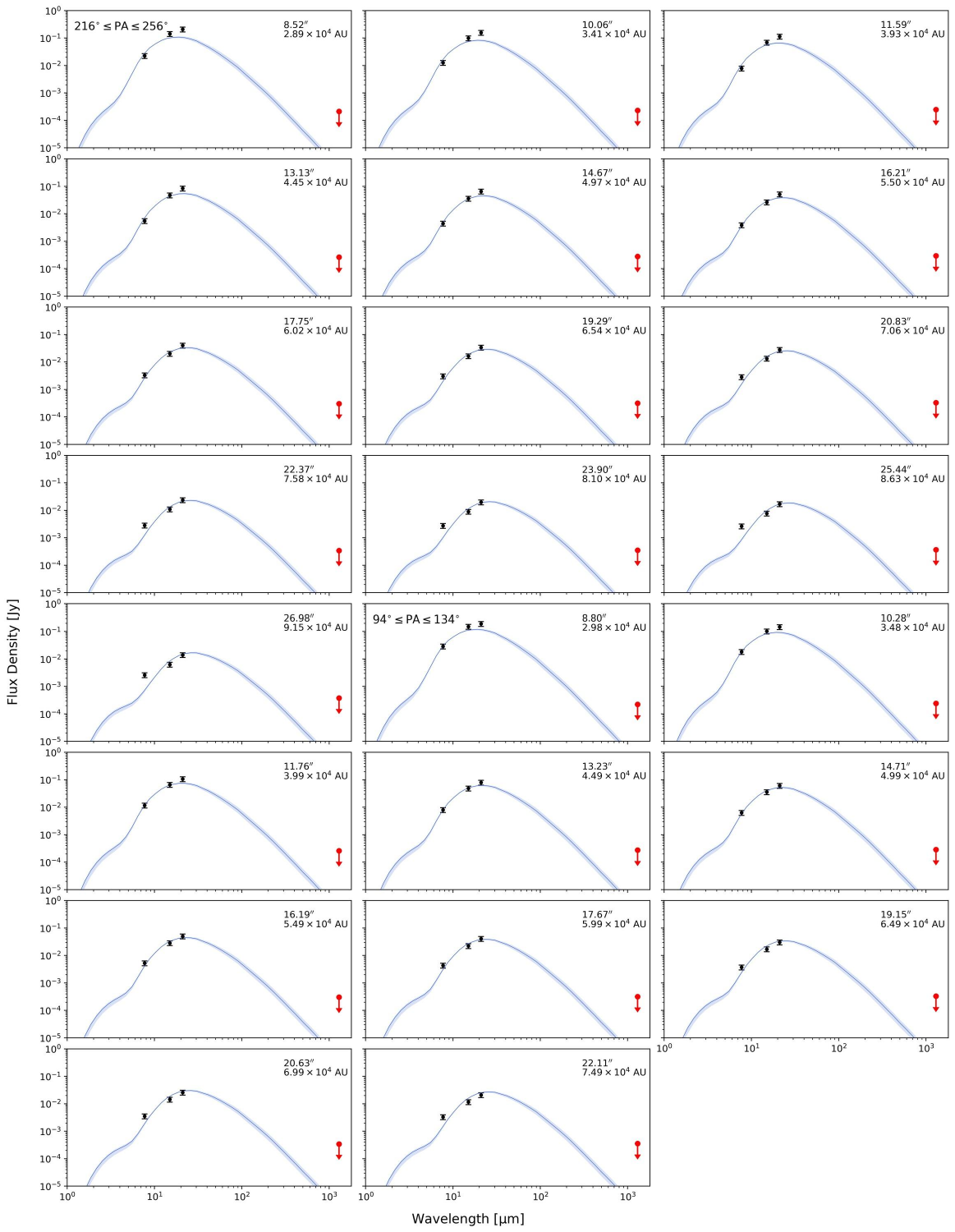}
    \caption{Same as Figure \ref{fig:SED_Tfree}, except for the emission model with stochastic heating described in Section \ref{sec:stochastic} for ACAR-type grains using the power law grain size distribution.}
    \label{fig:SED_DM_ACAR_P}
\end{figure*}

\FloatBarrier

\FloatBarrier
\section{Interpolation of probability density function}
\label{sec:optimal_transport}
It is not uncommon in physics for probability density function (PDF) to depend on some external parameters. For example, in this paper, the PDF for temperature depends on grain sizes, stellar photon flux and other conditions. 
During computation, it is often necessary to "interpolate" PDF over the external parameters as the external parameter space is discretely sampled. Such an interpolation can be achieved by considering smooth transformation of one PDF into another over the parameter space.

If the PDF only depends on one external variable, then such transformation is rather trivial. Let the external variable be $a$ and the random variable of the PDF be $x$. Then, the PDF can be denoted as $p(x \mid a)$. The cumulative distribution function (CDF) is defined as:
\begin{equation}
    \mathrm{CDF}(x \mid a) = \int_{-\infty}^{x} p(x^{\prime} \mid a) \, dx^{\prime}.
\end{equation}
Since CDF is monotonically increasing, its inverse function is well defined and is referred to as the quantile function:
\begin{equation}
    Q(y \mid a) = [\mathrm{CDF}(x \mid a)]^{-1},
\end{equation}
where $y$ is the cumulative probability. Then, the smooth transformation between PDFs is given by the linear interpolation of the quantile function over the external parameter space:
\begin{equation}
    Q(y \mid a) = \frac{a-a_1}{a_2-a_1} Q(y \mid a_1) + \frac{a_2-a}{a_2-a_1} Q(y \mid a_2),
\end{equation}
where $a_1 < a < a_2$. 

However, if the PDF depends on two or more external variables, i.e., $p(x \mid a, b, \cdots)$, then the smooth transformation between PDFs becomes much more complicated. It becomes a problem of optimal transport in mathematics. Fortunately, a PDF of temperature depending on one external parameter, grain size, is enough for this work.



\end{document}